\ifdefined\buildpart\RequirePackage{xr-hyper}\fi
\documentclass[pdflatex,sn-nature]{sn-jnl}

\usepackage{graphicx}
\usepackage{amsmath,amssymb,bm}
\usepackage{booktabs}
\usepackage{array}
\usepackage{placeins}
\usepackage{tikz}
\usetikzlibrary{arrows.meta,positioning,calc}
\usepackage{url}
\graphicspath{{figs/}}

\newcommand{\panel}[1]{{\sffamily\bfseries #1}}
\newcommand{\Dz}{D0}
\newcommand{\Do}{D1}
\newcommand{\Dt}{D2}
\newcommand{\Dh}{D3}
\newcommand{\dOZ}{D1$-$D0}
\newcommand{\dTO}{D2$-$D1}
\newcommand{\dHT}{D3$-$D2}
\unnumbered

\newif\ifmainpart\mainparttrue
\newif\ifsipart\siparttrue
\ifdefined\buildpart
\def\partmain{main}\def\partsi{si}
\ifx\buildpart\partmain\sipartfalse\fi
\ifx\buildpart\partsi\mainpartfalse\fi
\fi

\begin{document}
\ifmainpart

\title[The shape of individual differences in judgment]{The shape of individual differences in judgment: shared axes, personal weights, and when their spread crosses the consensus}

\author*[1]{\fnm{Hiroshi} \sur{Higashi}}\email{hgshrs@kansai-u.ac.jp}
\affil*[1]{\orgdiv{Faculty of Engineering Science}, \orgname{Kansai University}, \orgaddress{\city{Suita}, \state{Osaka}, \country{Japan}}}

\abstract{People differ in how they judge faces, artworks, photographs and jokes.
Previous work measured the size of these differences; we ask what shape they take.
We wrote three candidate shapes as nested generative models (same-sign reweighting of shared dimensions, a linear transform that allows reversal, and stimulus-specific offsets) and compared how well they predict people unseen in training, using ten public datasets (about 20,000 raters).
The models recovered a person's judgment space from 10--20 judgments and predicted judgments of unseen stimuli.
For faces, reweighting was sufficient, and weights opposing the population were not statistically detected.
For artworks, weights spread continuously along the shared abstract--figurative direction, and 28\% of raters fell on the opposite side from the population.
This tail was much smaller when a different rater sample judged the same artworks.
Individual differences thus differ in shape as well as size, and the shape depends on domain and raters.}

\maketitle

Even when people look at the same face, they differ in whether it looks trustworthy.
How large such individual differences are has been measured repeatedly.
For some face impressions, such as trustworthiness, the variance arising from perceivers rivals the variance arising from faces\cite{hehman2017,xie2019}, about half of the variance in attractiveness judgments is attributable to personal taste\cite{honekopp2006}, and these differences are shaped mainly by experience rather than by genes\cite{germine2015,sutherland2020}.
Preferences are also shared less for artworks and abstract images than for faces and landscapes\cite{vessel2010,vessel2018,leder2016,bignardi2024,schepman2015}.
All of these findings describe how much individual difference there is, as a proportion of variance.
However, even when the size of individual differences is the same, what they consist of can differ.
We consider three possibilities (Fig.~\ref{fig:concept}a--c).

First, individual differences may be differences in emphasis on a shared representation (Fig.~\ref{fig:concept}a).
Everyone agrees on which end of each evaluative axis is valued more highly, and people differ only in how much weight they give to each axis.
Each person's evaluation is then the population mean strengthened or weakened axis by axis; we call this the reweighting hypothesis.
Second, the distribution of weights on an axis may spread beyond the population consensus to the opposite side (Fig.~\ref{fig:concept}b).
For example, most people may value figurative works more highly while some value abstract works more highly, with no gap between the two.
For people in this tail, the population mean does not even indicate which end they value more highly.
This shape requires a person-specific linear transform of the shared axes (sign reversal and mixing between axes), so we call it the linear-transform hypothesis.
Third, there may be stimulus-specific personal evaluations that cannot be predicted from a stimulus's position on the shared axes (Fig.~\ref{fig:concept}c).
In that case, no low-dimensional shared representation can describe the individual differences.

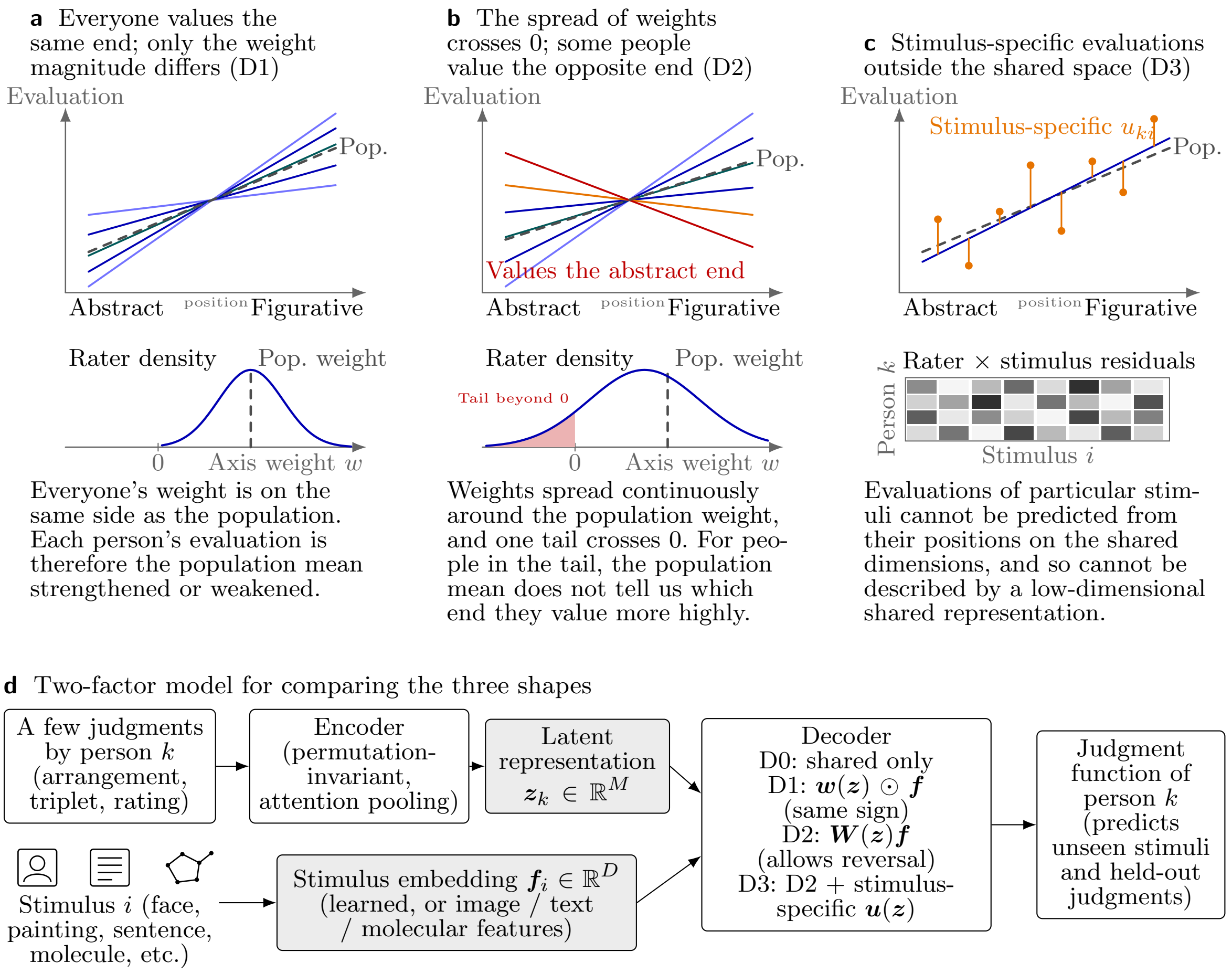
\begin{figure}[tbp]
\centering
\resizebox{\textwidth}{!}{%
\begin{tikzpicture}[>=Latex, font=\scriptsize, x=1cm, y=1cm, line cap=round,
  rater/.style={line width=0.7pt},
  pop/.style={line width=0.9pt, dashed, black!70},
  axis/.style={->, line width=0.5pt, black!60},
  ttl/.style={anchor=south west, inner sep=1pt, align=left, text width=4.6cm},
  note/.style={anchor=north west, inner sep=1pt, align=left, text width=4.6cm}]
\colorlet{c1}{blue!70!black}
\colorlet{c2}{blue!55}
\colorlet{c3}{teal!70!black}
\colorlet{c4}{orange!90!black}
\colorlet{c5}{red!75!black}
\newcommand{\topaxes}{%
  \draw[axis] (0,0) -- (3.9,0);
  \node[anchor=north west, inner sep=1pt, yshift=-1pt] at (0,0) {Abstract};
  \node[anchor=north east, inner sep=1pt, yshift=-1pt] at (3.9,0) {Figurative};
  \node[anchor=north, inner sep=1pt, black!60, align=center] at (1.95,-0.02) {\tiny position};
  \draw[axis] (0,0) -- (0,2.4) node[anchor=south, inner sep=1pt] {Evaluation};
}
\newcommand{\rline}[2]{\draw[rater, #2] (0.3, {1.2 + #1*(0.3-1.9)}) -- (3.5, {1.2 + #1*(3.5-1.9)});}
\newcommand{\pline}[1]{\draw[pop] (0.3, {1.2 + #1*(0.3-1.9)}) -- (3.5, {1.2 + #1*(3.5-1.9)});}
\newcommand{\botaxes}{%
  \draw[axis] (0,0) -- (3.9,0) node[anchor=north east, inner sep=1pt, yshift=-1pt] {Axis weight $w$};
  \draw[black!60] (1.2,0.05) -- (1.2,-0.05) node[anchor=north, inner sep=1pt] {0};
  \node[anchor=south west, inner sep=1pt] at (0,0.95) {Rater density};
}
\begin{scope}[shift={(0,0)}]
\node[ttl] at (-0.2,5.3) {\panel{a}\; Everyone values the same end; only the weight magnitude differs (\Do{})};
\begin{scope}[shift={(0.3,2.6)}]
\topaxes
\rline{0.12}{c2}\rline{0.28}{c1}\rline{0.45}{c3}\rline{0.58}{c1}\rline{0.7}{c2}
\pline{0.42}
\node[anchor=west, inner sep=1pt, black!70] at (3.5,{1.2+0.42*1.6}) {Pop.};
\end{scope}
\begin{scope}[shift={(0.3,0.6)}]
\botaxes
\draw[pop] (2.4,0) -- (2.4,1.05);
\node[anchor=south west, inner sep=1pt, black!70] at (2.45,0.95) {Pop.\ weight};
\draw[c1, line width=0.8pt, smooth, domain=1.25:3.7, samples=40] plot (\x, {1.0*exp(-((\x-2.4)^2)/(2*0.42^2))});
\end{scope}
\node[note] at (-0.2,0.2) {Everyone's weight is on the same side as the population. Each person's evaluation is therefore the population mean strengthened or weakened.};
\end{scope}
\begin{scope}[shift={(5.4,0)}]
\node[ttl] at (-0.2,5.3) {\panel{b}\; The spread of weights crosses 0; some people value the opposite end (\Dt{})};
\begin{scope}[shift={(0.3,2.6)}]
\topaxes
\rline{0.7}{c2}\rline{0.5}{c1}\rline{0.3}{c3}\rline{0.1}{c1}\rline{-0.12}{c4}\rline{-0.38}{c5}
\pline{0.32}
\node[anchor=west, inner sep=1pt, black!70] at (3.5,{1.2+0.32*1.6}) {Pop.};
\node[anchor=east, inner sep=1pt, c5] at (3.45,0.3) {Values the abstract end};
\end{scope}
\begin{scope}[shift={(0.3,0.6)}]
\botaxes
\begin{scope}
\clip (0,0) rectangle (1.2,1.2);
\fill[c5!30] plot[domain=0.05:1.2, samples=30] (\x, {1.0*exp(-((\x-2.1)^2)/(2*0.72^2))}) -- (1.2,0) -- (0.05,0) -- cycle;
\end{scope}
\draw[pop] (2.4,0) -- (2.4,1.05);
\node[anchor=south west, inner sep=1pt, black!70] at (2.45,0.95) {Pop.\ weight};
\draw[c1, line width=0.8pt, smooth, domain=0.05:3.7, samples=50] plot (\x, {1.0*exp(-((\x-2.1)^2)/(2*0.72^2))});
\node[anchor=south, inner sep=1pt, c5] at (0.4,0.5) {\tiny Tail beyond 0};
\end{scope}
\node[note] at (-0.2,0.2) {Weights spread continuously around the population weight, and one tail crosses 0. For people in the tail, the population mean does not tell us which end they value more highly.};
\end{scope}
\begin{scope}[shift={(10.8,0)}]
\node[ttl] at (-0.2,5.3) {\panel{c}\; Stimulus-specific evaluations\\outside the shared space (\Dh{})};
\begin{scope}[shift={(0.3,2.6)}]
\topaxes
\pline{0.42}
\rline{0.5}{c1}
\node[anchor=west, inner sep=1pt, black!70] at (3.5,{1.2+0.42*1.6}) {Pop.};
\foreach \xx/\uu in {0.5/0.45, 0.9/-0.35, 1.3/0.15, 1.7/0.55, 2.1/-0.5, 2.5/0.2, 2.9/-0.4, 3.3/0.35} {
  \draw[c4, line width=0.6pt] (\xx, {1.2+0.5*(\xx-1.9)}) -- (\xx, {1.2+0.5*(\xx-1.9)+\uu});
  \fill[c4] (\xx, {1.2+0.5*(\xx-1.9)+\uu}) circle (0.05);
}
\node[anchor=south west, inner sep=1pt, c4] at (0.35,1.95) {Stimulus-specific $u_{ki}$};
\end{scope}
\begin{scope}[shift={(0.3,0.6)}]
\node[anchor=south west, inner sep=1pt] at (0,0.95) {Rater $\times$ stimulus residuals};
\foreach \r/\row in {0/{0.15,0.6,0.05,0.8,0.3,0.1,0.7,0.2}, 1/{0.7,0.1,0.5,0.2,0.05,0.8,0.3,0.55}, 2/{0.2,0.4,0.9,0.1,0.6,0.3,0.05,0.75}, 3/{0.5,0.05,0.3,0.65,0.15,0.9,0.4,0.1}} {
  \foreach \v [count=\cc from 0] in \row {
    \fill[black!\fpeval{round(\v*90)}] ({0.1+\cc*0.42},{0.1+\r*0.2}) rectangle ++(0.38,0.17);
  }
}
\draw[black!60] (0.08,0.08) rectangle (3.5,0.9);
\node[anchor=north, inner sep=1pt, black!60] at (1.8,0.06) {Stimulus $i$};
\node[anchor=south, inner sep=1pt, black!60, rotate=90] at (-0.02,0.5) {Person $k$};
\end{scope}
\node[note] at (-0.2,0.2) {Evaluations of particular stimuli cannot be predicted from their positions on the shared dimensions, and so cannot be described by a low-dimensional shared representation.};
\end{scope}
\end{tikzpicture}}\\[4mm]
\begin{minipage}[t]{\textwidth}
\noindent\makebox[\textwidth][l]{\scriptsize\panel{d}\; Two-factor model for comparing the three shapes}\\[1mm]
\centering
\resizebox{\textwidth}{!}{%
\begin{tikzpicture}[>=Latex, font=\scriptsize, x=1cm, y=1cm,
  box/.style={draw, rounded corners=2pt, minimum height=7mm, inner sep=3pt, align=center, text width=2.5cm},
  lat/.style={box, fill=gray!15}]
\node[box]  (judg) at (1.4, 0.8) {A few judgments by person $k$\\(arrangement, triplet, rating)};
\node[box, text width=2.6cm] (enc) at (4.6, 0.8) {Encoder\\(permutation-invariant,\\attention pooling)};
\node[lat, text width=2.15cm] (z) at (7.4, 0.8) {Latent\\representation\\$\bm z_k \in \mathbb R^{M}$};
\begin{scope}[shift={(1.4,-0.5)}, scale=0.62, line width=0.5pt]
  \draw[rounded corners=1pt] (-1.9,-0.38) rectangle (-1.1,0.38);
  \draw (-1.5,0.09) circle (0.15);
  \draw (-1.78,-0.3) .. controls (-1.78,-0.08) and (-1.22,-0.08) .. (-1.22,-0.3);
  \draw[rounded corners=1pt] (-0.4,-0.38) rectangle (0.4,0.38);
  \foreach \yy/\ll in {0.2/0.5, 0.05/0.5, -0.1/0.5, -0.25/0.3} \draw (-0.25,\yy) -- ++(\ll,0);
  \draw (1.2,0.05) -- (1.5,0.28) -- (1.85,0.1) -- (1.8,-0.25) -- (1.4,-0.3) -- cycle;
  \draw (1.85,0.1) -- (2.1,0.3);
  \foreach \pt in {(1.2,0.05),(1.5,0.28),(1.85,0.1),(1.8,-0.25),(1.4,-0.3),(2.1,0.3)} \fill \pt circle (0.045);
\end{scope}
\node[anchor=north, inner sep=1pt, align=center] at (1.4,-0.8) {Stimulus $i$ (face,\\painting, sentence,\\molecule, etc.)};
\coordinate (stime) at (2.8,-0.95);
\node[lat, text width=4.4cm] (f) at (5.85, -0.95) {Stimulus embedding $\bm f_i \in \mathbb R^{D}$\\(learned, or image / text / molecular features)};
\node[box, text width=3.5cm] (dec) at (10.85, 0.05) {Decoder\\\Dz{}: shared only\\\Do{}: $\bm w(\bm z)\odot\bm f$ (same sign)\\\Dt{}: $\bm W(\bm z)\bm f$ (allows reversal)\\\Dh{}: \Dt{} $+$ stimulus-specific $\bm u(\bm z)$};
\node[box, text width=2.2cm] (out) at (14.5, 0.05) {Judgment function of person $k$\\(predicts unseen stimuli and held-out judgments)};
\draw[->] (judg) -- (enc);
\draw[->] (enc) -- (z);
\draw[->] (stime) -- (f);
\draw[->] (z.east) -- ([yshift=4mm]dec.west);
\draw[->] (f.east) -- ([yshift=-4mm]dec.west);
\draw[->] (dec) -- (out);
\end{tikzpicture}}
\end{minipage}
\caption{\textbf{Three shapes that individual differences in judgment can take, and the two-factor model used to compare them.}
\panel{a}--\panel{c} show three shapes that individual differences can take.
The top row of each panel shows how each person evaluates a stimulus as a function of its position on a shared axis (abstract--figurative, as an example) (evaluation functions, coloured lines), together with the population mean (dashed line).
Lines rising to the right represent people who value more figurative stimuli more highly, lines falling to the right represent people who value more abstract stimuli more highly, and the slope of a line is the weight $w$ on that axis.
The bottom row shows how the weight $w$ on that axis is distributed across raters.
\panel{a} The case in which everyone values the same end of the axis more highly (the weights have the same sign) and only the magnitude of the weight differs (reweighting hypothesis, \Do{}).
The weights are then distributed on the same side as the population weight.
Individual differences are therefore differences in emphasis on a shared representation.
\panel{b} The case in which the weights spread continuously along the dominant shared direction and, at the end of the distribution (the tail), the weights are opposite in sign to the population (linear-transform hypothesis, \Dt{}).
People who value the opposite end more highly are not a discrete minority but people in the tail of a continuous distribution.
For them, the population mean does not indicate which end is valued more highly.
\panel{c} The case in which there are stimulus-specific deviations $u_{ki}$ that cannot be expressed in the shared space (\Dh{}).
Person-specific structure then remains in the rater $\times$ stimulus residuals, so that individual differences cannot be described by a low-dimensional shared representation.
Variance decomposition cannot distinguish these three.
We therefore distinguish them by how well the nested generative models \Dz{}--\Dh{} predict unseen individuals.
\panel{d} Architecture of the model used for this purpose.
An encoder infers a person's latent representation $\bm z_k$ from a small number of that person's judgments.
A decoder generates that person's judgment function from this $\bm z_k$ and the shared stimulus embedding $\bm f_i$.
The person-dependent part of the decoder is one of four nested levels (\Dz{}--\Dh{}), and \Do{}, \Dt{} and \Dh{} correspond to the shapes in \panel{a}, \panel{b} and \panel{c}, respectively.}
\label{fig:concept}
\end{figure}

Existing methods cannot distinguish these possibilities.
Variance decomposition gives only the size of individual differences, and machine-learning models that predict individual ratings\cite{ren2017,li2020,yang2022,zhu2022,lin2023,yun2024,ryu2026,maerten2026} show that ratings can be predicted but not what differs between people.
The first two shapes were already formulated in three-way multidimensional scaling, as INDSCAL\cite{carroll1970} and IDIOSCAL\cite{carroll1972,tucker1972,carroll1974}, and later models of similarity either assume reweighting\cite{nosofsky1986,roads2019,roads2021,andrews2023} or describe only the population\cite{zheng2019,hebart2020,muttenthaler2022}.
These methods, however, need many judgments per person.
The shapes have therefore not been compared by how well they predict new individuals, nor applied to thousands of people who each judged only a few stimuli.

Here we write each possibility as a generative model and compare the models by how well they predict the judgments of people who did not appear in training.
Each stimulus has an embedding shared by everyone, and each person has a low-dimensional latent representation.
An encoder infers this latent representation from a small number of the person's judgments (amortised inference\cite{kingma2014}), and a decoder generates the person's judgment function from it (Fig.~\ref{fig:concept}d), extending a framework developed for decision-making tasks\cite{higashi2025,higashi2026}.
The person-dependent part of the decoder has four nested levels.
\Dz{} has no individual differences, \Do{} allows same-sign reweighting, \Dt{} allows a linear transform and hence reversal, and \Dh{} adds stimulus-specific offsets.
The level at which prediction of unseen individuals stops improving indicates the shape of individual differences.
Because the encoder needs only a few judgments, the framework applies to large public datasets in which each person judged a small fraction of the stimuli, such as the face arrangements of Xu et al.\cite{xu2026}.

We applied this framework to ten public datasets (about 20,000 people; Table~\ref{tab:datasets}): face arrangements, face attractiveness ratings, face similarity triplets, and ratings of artworks, photographs and jokes.
We also analysed odour pleasantness ratings and a dataset in which the same 129 raters rated art, fashion and landscapes\cite{hayashi2026}.
The model recovered a person's judgment space from 10--20 judgments and predicted judgments of stimuli that the person had never seen.
For faces, reweighting (\Do{}) was sufficient, and weights opposite in sign to the population were not statistically detected.
For artworks, individual weights spread continuously along the abstract--figurative direction, and the tail of this spread reached the opposite side from the population; the reversal-capable model (\Dt{}) captured this tail.
Photographs were intermediate between faces and artworks.
The size of the tail was a property of the rater sample rather than of the artworks, and falling in the tail was not a trait of individual raters.
The latent representation encoded what a person prefers but not self-reported demographic attributes.

\section*{Results}

\subsection*{A two-factor model and four nested hypotheses about individual differences}
Our model represents a judgment by two factors, the stimulus and the person (Fig.~\ref{fig:concept}d).
Each stimulus $i$ has an embedding $\bm f_i \in \mathbb R^{D}$ ($D = 8$), which is either learned from the judgments or projected from image, text or molecular features (Methods).
Each person $k$ has a latent representation $\bm z_k \in \mathbb R^{M}$ ($M = 4$), which a permutation-invariant encoder\cite{zaheer2017} outputs from the set of that person's judgments.
The decoder predicts the person's judgments from $\bm z_k$ and $\bm f_i$.
Its person-dependent part has four nested levels, each allowing individual differences more freedom than the one before.
\Dz{} is a population model that does not use $\bm z$.
\Do{} multiplies each dimension of the embedding by a positive weight, so people may differ in how much importance they give to each dimension but not in which direction they rate more highly (same-sign reweighting).
\Dt{} allows a free linear transform, so it can also express an evaluation that is opposite to that of the population along some direction (a reversal).
\Dh{} adds stimulus-specific offsets to \Dt{}.
All models were evaluated by 10-fold cross-validation at the level of individuals.
For people not used in training, we estimated $\bm z$ from a few estimation judgments and predicted the person's remaining judgments.
The reference is a population baseline that predicts from the average judgment of the people in the training folds.

\subsection*{An individual's face impression space can be recovered from arrangements of 20 faces}
We first tested whether an individual's face impression space can be recovered from a small number of judgments.
In Experiment 2 of Xu et al.\cite{xu2026}, each of 1,682 participants arranged 20 of 96 faces on the screen over multiple trials.
We estimated $\bm z$ from 70\% of each participant's trials and predicted the arrangement distances between faces in the remaining 30\% (held-out trials).
Accuracy was measured by the Spearman $\rho$ between the predicted and the actual distances.

The individual model outperformed the population (Fig.~\ref{fig:xu}).
Held-out $\rho$ rose from 0.295 for the population RDM (the population distances computed from the participants in the training folds) to 0.338 for \Do{}.
The participant-level difference was $+0.043$ relative to the population RDM (95\% CI [0.037, 0.050], $t(1681) = 14.0$, $d_z = 0.34$, $P = 2 \times 10^{-42}$) and $+0.039$ relative to the \Dz{} population model (95\% CI [0.035, 0.044], $t(1681) = 16.3$, $d_z = 0.40$, $P = 9 \times 10^{-56}$).
This accuracy exceeded the split-half reliability of the person's own estimation trials (0.309).
Baselines that use the person's own data without the model were either weaker than \Do{} or, in the case of the shrinkage blend of the person's own distances with the population distances, did not extend to face pairs absent from the estimation trials (Supplementary Note~1).
Advancing the model to \Dt{} (0.341) or \Dh{} (0.337) left the accuracy almost unchanged.
The estimated $\bm z$ was stable within a person: the $\bm z$ estimated separately from the estimation trials and from the held-out trials agreed well (intraclass correlation 0.77).

\begin{figure}[tbp]
\centering
\includegraphics[width=\textwidth]{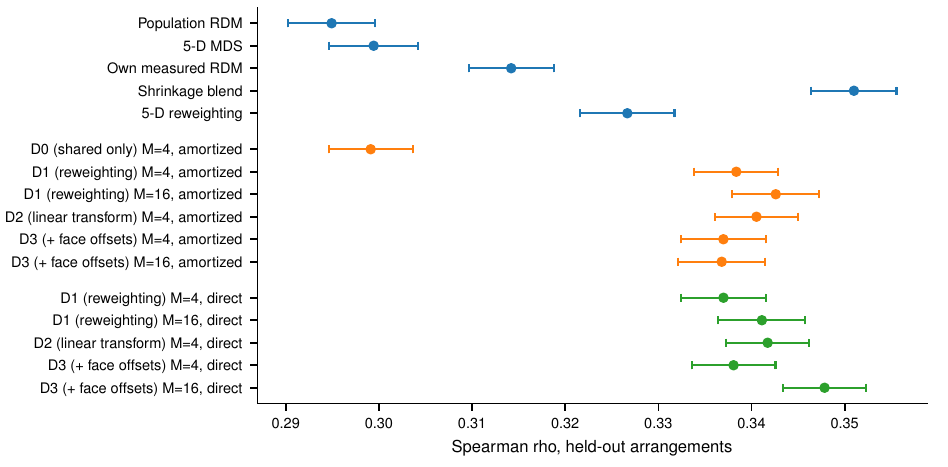}
\caption{\textbf{Recovery of an individual's face impression space from arrangements of 20 faces (Xu et al.\ Experiment 2, 96 multiracial faces, 1,682 participants).}
The prediction accuracy of each method is shown.
Values are Spearman $\rho$ on the held-out trials (trials not used for estimation) of fold-out participants (mean $\pm$ s.e.m., averaged over seeds).
From top to bottom: methods that ignore individual differences (the population RDM, Population RDM; the 5-dimensional MDS of the population, 5-D MDS); methods that use the person's own data without a model (the person's own estimation RDM, Own measured RDM; the blend of the person and the population, Shrinkage blend; the 5-dimensional reweighting, 5-D reweighting; blue up to this point); and our models \Dz{}--\Dh{} with $\bm z$ estimated by amortised inference (orange) and by direct optimisation (green).
\Do{} outperforms the population RDM and \Dz{}, but advancing to \Dt{} or \Dh{} leaves the accuracy almost unchanged.
}
\label{fig:xu}
\end{figure}

\subsection*{Recovery replicates across tasks, stimulus sets and domains, and transfers to unseen stimuli}
Next, we applied the same framework to the remaining datasets, which differ from the Xu data in task, stimulus set and domain (Table~\ref{tab:datasets}, Fig.~\ref{fig:gen}a; numbers in Supplementary Table~\ref{tab:gen}, details in Supplementary Notes~2--4).
We call the improvement of the \Do{} reweighting model over the \Dz{} population model (\dOZ{}) the individualisation gain.

\begin{table}[tbp]
\centering
\caption{\textbf{Public datasets analysed.}
The number of raters is the number used in the analysis (the total in the distributed data is given in parentheses).
The ``unseen stimuli'' entry indicates whether transfer to stimuli that the person did not judge was evaluated.
In the versions that obtain the embedding from image, text or molecular features, those stimuli were not used to train the model either.}
\label{tab:datasets}
\small
\setlength{\tabcolsep}{4pt}
\begin{tabular}{p{3.0cm}p{4.5cm}p{2.0cm}p{1.6cm}p{2.8cm}}
\toprule
Dataset & Task and stimuli & Raters & Per rater & Stimulus features (unseen stimuli) \\
\midrule
Xu et al.\ Experiment 1 \cite{xu2026} & Multi-arrangement of 20 of 100 White faces (about 60 trials); the retest task used 40 faces, overlapping with the person's 20 faces by 8 faces on average & 1,751 & 20 faces & Learned embedding (retest task) \\
Xu et al.\ Experiment 2 \cite{xu2026} & Multi-arrangement of 20 of 96 White, Black and Asian faces (about 50 trials); the retest was within the person's 20 faces & 1,682 & 20 faces & Learned embedding (none) \\
London Set \cite{debruine2017} & Attractiveness ratings of 102 faces (1--7, every rater rated every face) & 2,501 (2,513) & 102 faces & CLIP image features (yes) \\
AVFS \cite{andrews2023} & Triplet odd-one-out judgments of face similarity. Main set of 4,921 faces (FFHQ\cite{karras2019}), novel set of 56 faces, 1,000 common triplets $\times$ 24 judgments & main 1,283, novel 504 (1,645) & 60--260 triplets & CLIP (yes: 56 novel faces) \\
FaceSim3D \cite{hofmann2025} & Triplets of 100 faces (CFD\cite{ma2015}), 2D static and 3D dynamic & 1,219/1,053 (2,354) & median 183 & None \\
SCUT-FBP5500 \cite{liang2018} & Attractiveness ratings of 5,500 faces (4,000 Asian, 1,500 White) (1--5, every rater rated every face) & 60 & 5,500 faces & CLIP (yes) \\
LAPIS \cite{maerten2025} & Aesthetic ratings of 11,723 artworks (figurative and abstract, 26 styles) (0--100) & 553 (555) & median 479 & CLIP (yes) \\
AADB \cite{kong2016} & Aesthetic ratings of 9,958 photographs (Flickr) (0--5, 5 raters per photograph) & 68 (190) & median 340 & CLIP (yes) \\
EVA \cite{kang2020} & Aesthetic ratings of 4,070 photographs (AVA\cite{murray2012}, 6 content categories) (0--10, 30 or more raters per photograph) & 175 (1,094) & median 326 & CLIP (yes) \\
Jester \cite{goldberg2001} & Funniness ratings of 150 jokes ($-10$--$+10$) & 11,067 (54,905) & median 87 & Sentence embeddings (weak) \\
Keller \& Vosshall \cite{keller2016} & Pleasantness ratings of 480 odour molecules $\times$ 2 concentrations (0--100) & 55 & about 800 & RDKit descriptors (weak) \\
XPASS-Vis \cite{hayashi2026} (public, password required) & Aesthetic ratings by the same raters of art (2,345 LAPIS artworks), 2,082 fashion images and 2,099 landscape images (0--6) & 129 & 184--247 per domain & CLIP (yes) \\
\bottomrule
\end{tabular}
\end{table}

\begin{figure}[tbp]
\centering
\begin{minipage}[t]{\textwidth}
\panel{a}\\[1mm]
\includegraphics[width=\textwidth]{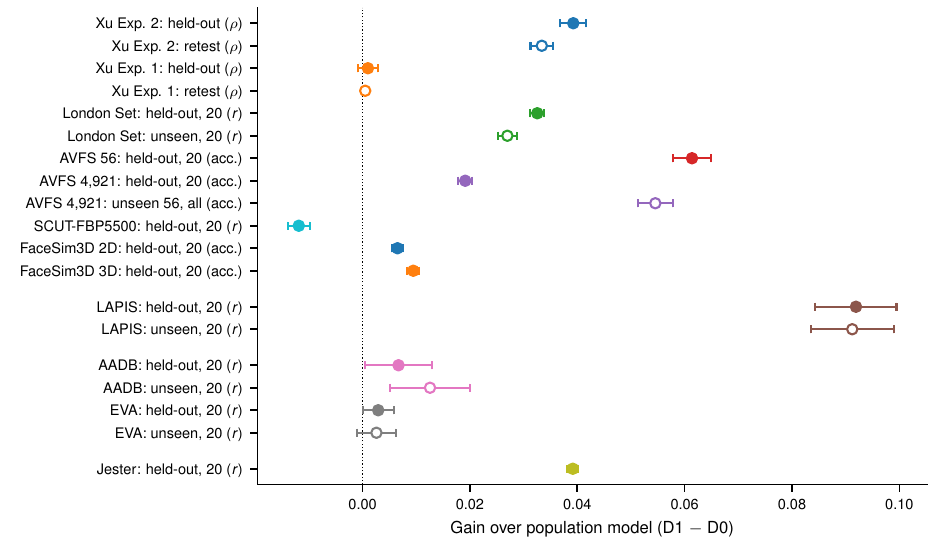}\\[2mm]
\panel{b}\\[1mm]
\includegraphics[width=\textwidth]{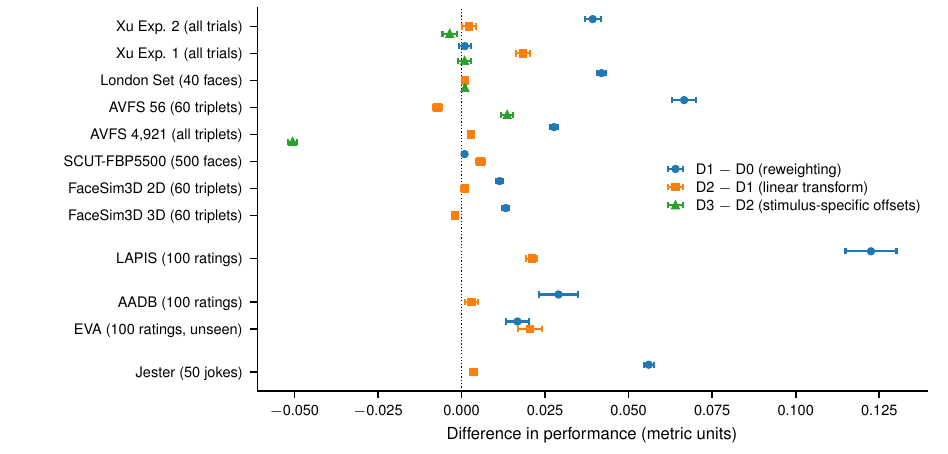}
\end{minipage}
\caption{\textbf{Recovery of individual judgment spaces replicates across datasets, and the shape of individual differences differs between datasets.}
\panel{a} Individualisation gain.
The participant-level gain of the \Do{} reweighting model over the \Dz{} population model is shown for each dataset and evaluation (mean $\pm$ s.e.m., averaged over seeds).
Filled markers show the gain on held-out judgments of the person's own stimuli.
Open markers show the gain on stimuli that the person did not judge (in the rows that use image features, stimuli on which the model was not trained either).
The row labels give, in order, the dataset, the evaluation (held-out: held-out judgments of the person's own stimuli; unseen: unseen stimuli; retest: retest task), the number of judgments used for estimation, and the metric (Spearman $\rho$, Pearson $r$, triplet accuracy acc.).
To make the datasets easier to compare, the number of judgments was all estimation trials for Xu and was fixed at 20 for the others.
For many datasets, therefore, the gain at the largest number of judgments, shown in \panel{b}, is larger than the value shown here.
\panel{b} Differences between the four model levels.
The participant-level differences on held-out judgments are shown for each dataset (mean $\pm$ s.e.m., averaged over seeds).
For visibility, the series in the same row are slightly offset vertically.
In the legend, \dOZ{} is the difference due to reweighting of the shared dimensions, \dTO{} is the difference due to adding the reversal-capable linear transform, and \dHT{} is the difference due to adding stimulus-specific offsets.
Unlike in \panel{a}, the number of judgments used for estimation was the largest available for each dataset (in parentheses in the row labels).
This is because differences in the shape of individual differences appear more clearly with more judgments.
The EVA values are for unseen photographs, and \Dh{} was trained only for Xu, London and AVFS.
The AADB row with 100 ratings gives the values for the 55 raters for whom 100 estimation ratings were available.
The rows are in the same order in the two panels: from the top, faces, artworks, photographs and jokes.
The learning curves for each dataset are shown in Supplementary Figs.~\ref{sfig:gen} and \ref{sfig:curves}.
}
\label{fig:gen}
\end{figure}

For faces, a gain was obtained even when the task was changed.
For attractiveness ratings (London Set, 2,501 raters), estimating $\bm z$ from 20 ratings raised the Pearson $r$ on the remaining faces from 0.590 for the population mean to 0.622 ($+0.033$, 95\% CI [0.030, 0.035], $t(2500) = 25.3$, $d_z = 0.51$, $P < 10^{-100}$).
For similarity triplets (AVFS novel set, 504 raters), estimating $\bm z$ from 20 triplets raised the accuracy from 0.588 to 0.650 ($+0.061$).

The individual models also transferred to stimuli that the person had never seen (open markers in Fig.~\ref{fig:gen}a).
With embeddings built from image features, we predicted stimuli that were used neither to train the model nor to estimate the person's $\bm z$.
On such unseen faces, $r$ rose from 0.416 to 0.441 in the London Set ($+0.025$, 95\% CI [0.022, 0.028], $t(2482) = 14.6$, $d_z = 0.29$, $P = 3 \times 10^{-46}$), and accuracy rose from 0.574 to 0.629 in AVFS ($+0.055$, 95\% CI [0.048, 0.061], $t(597) = 16.9$, $d_z = 0.69$, $P = 10^{-52}$).
For artworks (LAPIS) and photographs (AADB, EVA), the gain on unseen images was the same as the gain on the rater's remaining ratings of training images.
The gain was small where raters closely agreed with one another (FaceSim3D and SCUT-FBP5500, $+0.01$ or less), and it was not detected for odour pleasantness (Keller \& Vosshall\cite{keller2016}, 55 raters; at most $+0.006$, n.s.).

\subsection*{The shape of individual differences differs between datasets}
To examine the shape of individual differences, we compared across datasets the level at which prediction saturates (Fig.~\ref{fig:gen}b).
If prediction saturates at \Do{}, same-sign reweighting is sufficient.
If \Dt{} improves it further (\dTO{} positive), there are individual differences that same-sign reweighting cannot express.

For faces, prediction saturated at \Do{} irrespective of the task.
In Xu Experiment 2, adding \Dt{} gave $+0.002$ (95\% CI [$-0.002$, 0.006], $t(1681) = 1.01$, $d_z = 0.02$, $P = 0.31$) and further adding \Dh{} gave $-0.004$ (95\% CI [$-0.008$, 0.001], $t(1681) = -1.66$, $d_z = -0.04$, $P = 0.10$).
This saturation is hard to explain by insufficient statistical power.
With simulated participants whose split-half reliability was matched to the value measured in Experiment 2 (0.32), \Dt{}-type individual differences of a size comparable to the \Do{}-type ones were detected as \dTO{} $= +0.009$ (95\% CI [0.006, 0.013]; Supplementary Note~5).
In the London Set, AVFS and FaceSim3D, \dTO{} ranged from $-0.009$ to $+0.003$.
Two face datasets showed a small advantage of \Dt{} (Xu Experiment 1, $+0.018$; SCUT with 500 faces, $+0.006$; Supplementary Notes~5 and 2), and \Dh{} did not improve prediction of unseen faces in any dataset.

For artworks, in contrast, \Dt{} consistently outperformed \Do{}.
In LAPIS (553 raters), the individualisation gain was about three times as large as for faces ($+0.092$ with 20 ratings, $+0.123$ with 100 ratings).
On top of this, \dTO{} was positive and increased with the number of ratings ($+0.011$ with 20 ratings, 95\% CI [0.007, 0.015], $t(552) = 5.61$, $d_z = 0.24$, $P = 3 \times 10^{-8}$; $+0.021$ with 100 ratings, 95\% CI [0.018, 0.024], $t(550) = 12.2$, $d_z = 0.52$, $P = 3 \times 10^{-30}$), and the same held for unseen artworks.

Photographs lay between faces and artworks, and the two photograph datasets diverged.
In EVA (175 raters), \dTO{} on unseen photographs was $+0.020$ against a \dOZ{} of $+0.017$ (100 ratings, 95\% CI [0.013, 0.028], $t(174) = 5.67$, $d_z = 0.43$, $P = 6 \times 10^{-8}$).
In AADB (68 raters), \dTO{} was only $+0.003$ (95\% CI [$-0.001$, 0.007], $t(54) = 1.47$, $d_z = 0.20$, $P = 0.15$) against a \dOZ{} of $+0.029$.
The number of raters does not explain this divergence, and its cause cannot be determined from the present data (Supplementary Note~2).

For jokes (Jester, 11,067 raters), shared preferences were the weakest of all datasets (predictive power of the population mean 0.34), yet \dTO{} was small, at $+0.004$--$0.007$ (Supplementary Note~3).
Weak shared preferences alone therefore do not make a reversal-capable model necessary.

\subsection*{A continuous spread of individual weights along a shared direction produces the advantage of the reversal-capable model}
What shape of individual differences gives \Dt{} its advantage?
The model comparison alone cannot answer this question.
Because \Do{} learns its basis, it can absorb part of a reversal, so a small \dTO{} does not mean that there is no reversal (Supplementary Note~6).
We therefore carried out a sign check that does not use the encoder or the decoder (Fig.~\ref{fig:sign}, Methods).
The sign check estimates each rater's weight directly from that rater's own ratings and asks whether it points in the same direction as the population weight.
The main measure is the loss under the \Do{} constraint: the drop in predictive power when the rater's weight is restricted to the population weight multiplied by a non-negative factor in each dimension.
A large positive loss means that the rater's ratings cannot be explained by strengthening or weakening the population weight.
We also report the proportion of raters whose weight has a negative cosine with the population weight, and the reversed energy, the fraction of a rater's weight that lies in dimensions opposite in sign to the population.

\begin{figure}[tbp]
\centering
\begin{minipage}[t]{0.48\textwidth}
\panel{a}\\[1mm]
\includegraphics[width=\textwidth]{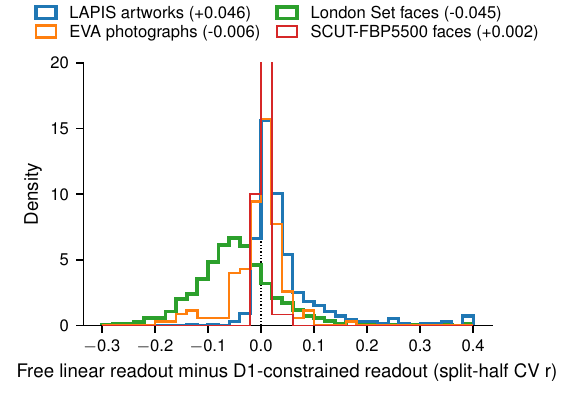}
\end{minipage}\hfill
\begin{minipage}[t]{0.48\textwidth}
\panel{b}\\[1mm]
\includegraphics[width=\textwidth]{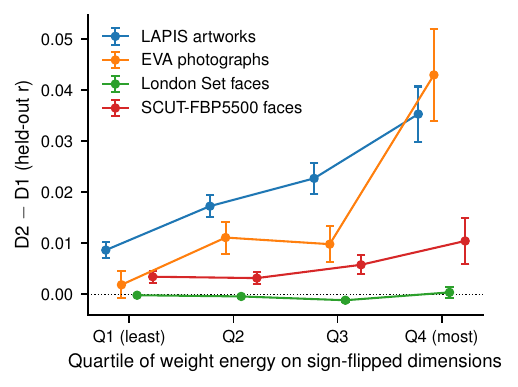}
\end{minipage}\\[3mm]
\begin{minipage}[t]{0.48\textwidth}
\panel{c}\\[1mm]
\includegraphics[width=\textwidth]{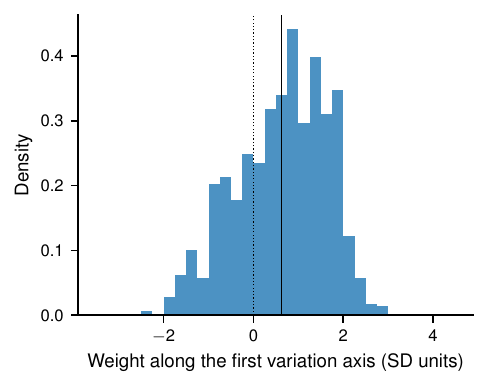}
\end{minipage}\hfill
\begin{minipage}[t]{0.48\textwidth}
\panel{d}\\[1mm]
\includegraphics[width=\textwidth]{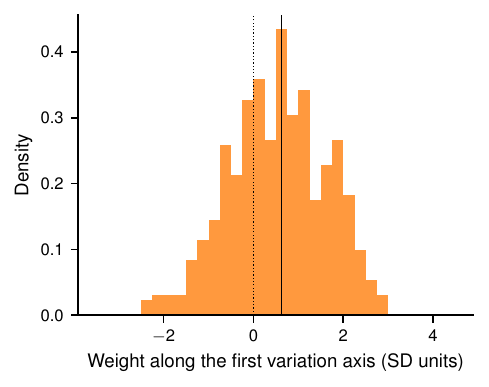}
\end{minipage}
\caption{\textbf{A continuous spread of individual weights along a shared direction produces the advantage of the reversal-capable model (sign check).}
\panel{a} Distribution across raters of the loss under the \Do{} constraint.
A positive value means that strengthening or weakening the population weight dimension by dimension cannot fully account for that rater's ratings.
The data are LAPIS artworks (553 raters, median 383 ratings), EVA photographs (175 raters, median 256), London Set faces (2,501 raters, 82 faces) and SCUT-FBP5500 faces (60 raters, 4,400 faces).
The numbers in the legend are means.
The peak of SCUT near 0 (density 38) is truncated at 20 in the plot.
The distributions for the two face datasets (London, SCUT) lie at or below 0 or very close to 0, whereas the LAPIS distribution extends to the right of 0, and EVA lies in between.
\panel{b} \dTO{} of the amortised models (held-out ratings; mean $\pm$ s.e.m.) when raters are divided into quartiles of reversed energy.
For visibility, the series in the same quartile are slightly offset horizontally.
In LAPIS and EVA, \dTO{} increases with reversed energy, whereas in London it stays at 0, and in SCUT it is at most $+0.010$.
\panel{c} Distribution of the raters' scores on the axis along which the individual weights vary most, for LAPIS (the 20 fold models are pooled).
The solid line shows the projection of the population weight (0.62 SD) and the dotted line shows 0; raters to the left of the dotted line are on the opposite side from the population.
The distribution is unimodal, and its tail crosses 0.
\panel{d} Results of the same analysis for EVA (30 fold models).}
\label{fig:sign}
\end{figure}

For faces, no reversal was statistically detected.
In the London Set, only 1.6\% of raters had a negative cosine, and imposing the \Do{} constraint did not reduce predictive power (loss $-0.045$; green in Fig.~\ref{fig:sign}a).
In SCUT-FBP5500, in which 60 raters each rated 5,500 faces and the weights can be estimated very stably, no rater had a negative cosine and the loss was $+0.002$ (red in Fig.~\ref{fig:sign}a).

For artworks (LAPIS), in contrast, imposing the \Do{} constraint reduced predictive power by 0.046 (s.e.m.\ 0.003).
The loss exceeded 0.02 for 53\% of raters (blue in Fig.~\ref{fig:sign}a), and 9.6\% of raters had a negative cosine.
The sign check agreed with the model comparison: \dTO{} increased monotonically across quartiles of reversed energy, from $+0.009$ to $+0.035$ (blue in Fig.~\ref{fig:sign}b).

Are these raters a discrete minority, or the tail of a continuous distribution?
To find out, we took the direction along which the weights vary most between raters (the first variation axis) and examined how the raters are distributed along it (Fig.~\ref{fig:sign}c).
This axis accounted for 64\% of the between-rater variance of the weights, and a rater's score on it tracked that rater's preference for abstract over figurative art (Spearman $\rho = -0.93$).
The axis overlapped substantially with the population weight ($|\cos| = 0.59$), so the individual weights were spread along a direction of preference common to many raters (the shared direction).
The distribution of scores was unimodal (bimodality coefficient 0.49; it would exceed 0.555 if there were two modes).
The population weight lay only 0.62 standard deviations from 0, so the tail of the distribution crossed 0, and 28\% of raters had scores opposite in sign to the population.
The 9.6\% of raters with a negative cosine were those at the far end of this tail, not a discrete subgroup.

Simulated raters whose structure was known in advance supported this reading (Methods, Supplementary Note~6).
Simulated raters whose weights were spread continuously along the abstract axis reproduced a \dTO{} of a size comparable to LAPIS ($+0.032$; $+0.021$ in LAPIS) and its increase across quartiles of reversed energy.
With simulated raters of whom only 10\% were discretely reversed, \dTO{} was small and flat across quartiles, although the loss under the \Do{} constraint still detected the reversed raters.
The advantage of \Dt{} therefore reflects a continuous spread that same-sign reweighting cannot fully express, not the mere presence of reversals.

Photographs (EVA) showed a spread of the same type, but weaker and distributed over more axes.
No rater had a negative cosine and the mean loss under the \Do{} constraint was $-0.006$, but the loss exceeded 0.02 for 25\% of raters (orange in Fig.~\ref{fig:sign}a), and \dTO{} was large only in the top quartile of reversed energy ($+0.043$; orange in Fig.~\ref{fig:sign}b).
The first variation axis accounted for 40\% of the variance, and 29\% of raters were on the opposite side from the population (Fig.~\ref{fig:sign}d).
These raters preferred still lifes and close-ups to the photographs of landscapes and architecture that the majority preferred.

In summary, the proportion of raters whose loss under the \Do{} constraint exceeded 0.02 increased from faces (SCUT 3\%, London 16\%) through photographs (EVA 25\%) to artworks (LAPIS 53\%), and \dTO{} increased in the same order.
What separated the datasets, however, was neither the domain as such nor how strong the shared preferences are.
It was how many raters, in the tail of the spread of weights along the shared direction, had a weight opposite in sign to the population.

\subsection*{The size of the tail depends on the sample of raters and is not a trait of individual raters}
Is the size of this tail determined by the artworks, or by who rated them?
In XPASS-Vis\cite{hayashi2026}, 129 university students in Japan and China rated the same 2,345 artworks as the LAPIS test split, together with fashion and landscape images (Table~\ref{tab:xpass}).

\begin{table}[tbp]
\centering
\caption{\textbf{Comparison of a different sample rating the same artworks (XPASS-Vis) with LAPIS.}
Values are Pearson $r$ on the rater's remaining ratings of training images, computed for fold-out raters (100 estimation ratings, averaged over seeds).
``Subsampled'' is a control for examining the effect of the differences in the number of raters and the number of ratings.
The LAPIS raters were randomly reduced to 129 raters (with all their ratings) or to 129 raters $\times$ 227 ratings (mean of two samples).
``Transfer'' is a model trained on one rater sample and applied to the raters of the other.
For the transfer to LAPIS raters, we used the ratings of the same 2,345 artworks, with 50 estimation ratings (\dOZ{} $+0.086$ on unseen artworks).
For XPASS-Vis fashion and landscapes, \Dz{} does not reach the image-mean baseline, and the inadequacy of the stimulus representation limits predictive power.
The results of the sign check (XPASS-Vis art/fashion/landscapes) were 3.1/0/0.8\% of raters with a negative cosine and a loss under the \Do{} constraint of $+0.004$/$-0.030$/$-0.006$.
n.s.: $P > 0.05$.
}
\label{tab:xpass}
\footnotesize
\setlength{\tabcolsep}{4pt}
\begin{tabular}{p{4.6cm}p{1.4cm}p{1.3cm}p{1.0cm}p{1.0cm}p{1.0cm}p{1.4cm}p{1.4cm}}
\toprule
Analysis & Raters & Image mean & \Dz{} & \Do{} & \Dt{} & \dOZ{} & \dTO{} \\
\midrule
LAPIS, full & 551 & 0.436 & 0.432 & 0.554 & 0.575 & $+0.123$ & $+0.021$ \\
LAPIS, subsampled (129 raters, all ratings) & 129 $\times$ 2 & 0.313 & 0.413 & 0.529 & 0.548 & $+0.116$ & $+0.019$ \\
LAPIS, subsampled (129 raters $\times$ 227 ratings) & 129 $\times$ 2 & --- & 0.438 & 0.530 & 0.552 & $+0.093$ & $+0.022$ \\
XPASS-Vis art (own model) & 129 & 0.478 & 0.473 & 0.501 & 0.508 & $+0.028$ & $+0.007$ \\
Transfer: XPASS-Vis raters $\leftarrow$ LAPIS model & 129 & 0.477 & 0.466 & 0.508 & 0.516 & $+0.043$ & $+0.008$ \\
Transfer: LAPIS raters (same artworks) $\leftarrow$ XPASS-Vis model & about 507 & --- & --- & --- & --- & $+0.085$ & $+0.015$ \\
XPASS-Vis fashion & 129 & 0.430 & 0.350 & 0.352 & 0.362 & $+0.002$ n.s. & $+0.010$ \\
XPASS-Vis landscapes & 129 & 0.483 & 0.444 & 0.453 & 0.464 & $+0.009$ & $+0.011$ \\
\bottomrule
\end{tabular}
\end{table}

Although the artworks were the same, individual differences were much smaller in this sample.
With 100 ratings, \dOZ{} was $+0.028$ and \dTO{} was $+0.007$, only a quarter to a third of the values in LAPIS ($+0.123$, $+0.021$).
Only 3.1\% of raters had a negative cosine, and the loss under the \Do{} constraint was $+0.004$ (s.e.m.\ 0.004).
Fewer raters or fewer ratings did not explain the difference: when LAPIS was subsampled to the same scale, the gain remained large (\dOZ{} $+0.093$ to $+0.116$).
Nor did the population on which the model was trained: a model trained on the LAPIS raters gave the XPASS-Vis raters a gain of only $+0.043$, whereas a model trained on the XPASS-Vis raters gave the LAPIS raters a gain of $+0.085$.
The size of the tail observed in LAPIS is therefore not a constant intrinsic to the artworks, but a property of one particular sample of raters (Supplementary Note~6).

Because the same raters rated three domains, we could also ask whether falling in the tail is a trait of the individual rater.
It was not.
The per-rater \dOZ{}, \dTO{}, loss under the \Do{} constraint and reversed energy were all uncorrelated across domains (Spearman $|\rho| \le 0.17$, all $P > 0.05$).
In contrast, the prediction accuracy of \Do{} ($\rho = 0.29$--$0.40$) and the cosine with the population weight (0.19--0.29) did correlate across domains, and the similarity structure between raters was shared across domains (Mantel $\rho = 0.31$--$0.46$, $P \le 0.006$).
How closely a rater follows the population is thus a trait preserved across domains, whereas reading a shared direction the opposite way (having a weight along it whose sign is opposite to the population) belongs to the combination of stimulus domain and rater sample.

\subsection*{The latent representation encodes what a person prefers, not who the person is}
Finally, we asked whether the latent representation $\bm z$ represents the content of a person's preferences or attributes such as sex and age (Supplementary Note~7, Supplementary Fig.~\ref{sfig:effect}c).
In the London Set, the largest axis of individual differences is a preference for male over female faces.
The $\bm z$ of \Do{} estimated from 40 faces predicted this preference with a cross-validated $R^2$ of 0.71, but classified the four groups defined by the rater's sex and sexual orientation with an accuracy of only 0.33 (chance 0.25).
In LAPIS, $\bm z$ predicted the preference for abstract artworks with an $R^2$ of 0.82--0.85, but did not encode interest in art, age or sex.
In Xu Experiment 2 and AVFS, the perceiver's race, region of ancestry, sex and age could hardly be decoded from $\bm z$.
Thus, across the four datasets, $\bm z$ captured what the person prefers, but not demographic attributes or self-reported interests.

\section*{Discussion}
This study asked not how much people differ in their judgments but what shape those differences take.
We arranged hypotheses about the shape as nested models and compared them by how well they predicted the judgments of people never seen before.
For faces, raters who rated the opposite end of a shared axis more highly were not statistically detected, and only the magnitude of the weights differed between people.
For artworks, individual weights spread continuously and unimodally along the dominant shared direction (abstract--figurative).
This spread was wide (its standard deviation was 1.6 times the projection of the population weight onto this direction), so the tail of the distribution crossed 0 and the weights of some raters were opposite in sign to the population.
Photographs showed the same type of spread more weakly.
The number of raters in the tail was not a constant determined by the artworks: it differed between rater samples that judged the same artworks.
All of these analyses rest on recovering a person's judgment space from 10--20 judgments.
The \Do{} reweighting model significantly outperformed the population model in eight of the ten public datasets (about 20,000 people; in the two photograph datasets it did so with 100 ratings but not with 20; Supplementary Table~\ref{stab:stats}).
Of the remaining two, in Xu Experiment 1 only the \Dt{} linear-transform model exceeded the population model, and in SCUT-FBP5500, in which consensus dominated, there was no gain.
For face similarity, the individual model predicted a person better than the majority vote of about 23 other people (Supplementary Note~8).

For faces, same-sign reweighting of shared dimensions described the individual differences across tasks (arrangement, rating, triplet) and stimulus sets.
Even in dense data in which 60 people each rated 5,500 faces, not a single rater read a shared axis in the opposite direction.
This conclusion rests on the sign check, which examines each rater's weights directly.
It does not rest on \dTO{} $\approx 0$, which a model that learns its basis can produce even when reversals exist.
Faces are known to be evaluated along a small number of dimensions, such as trustworthiness and dominance, with high consensus\cite{oosterhof2008,todorov2015,lin2021}.
Our result adds that individual differences in face impressions are differences in emphasis among these dimensions: neither the dimensions themselves nor the end that is rated more highly differ between people.
This is the form that attention-weighted embeddings\cite{roads2019,roads2021} and the generalized context model\cite{nosofsky1986} assume, and it proved sufficient under the stringent test of predicting unseen individuals from a small number of judgments.
Two datasets showed a small advantage of the linear transform (Xu Experiment 1 and SCUT with 500 faces), which we cannot yet interpret, and stimulus-specific offsets did not transfer to unseen faces.

The advantage of the reversal-capable model for artworks is consistent with reports that preferences for art are shared less than preferences for faces\cite{vessel2010,vessel2018,leder2016,chen2025}.
Our results restate the reason: not that sharing is low, but that individual weights spread continuously along the dominant shared direction until the tail of the spread reaches the opposite side from the population.
Two findings support this restatement.
First, weak shared preferences alone do not determine the shape: jokes had the weakest shared preferences, yet the reversal-capable model gained little.
Second, the tail in LAPIS was not a discrete group: the distribution of weights was unimodal with its centre close to 0, so its natural tail crosses 0 without any need to assume a separate group.
The same held for photographs, where the tail lay along a preference for landscapes and architecture, a domain regarded as highly shared\cite{vessel2010,vessel2018}.
The loss under the \Do{} constraint, which needs no amortised inference, therefore serves as a practical indicator of which model a dataset requires, whatever the domain.

The size of the tail also depended on who was rating.
University students in Japan and China who rated the LAPIS artworks showed almost no tail, and whether a rater fell in the tail was uncorrelated across art, fashion and landscapes.
Falling in the tail thus belongs to the combination of stimulus domain and rater sample.
Our results therefore do not show that faces and art are intrinsically different domains; they show that different shapes of individual differences were observed for different combinations of domain and raters.
Whether the difference between the two samples derives from culture, recruitment method, age composition or the rating scale cannot be determined from a comparison of two samples.

Across datasets, the latent representation captured what a person prefers (male or female faces, abstract or figurative art, portraits) but not who the person is (race, gender, age, interest in art).
This does not contradict previous studies that reported relationships between demographic attributes and face impressions\cite{xie2019,hehman2017}.
It means that attributes explain little of the variance and that the main axes of individual differences are orthogonal to them.
It also complements, from the side of the perceiver, the report of Xu and colleagues that stereotypes about the race of faces depend on the stimulus set\cite{xu2026}.
Research on individual differences may therefore do better to take as its unit the latent representation estimated from judgments than to divide people into groups by self-reported attributes.

This study connects four lines of research (Supplementary Table~\ref{stab:related}, Supplementary Note~9).
The reweighting and linear-transform hypotheses descend from INDSCAL and IDIOSCAL\cite{carroll1970,carroll1972,tucker1972,carroll1974}; we rewrote them as generative models that amortised inference makes applicable to thousands of people with 20 judgments each.
Unlike psychological embeddings that are conditioned on the annotator\cite{andrews2023} or defined for the population\cite{zheng2019,hebart2020,muttenthaler2022,hebart2023}, our model handles people who do not appear in training and compares shapes other than reweighting.
Personalised aesthetic assessment and facial attractiveness prediction\cite{ren2017,whitehill2008,li2020,yang2022,zhu2022,lin2023,yun2024,ryu2026,maerten2026} ask how accurately ratings can be predicted, whereas we ask what differs between people.
And where variance decomposition and twin studies\cite{honekopp2006,hehman2017,xie2019,martinez2020,germine2015,sutherland2020} quantify how much people differ, we describe the shape of the difference.
The individualisation gain is small in units of correlation (0.02--0.04 for faces), but it exceeded the rater's own split-half reliability and the crowd consensus (Supplementary Note~8).

Four limitations should be kept in mind (details and further limitations are given in Supplementary Note~10).
First, this study is a reanalysis of public data and not a preregistered confirmatory experiment.
The four levels of hypotheses were formulated after analysing the first dataset, and the interpretation of the structure for artworks after seeing those results; the synthetic control and the second rater sample test that interpretation but do not replace preregistration.
Second, the size of the tail is a property of the sample.
The values for LAPIS (9.6\% of raters with a negative cosine, loss $+0.046$) should not be read as population constants: for the same artworks, the values in XPASS-Vis were 3\% and $+0.004$.
Third, model comparison alone cannot determine whether reversals are present, because a learned basis can absorb them.
We therefore relied on the sign check, whose loss measure shifts with the choice of basis without changing the ordering of the datasets.
Fourth, the evidence outside faces is limited to one artwork dataset with one additional rater sample, two photograph datasets whose results diverged and one joke dataset, and there are no data in which the same participants rated both faces and artworks.
We make no causal or neural claims.
Because a latent representation that predicts a person's judgments could be misused in hiring or surveillance, we release only the code and aggregate-level results, not the perceivers' latent representations.
Which conclusions replicated across datasets and controls, and which remain exploratory, is set out in Supplementary Note~11.

In conclusion, individual differences in judgment differ not only in size but also in shape.
Face judgments were described by same-sign reweighting of shared axes, whereas for artworks the weights spread continuously along the dominant shared direction until their tail lay on the opposite side from the population, to an extent that depended on the rater sample.
This shape can be measured separately from the size of the variance and from a small number of judgments per person, and it appeared as a continuous spread rather than as discrete subgroups.

\section*{Methods}

\subsection*{Ethics}
This study is a secondary analysis that re-analyses data that had already been collected, and no new data were collected.
Two kinds of data were used: anonymised public data, and anonymised data that are publicly distributed under password protection (XPASS-Vis).
The author had no access to information that identifies individuals.
At the author's institution, such secondary analyses of anonymised existing data are not subject to review by an ethics committee.
According to the original sources, ethics review and consent in the original data collection were as follows.
Xu et al.\cite{xu2026}, Keller \& Vosshall\cite{keller2016}, LAPIS\cite{maerten2025} and XPASS-Vis\cite{hayashi2026} report approval by an ethics committee and informed consent from the participants.
AVFS\cite{andrews2023} reports that the annotators gave consent (and states that no ethical review was conducted), and FaceSim3D\cite{hofmann2025} reports that the participants gave consent.
The London Set\cite{debruine2017} reports signed consent from the people photographed.
The original sources of SCUT-FBP5500\cite{liang2018}, AADB\cite{kong2016} and Jester\cite{goldberg2001} contain no statement on ethics review or consent, and none is given for the raters of the London Set.

\subsection*{Datasets}
We analysed datasets that met the following three conditions.
The first condition was that the rating or judgment data are public or can be obtained on request.
The second condition was that each rating carries a rater ID, so that the data can be analysed rater by rater.
The third condition was that there are enough ratings per person to be used both for estimating the person's latent representation and for evaluating it.
The datasets that met these conditions are summarised in Table~\ref{tab:datasets}.
Data that were considered as candidates but did not meet the conditions are described at the end of this section.
Below, we first summarise the licences and then describe, for each dataset, its content, how the raters used in the analysis were selected, and the basic properties of the data.
The licences are as follows.
The London Set is CC BY, AVFS is CC BY-NC-SA 4.0, FFHQ is CC BY-NC-SA 4.0 (the images are under their individual Flickr licences), FaceSim3D is CC BY 4.0, LAPIS is CC BY-NC 4.0, and EVA is CC0.
SCUT-FBP5500 and AADB may be used for research purposes only, and Jester may be used for research provided that the paper is cited.
The data of Xu et al.\ are public on OSF (nm7y8), and the data of Keller \& Vosshall are public as supplementary files of the paper.

\paragraph{Xu et al.\ 2026.}
The data of Xu et al.\ come from a task in which participants arranged faces on the screen.
From OSF (nm7y8), we obtained the trial-level arrangement coordinates (per-participant JSON files for Experiment 1 and annotation CSV files for Experiment 2), the participants' attributes (gender, age, race, whether Hispanic, and native language), and the free-text responses to the post-experiment questionnaire of Experiment 1.
In Experiment 1, 1,751 people each arranged 20 faces assigned from 100 White faces.
In Experiment 2, 1,683 people (433 university students, 1,250 online) each arranged 20 faces assigned from 96 faces.
Of these 96 faces, White, Black and Asian faces each make up one third.
In both experiments, 3--8 faces were arranged in a circular arena on each trial ``so that faces with similar personalities are close together'' (61 trials on average in Experiment 1 and 49 trials on average in Experiment 2).
To analyse the same data as the original paper, we re-implemented the trial and participant exclusion rules of the original paper and confirmed that the remaining trials and participants matched those of the authors exactly.
In addition, in Experiment 2 we excluded one person with fewer than 2 valid trials, so that 1,682 participants were analysed.
Before starting the analysis, we also confirmed that we could reproduce the population RDM, the 5-dimensional MDS and the within-participant consistency measures of the original paper.
Both experiments include a retest task in which the same participants performed the arrangement once more on a later day.
The retest task of Experiment 1 was performed with 40 faces, and these overlapped with the 20 faces that the person had arranged in the main task by only 8 faces on average, so most of the face pairs in the retest task are pairs that the person had never arranged (unseen pairs).
In contrast, the retest task of Experiment 2 was performed within the person's own 20 faces.
The mean split-half reliability, which indicates how stable the participants' judgments are, was about 0.20 in Experiment 1 and 0.323 in Experiment 2 (value over all trials; Spearman--Brown corrected 0.456), and the mean retest $\rho$ was 0.246/0.390.
Furthermore, to check whether the learned face embedding corresponds to external ratings, we used the ratings of Lin et al.\ on 100 trait words\cite{lin2021} for the 100 faces of Experiment 1.

\paragraph{London Set.}
The Face Research Lab London Set\cite{debruine2017} consists of attractiveness ratings (1--7) by 2,513 people of 102 faces (frontal, neutral).
We excluded 12 people whose ratings had no variance, because individual differences cannot be examined for them, and analysed the remaining 2,501 people.
These data are accompanied by the raters' gender and sexual orientation.
As a guide to the size of individual differences, we examined the residual (the rating after removing the rater mean and the face mean); the residual accounted for 47\% of the variance.
This residual was not only noise: when the raters were split into halves, the principal components of the residual agreed (cosines 0.99, 0.97, 0.96, 0.92).

\paragraph{AVFS.}
A View From Somewhere\cite{andrews2023} (Zenodo 7878655) consists of odd-one-out judgments of face similarity, in which the odd one out is chosen from three faces, and comprises three sets.
The first, the main set, consists of 638,180 unique triplets of 4,921 faces from FFHQ\cite{karras2019} (one judgment each, 1,645 people, median 260 judgments), of which we analysed the 1,283 people with 100 or more judgments.
The second, the novel-stimulus set, consists of all 27,720 triplets of 56 faces, judged by 632 people (about 127 judgments per person), and of these we analysed the 504 people with 60 or more judgments.
The third, the shared set, consists of 1,000 triplets each judged by 24 people.
These data are accompanied by the annotators' gender, age and ancestry region.
Image features were obtained from the FFHQ thumbnails (128 $\times$ 128).

\paragraph{FaceSim3D.}
FaceSim3D\cite{hofmann2025} (OSF 4a2jf) also consists of triplet judgments of faces.
In these data, 100 faces from the Chicago Face Database\cite{ma2015} were reconstructed in three dimensions with DECA, and triplet judgments were collected in a 2D static condition and in a 3D dynamic condition.
To remove the influence of inattentive participants, we first excluded the trials on which a catch trial was failed, and then took as the target of analysis the participants with a catch failure rate of 0.2 or less and 20 or more trials (1,267 people in 2D and 1,087 people in 3D).
Of these, the 1,219/1,053 people with 100 or more triplets were used for model training and evaluation.
Note that 98\% of the participants are White.

\paragraph{SCUT-FBP5500.}
SCUT-FBP5500\cite{liang2018} is a dense dataset without missing values in which all 60 people rated 5,500 faces (2,000 Asian female, 2,000 Asian male, 750 White female, 750 White male) on a 1--5 scale.
Of the 60 people, 7 rated all faces a second time, so the stability of the ratings can be examined.
The retest correlation of the residuals of single ratings was low, at 0.12, but this is due to the noise that accompanies integer ratings on a 1--5 scale.
In contrast, the similarity that expresses how alike the per-rater residual profiles are between raters had a high split-half reliability of 0.975, so that individual differences can be measured stably when many faces are pooled.
In addition, the between-rater similarity of the residuals correlated at $r = 0.67$ between Asian faces and White faces, so that individual differences were preserved across face race.

\paragraph{LAPIS.}
LAPIS\cite{maerten2025} (Hugging Face asmaerten/LAPIS) consists of aesthetic ratings (0--100) by 555 people of 11,723 artworks (figurative and abstract, 26 styles, 7 genres).
The median number of ratings per person is 479 artworks, and each artwork was rated by 24 people.
These data are accompanied by the raters' age, nationality, gender, education and interest in art (VAIAK).
Of these raters, we analysed the 553 people with 100 or more ratings.
The residual accounted for 62\% of the variance, so individual differences are large in these data.
These individual differences are stable: each rater's abstract--figurative preference (the rating of abstract art minus the rating of figurative art) agreed at $r = 0.96$ when the artworks were split into halves.

\paragraph{AADB.}
AADB\cite{kong2016} consists of aesthetic ratings of photographs.
So that raters could be distinguished, we used the raw MTurk data (with WorkerId, 5,018 HITs $\times$ 10 photographs) after reformatting them.
The reformatted data consist of 50,180 ratings (0--5, 5.0 people per photograph) of 9,958 photographs.
Of these raters, we analysed the 68 people with 100 or more ratings, and these 68 people account for 94\% of the ratings.
The residual accounted for 46\% of the variance (image mean 46\%).
For the images we used the distributed 256 $\times$ 256 version.

\paragraph{EVA.}
EVA\cite{kang2020} also consists of aesthetic ratings of photographs, comprising the ratings (0--10, 30 or more people per photograph) by 1,094 people of 4,070 photographs (6 content categories) from AVA\cite{murray2012}.
These data are accompanied by the raters' age, region, photographic experience and gender.
Of these raters, we analysed the 175 people with 100 or more ratings (median 326 ratings).
The residual accounted for 49\% of the variance (image mean 26\%).

\paragraph{Jester.}
Jester Dataset 3\cite{goldberg2001} consists of funniness ratings ($-10$--$+10$) by 54,905 people of 150 jokes.
Of these raters, we analysed the 11,067 people who rated 50 or more jokes (median 87 jokes).
The residual accounted for 60\% of the variance, and the joke mean for 8\%.

\paragraph{Keller \& Vosshall.}
The data of Keller \& Vosshall\cite{keller2016} consist of ratings of odour pleasantness, and from Supplementary File 1 we used the ratings (0--100) by 55 people of 480 molecules $\times$ 2 concentrations.
Of these, 20 molecules were presented repeatedly, so the stability of the ratings can be examined.
The concentrations were given at dilutions that differ between molecules.
For the main analysis, we used the 421 molecules presented at a dilution of 1/10 or 1/1,000 (55 $\times$ 421 stimuli, 90\% observed).
To confirm that the results do not depend on this choice, we also analysed the 55 $\times$ 766 stimuli obtained by treating each combination of molecule and concentration band (1/10 or 1/1,000 versus 1/100,000 or 1/10,000,000) as a separate stimulus (479 molecules whose structure could be retrieved), and confirmed the same tendency.
The residual accounted for 68\% of the variance, and the retest correlation of the residuals over the repeated presentations was 0.31.

\paragraph{XPASS-Vis.}
XPASS-Vis\cite{hayashi2026} can be downloaded by anyone, but the files are password-protected.
We contacted its authors and received the password.
In these data, 129 people, mainly university students in Japan and China (87 Japan, 41 China, 1 other; mean age 29 years), gave 184--247 ratings to stimuli in each of 3 domains.
The 3 domains are art (the 2,345 artworks of the LAPIS test split), fashion (2,082 images) and landscapes (2,099 images).
There are 87,836 ratings in total, and of the several rating items we used the overall aesthetic evaluation (0--6).
These data are accompanied by the Big Five measured with the TIPI, age, and interest in each domain.
In each domain 12 stimuli were presented twice, and the within-person retest correlation was 0.59--0.61.
To keep the image embeddings consistent with the other data, we reused the LAPIS CLIP embeddings for the art images and embedded the fashion and landscape images with the same CLIP ViT-B/32.

These data were used for two purposes.
The first purpose was to examine, by comparison with LAPIS, whether the results change when the same artworks are rated by a different rater sample.
For this we performed two kinds of analysis.
One was a comparison with controls in which LAPIS was subsampled to the same scale as XPASS-Vis (2 samples each), by which we checked whether the difference between the two datasets can be explained by sample size alone.
The subsampling was done in two ways: reducing the raters to 129 people (keeping all their ratings), and reducing the data to 129 people $\times$ 227 ratings.
The other was between-population transfer, in which a model trained on one population is applied to the raters of the other population.
Specifically, we applied the model trained on LAPIS to the XPASS-Vis raters and, conversely, applied the model trained on XPASS-Vis art to the LAPIS raters.
For the latter, we used the ratings that LAPIS raters gave to the same 2,345 artworks (60 or more ratings).
The second purpose was to examine whether the individual differences of the same raters are common across domains.
Two methods were used for this between-domain comparison.
These were the Spearman correlation of per-rater measures and the Mantel test on the between-rater correlation matrices.
For the Mantel test, we used the rater pairs with 10 or more stimuli in common (825 pairs).
Note that the above conclusions did not change when the rating item was replaced by Like or Beautiful.

\paragraph{Data that did not meet the conditions.}
Of the candidate data, the following did not meet the conditions above and were therefore not analysed.
Those that did not meet the first condition (being obtainable) were the multi-domain data of Vessel et al.\cite{vessel2018}, the art ratings of Iigaya et al.\cite{iigaya2021} and PARA\cite{yang2022}.
The first two are not public, and PARA, a set of photograph ratings accompanied by the Big Five, is available on application only and could not be obtained at the time of the analyses of this study.
Those that did not meet the second condition (having rater IDs) were One Million Impressions\cite{peterson2022} and the Minimum Semantic Content dataset\cite{penacchio2026}.
The one that did not meet the third condition (number of ratings per person) was the architectural interior data of Coburn et al.\cite{coburn2020}, in which there were 13 images rated per person.

\paragraph{Screening before modelling.}
Before fitting the models, we checked in two ways whether each dataset contains stable individual differences.
First, for each dataset we decomposed the variance of the ratings into the rater mean, the stimulus mean and the residual, and computed the split-half reliability of the residual preference profiles.
Second, for the datasets with many raters, we computed how much a model-free collaborative filter exceeds the population mean (the gain) and used this as a reference value.
This collaborative filter is a ridge regression from the $n_{\mathrm{est}}$ residuals of a held-out-fold rater onto the residual profiles of the training raters.
The reference values were $+0.04$/$+0.06$ for London (20/50 ratings), $+0.01$--$0.03$ for LAPIS, and $+0.06$/$+0.10$/$+0.13$ for Jester (20/50/100 jokes).
In contrast, they were $+0.01$ or less for AADB, EVA, SCUT and Keller.
This method works only when there are many training raters, so this reference is insensitive for datasets with a few dozen raters.

\subsection*{Stimulus embeddings}
The model represents each stimulus $i$ by a $D$-dimensional vector $\bm f_i \in \mathbb R^{D}$ (the stimulus embedding, $D = 8$).
The stimulus embedding was constructed in one of the following three ways, depending on the data.
First, for the Xu arrangements, the rating version of London, the novel set of AVFS and FaceSim3D, the stimulus embedding was learned directly from the judgments.
Of these, for Xu it was initialised by classical MDS of the population RDM of the training folds.
Second, in the image version, the stimulus embedding was built from image features as $\bm f_i = \bm P \bm e_i$.
Here $\bm e_i$ is obtained by principal component analysis, on the training stimuli, of the image features of CLIP ViT-B/32 (laion2b\_s34b\_b79k\cite{radford2021,cherti2023}) (32 components for faces, 64 components for LAPIS, AADB and EVA), and $\bm P$ is a learned linear projection.
In this form, the embedding is determined by the image features alone, so an embedding can also be given to stimuli that were not used in training.
Third, in the hybrid version (London, SCUT, Jester, Keller), the two were added together as $\bm f_i = \bm g_i + \bm P \bm e_i$.
Here $\bm g_i$ is an item embedding that is learned freely for each stimulus, and it was dropped out with probability 0.5 (0.2 for SCUT) during training.
Owing to this dropout, the model becomes able to predict even without $\bm g_i$, so unseen stimuli ($\bm g_i = \bm 0$) can also be handled.
For stimuli other than images, the following were used as the features $\bm e_i$.
For jokes, we used the sentence embeddings of all-mpnet-base-v2\cite{reimers2019}.
For odour molecules, we used 181 two-dimensional descriptors computed with RDKit\cite{rdkit} from the PubChem SMILES (after removing constant and missing columns, applying a log1p transform and standardising).

\subsection*{Decoder: four levels of individual differences}
The decoder is a model that receives the latent representation $\bm z_k \in \mathbb R^{M}$ ($M = 4$) of person $k$ and predicts that person's judgments.
To compare four hypotheses about the shape of individual differences, we divided into four levels which part of the model $\bm z$ can change.
The levels are nested, each containing the preceding one.
\Dz{} is a shared model that does not use $\bm z$ and makes the same prediction for everyone.
\Do{} changes the weight of each dimension from person to person but keeps the sign of the weights the same for everyone.
Specifically, the dimension weights were set to $\bm w(\bm z) = \bm w_0 \odot \mathrm{softplus}(\bm h(\bm z)) > 0$.
\Dt{} removes the constraint on the sign of the weights and allows a free linear direction $\bm w_0 + \bm W(\bm z)$.
In the arrangement and triplet tasks, this corresponds to a general metric $\bm A(\bm z) \bm f$ ($\bm A(\bm z) \in \mathbb R^{D \times D}$).
\Dh{} adds to \Dt{} stimulus-specific offsets $\bm u(\bm z)$ (in $\mathbb R^{F}$ for ratings and in $\mathbb R^{F \times D}$ for arrangements and triplets; $F$ is the number of stimuli) and allows stimulus-by-stimulus individual differences that cannot be expressed by the shared dimensions.
These $\bm h$, $\bm W$, $\bm A$ and $\bm u$ are all small multilayer perceptrons that take $\bm z$ as input (hypernetworks\cite{ha2017}).

The hypernetworks had the same architecture for all datasets and levels.
Each is a fully connected network with one hidden layer (64 units, GELU activation).
Its input is the $M$-dimensional $\bm z$, and a linear output layer outputs the parameters of that person's prediction model.
The number of output parameters depends on the task and the level.
For ratings, the prediction model is the linear readout $\bm w(\bm z) \cdot \bm f_i + b(\bm z)$.
The hypernetworks of \Do{} and \Dt{} output $D$ values ($\bm h(\bm z)$ or $\bm W(\bm z)$).
In \Dh{}, a further hypernetwork outputs $F$ stimulus-specific offsets, which are added to the predicted ratings.
The intercept $b(\bm z)$ is output by a separate hypernetwork of the same architecture (one output).
$\bm w_0$ is a learned parameter shared by everyone, and the readout of \Dz{} is $\bm w_0 \cdot \bm f_i + b_0$.
For arrangements and triplets, the prediction model is a metric on the embedding space.
The hypernetwork of \Do{} outputs $D$ positive dimension weights (for arrangements there is no $\bm w_0$, and $\mathrm{softplus}(\bm h(\bm z))$ is used directly as the weight).
The hypernetwork of \Dt{} outputs the $D^2 = 64$ elements of the $D \times D$ matrix $\bm A(\bm z)$.
In \Dh{}, a further hypernetwork outputs a $D$-dimensional offset for each stimulus ($F \times D$ values in total), which is added to the embedding.
The output layers were initialised so that the prediction model is close to a model without individual differences at the start of training.
Specifically, for \Do{} with ratings and triplets, a constant was added to the output so that the multiplier $\mathrm{softplus}(\cdot)$ starts at 1.
For \Dt{} with arrangements and triplets, $\bm A(\bm z)$ starts near the identity matrix, and for \Dh{} the offsets start near 0.
Apart from the sign constraint imposed by the softplus in \Do{}, no constraint or regularisation was applied to the outputs of the hypernetworks.

The way in which the decoder output is compared with the actual judgments (the readout and the loss) was defined for each task as follows.
For arrangements, the Euclidean distances between weighted embeddings were compared with the on-screen distances within a trial, and the within-trial Pearson correlation was used as the loss.
The correlation was used so that the scale of the distances may differ from trial to trial, which follows inverse MDS\cite{kriegeskorte2012}.
For triplets, the loss was the cross-entropy with respect to the softmax of the three pairwise similarities (inner product or general metric).
For ratings, the ratings $z$-scored within each rater were predicted by the linear readout $\bm w(\bm z) \cdot \bm f_i + b(\bm z)$, and the squared error was used as the loss.

\subsection*{Encoder and training}
The $\bm z$ used by the decoder is estimated by the encoder from the person's judgments.
That is, the encoder is a network that takes the set of one person's judgments as input and outputs $\bm z$.
The input to the encoder consists of per-item features, namely the concatenation of the stimulus embedding $\bm f$, the judgment and their product (for arrangements, the distance of the face pair and its residual from the population prediction).
So that the same $\bm z$ is obtained regardless of the order or number of the judgments, we used for the encoder a permutation-invariant network\cite{zaheer2017} that aggregates the per-item features by attention pooling.
This network consists of three stages.
In the first stage, the per-item features are passed through two fully connected layers (the activation function is GELU, and the hidden dimension is 128 for Xu and 64 otherwise).
In the second stage, a single linear layer assigns a score to each item from this output, and a weighted sum is taken with weights given by the softmax over the items of the same person.
In the third stage, the aggregated vector is mapped to the $M$-dimensional $\bm z$ (a fully connected network with one hidden layer for Xu, and a single linear layer otherwise).

The encoder and the decoder (and the learned stimulus embeddings) were trained jointly.
In training, the model was made able to predict other judgments of the same person from a $\bm z$ estimated from a small number of judgments.
To this end, in every epoch each person's judgments were randomly split into an estimation subset (5--55\%; 30--80\% for Xu) and a target subset, $\bm z$ was obtained from the estimation subset, and the loss on the target subset was minimised.
As the form of the encoder, we first tried a variational encoder with KL regularisation\cite{kingma2014}, but it underwent posterior collapse (the between-participant variance of $\bm z$ became 0 and the decoder ignored $\bm z$).
We therefore adopted a version in which an L2 penalty on $\bm z$ ($10^{-3}$) is imposed on a deterministic encoder.

To check whether the estimation by the encoder (amortised inference) is good enough, we also evaluated ``direct inference'' as a comparison.
In direct inference, the trained decoder is fixed and $\bm z$ is optimised directly by a gradient method so as to fit the same estimation data (Adam, 200 steps, learning rate 0.05, L2 on $\bm z$ $10^{-2}$).
The comparison showed that amortised inference was equivalent to direct inference with 20--50 or more judgments and was better than direct inference with fewer judgments (for example, with 5 faces in London, \Do{} $+0.061$ and \Dt{} $+0.172$).

The hyperparameters were as follows.
For Xu, we used 100 epochs, a batch of 32 people, a learning rate of $5 \times 10^{-3}$ (cosine schedule) and MDS initialisation.
For the rating and image versions, we used 300 epochs (600 for Keller and SCUT), a learning rate of $3 \times 10^{-3}$, a hidden layer of 64 and an L2 penalty on $\bm P$ of $10^{-3}$ ($10^{-4}$ for LAPIS).
The L2 penalty on the item embeddings was $10^{-3}$ per item.

\subsection*{Cross-validation and evaluation}
All model evaluation was performed on raters who were not used in training.
To this end, the raters were split into 10 folds (for Xu, stratified by experiment, gender, race and source), and the held-out-fold raters were not used at all in training the model, including the stimulus embeddings.
The population baseline that serves as the reference for comparison (the prediction from the average judgment of the population) was also, for the same reason, built from the raters of the training folds only.
This is because, if it were built from everyone including the person being predicted, that person's judgments would be mixed into the population average and the prediction accuracy of the population baseline would come out about 0.03 higher than it actually is.

In the versions that use image, text or molecular features, the stimuli were also split into 5 folds in order to evaluate transfer to stimuli that were not used in training (the stimulus fold is the rater fold mod 5).
In this case, training was performed on training raters $\times$ training stimuli only, and the ``unseen'' evaluation was performed on the stimuli that were not used in training.
Because unseen stimuli have no population mean rating, the population baseline on unseen stimuli was a ridge regression from the stimulus features to the population mean, fitted on the training stimuli.

The held-out-fold raters were evaluated by the following procedure.
First, $\bm z$ was estimated from $n_{\mathrm{est}}$ estimation judgments among that rater's judgments, and then we examined how well the remaining judgments could be predicted.
For Xu, the held-out 30\% was fixed.
The measure of predictive performance is Pearson $r$ for ratings, Spearman $\rho$ for arrangements and accuracy for triplets.
For the learning curves, which show how prediction changes with the number of judgments, the held-out set was fixed and only $n_{\mathrm{est}}$ was varied.
The number of seeds was 3 for Xu (2 for $M = 16$), 3--5 for London (5 for the learned-embedding version), 2 for AVFS (5 for the novel 56-face set), 2 for LAPIS, 3 for AADB, EVA and SCUT, and 2 for Jester.
For comparisons between models, the measure was first averaged over seeds for each rater, and the between-model difference in these values (the paired difference) was tested with a paired $t$-test (two-sided).
Unless stated otherwise, the $P$ values in the text and tables are from this $t$-test.
The results of the Wilcoxon signed-rank test, which does not assume normality of the differences, are given alongside in Supplementary Table~\ref{stab:stats}.
For the Xu data, to confirm that the number of latent dimensions did not limit the results, \Do{} and \Dh{} were also trained with $M = 16$ (2 seeds), with $\bm z$ obtained both by amortised inference and by direct optimisation.
The s.e.m.\ was computed across raters.

The baselines with which the models were compared are as follows.
These are the population RDM or mean, a shrinkage blend of the person's own estimation data and the population (with the mixing ratio selected on the training participants), the image ridge regression for unseen stimuli, the split-half reliability of the person's own estimation data (a ceiling, Xu), and the leave-one-out majority vote of the other annotators on the shared triplets (AVFS).

\subsection*{Controls with simulated participants}
To interpret the results of the model comparison, we need to know whether this method can detect a given shape of individual differences when it is actually present.
We therefore generated simulated participants whose shape of individual differences is known in advance and analysed them with the same procedure as the real data.
Simulated participants were created for two purposes.

The first purpose was to examine whether \Do{}-type and \Dt{}-type individual differences can be detected in the arrangement task of Xu.
To this end, simulated participants were generated on the fold-0 \Dz{} embeddings of Xu Experiments 1 and 2, with the same trial structure as the real participants.
The individual differences of the simulated participants were of the \Do{} type (log-normal dimension weights, SD 0.6) or the \Dt{} type ($\bm A_k = \bm I + \epsilon \sum_m z_{km} \bm B_m$, rank 4).
To equate the size of individual differences between the two types, $\epsilon$ was calibrated so that ``the Spearman $\rho$ between the individual's true RDM and the population RDM'' matched that of the \Do{} type (0.865--0.879).
The noise was calibrated so that the split-half reliability of the simulated participants matched that of the real participants (0.22 in Experiment 1 and 0.32 in Experiment 2).
Split-half reliability here is the Spearman $\rho$ between the RDMs obtained from a random 70\% and the remaining 30\% of the trials, and the same definition was applied to the real participants.
The \Do{} and \Dt{} models were trained on these data and evaluated on the held-out folds ($M = 8$, 100 epochs, 2 seeds).
To know the upper bound when there is no estimation error, we also evaluated an oracle that was given the true $\bm z$.
Separately from this, we also confirmed, with simulated participants having known weights, that the encoder can recover the weights used for generation (canonical correlations 0.89/0.77).

However, this \Dt{}-type control is a continuous linear transform and does not include discrete reversal (only some people reading an axis in the opposite direction).
As the second purpose, we therefore evaluated with separate simulated raters how far this method can detect and distinguish discrete reversal and continuous spread.
To bring this control close to the real data, we generated 500 simulated raters $\times$ 480 ratings with known structure on the actual CLIP features of LAPIS and the fold-0 population weights.
The structure of the simulated raters had two conditions.
In the ``continuous'' condition, each rater's weight was the sum of a normal distribution along one particular axis and a small orthogonal component.
This axis lies along the abstract--figurative direction and makes $|\cos| = 0.59$ with the population weight (the projection of the population is 0.62 times the SD of the raters).
In the ``discrete'' condition, 10\% of the raters reflected the abstract component of the population weight.
In both conditions, the magnitude of the noise was calibrated to the consensus (0.439) and the split-half reliability of the weights (0.875) of LAPIS measured with the same code.
On these simulated data, \Dz{}--\Dt{} were trained with the same procedure as for LAPIS (10 rater folds $\times$ 5 stimulus folds, 2 seeds), and the sign check of the next section was performed.

\subsection*{Sign check (analysis of reversal) and axes of weight variation}
When \Dt{} exceeds \Do{}, the model comparison alone cannot tell whether the reason is that there are raters whose weights are opposite in sign to the population (reversal).
We therefore obtained the weights of each rater directly, separately from the amortised model, and compared their signs with the population weight.
We call this analysis the sign check.

First, the weights of each rater were obtained as follows.
On the embedding basis of each \Dt{} fold model (with the dimensions standardised), for each held-out-fold rater we ridge-regressed the rater's $z$-scored ratings of all of his or her training stimuli and obtained a weight vector $\bm w_k$.
Next, this $\bm w_k$ was compared with the population weight $\bm w_{\mathrm{pop}}$ obtained by pooling the training raters.
The following four measures were used for the comparison.
(i) The proportion of raters with $\cos(\bm w_k, \bm w_{\mathrm{pop}}) < 0$, that is, the proportion of raters whose overall weight direction is opposite to the population.
(ii) The proportion $U$ of the weight energy that lies on dimensions whose sign is opposite to $\bm w_{\mathrm{pop}}$.
Because $U$ is positive from estimation noise alone, we also computed a noise-corrected version based on the product of two estimates $\bm w_a$ and $\bm w_b$ from split halves of the ratings, $\sum_{j \in \mathrm{flip}} w_{aj} w_{bj} / \sum_{j} w_{aj} w_{bj}$ (the expected value for a dimension whose true weight is 0 is 0).
Here $w_{aj}$ and $w_{bj}$ are the components of $\bm w_a$ and $\bm w_b$ on the $j$-th dimension, and $\mathrm{flip}$ is the set of dimensions on which the sign of $\bm w_k$ is opposite to that of $\bm w_{\mathrm{pop}}$.
(iii) The loss under the \Do{} constraint, that is, the difference in Pearson $r$, in split-half cross-validation, between a free linear readout and a readout under the \Do{} constraint ($\bm w_{\mathrm{pop}} \odot \bm s$, $\bm s \ge \bm 0$, non-negative least squares).
If this difference is positive, the constraint that keeps the sign of the weights the same as the population makes the prediction for that rater worse.
(iv) The Spearman correlation between these measures and the per-rater \dTO{} of the amortised model, and the mean \dTO{} in each quartile of $U$.
Measure (iv) serves to check whether the advantage of \Dt{} is larger for raters whose signs are more reversed.
The numbers of estimation ratings for \dTO{} are 100 for LAPIS, 100 for EVA, 40 for London, 500 for SCUT and 50 for Jester.

Two cautions apply to the reading of these measures.
First, the sign of each dimension changes with the learned (non-orthogonal) basis.
However, \Do{} itself is defined in that basis, and measure (i), which does not change under rotation of the basis, gave the same conclusion, so this point does not affect the conclusion.
Measure (iii) is also a dimension-wise constraint and its value can therefore change with the choice of basis, but the conclusion held when the basis was changed (Supplementary Note~6).
Second, \Do{} can absorb discrete reversal by duplicating a dimension with the same content with the opposite sign.
In that case the advantage of \Dt{} becomes smaller, so in deciding whether reversal is present we treated (iii), which does not depend on amortised inference, as primary and (i) as secondary.

Next, we examined in which directions and how the raters' weights are scattered (the structure of the weight variation).
To this end, in the basis of each \Dt{} fold model, we eigendecomposed the between-rater covariance of $\bm w_k$.
The following quantities were computed for each fold model and their medians were reported (Fig.~\ref{fig:sign}c,d show the distributions pooled over fold models).
These are the variance ratio of the first eigenvector, the effective number of axes (participation ratio), the $|\cos|$ between the first eigenvector and the population weight, the projection of the population weight onto the axis divided by the SD of the raters' scores, the proportion of raters on the opposite side from the population along the axis, and the bimodality coefficient (bimodal above 0.555).
In addition, we examined what content this axis corresponds to.
For LAPIS, we computed the Spearman correlation between the scores on the axis and the abstract--figurative preference, and the correlation between the loadings and the correlation of each dimension with abstract preference.
For EVA, we computed the correlations between the scores on the axis and the residual preferences for the 6 content categories.

We also examined whether the duplication described above actually occurs, that is, whether \Do{} can reverse a content axis on the learned basis.
First, the embedding projection $\bm P$ of the \Do{} fold models was expressed in the principal component basis of the stimuli, and the contribution of each dimension to a content direction $\bm u$, $a_j = w_{0j}(\bm P_j \cdot \bm u)$, was computed.
Next, these contributions were divided by the sign of $w_{0j}$, and we tallied the capacity that can read out the content direction in the same direction as the population and the capacity that can read it out in the opposite direction.
Furthermore, for the held-out-fold raters (estimated from 100 ratings), we computed the proportion for whom the readout of the content direction actually given by \Do{} and \Dt{} was opposite in sign to the population.

Finally, we examined what the raters in the tail of EVA evaluate in the opposite direction to the population, using the residual preferences for each of the 6 content categories.
The categories are animals, architecture and urban scenes, people, natural landscapes, still life and close-ups (photographs whose subject is arranged objects, or photographs of a subject taken from close by), and other (confirmed by visual inspection of the images).
The raters' residual preferences were compared between the top quartile of $U$ (44 people) and the rest (131 people) with the Mann--Whitney test.

\subsection*{Alternative expressions of effect size}
The gain in correlation over all stimuli can make the effect of individualisation look small.
This is because, for many stimuli, the population mean already gives a good prediction, and individualisation is useful for the part that the population mean cannot predict.
Therefore, in addition to the gain in correlation, we reported two measures restricted to the part that the population mean cannot predict, alongside the values for all stimuli.

(1) The $r$ on contested stimuli.
This is the $r$ computed only on the stimuli whose per-stimulus SD of the training raters' ratings (after removing the rater mean) falls in the top third.
Only the training raters are used to select the stimuli, and no information from the held-out-fold raters is used.

(2) The discrimination accuracy for tied pairs (tied-pair accuracy).
This is the proportion of cases in which the model correctly predicted which of two stimuli the person rates higher, for two stimuli that are almost tied in the population mean.
Specifically, among the pairs of the person's held-out stimuli, we used those for which the difference in the population mean was less than 0.1 SD and the person's ratings differed.
We computed the proportion of these pairs for which the model predicted the person's ordering correctly (0.5 if the predictions were tied).
A maximum of 4,000 pairs per person was used, and raters with 20 or more pairs were included.
On this measure, the population model is about 0.5 by construction.

Depending on the dataset, the treatment of these measures was changed as follows.
For AADB, where there are few ratings per stimulus, the per-stimulus SD cannot be computed stably, so (1) was not defined.
EVA was reported on the side of unseen photographs.
This is because, on the side of the rater's own training photographs, the population baseline based on the image mean of 30 people is stronger than the 8-dimensional image model.
London was, for this analysis only, retrained with a re-implementation for dense matrices (the hybrid version) (slightly lower than the rating version, at 0.589 $\to$ 0.607 with 20 faces).

\subsection*{Interpretation of the latent representation and decoding of attributes}
To examine what the latent representation $\bm z$ represents, we examined whether the raters' attributes and preferences can be predicted from $\bm z$.
In this analysis, $\bm z$ was obtained by encoding all participants with each fold model.
For prediction we used ridge regression (continuous variables, $R^2$) or logistic regression (categories, balanced accuracy) with 10-fold cross-validation within each model, and compared the result with chance level by a permutation test (30 permutations).
For Xu Experiment 2, because individual differences in the reliability of the judgments could produce apparent prediction, we added a $\bm z$ residualised by regression on the split-half reliability and a control that used the reliability alone.

In addition, we examined the following three points.
First, the in-group effect was examined with the mean of the decoded dissimilarities ($z$-scored within participant) within same-race face pairs.
Specifically, we tested whether this mean changes depending on whether the race of the perceiver and the race of the faces match.
Second, the correspondence between the learned face embedding and the 4 factors of Lin et al.\cite{lin2021} was evaluated by regression from the fold-0 \Dz{} embedding to each factor score.
The measures are the $R^2$ of 10-fold cross-validated ridge regression and the canonical correlation.
Third, the stability of $\bm z$ was evaluated by the mean over dimensions of the ICC(2,1)\cite{shrout1979} of the $\bm z$ estimated separately from the estimation trials and from the held-out trials.

\subsection*{Sensitivity analyses}
To confirm that the main results do not change with the model settings, we checked robustness by varying $D \in \{5, 8, 16\}$, $M \in \{2, 4, 8\}$ and the seed (1 seed $\times$ 10 folds for Xu; 5 seeds for London and AVFS).
For \Do{} in Xu Experiment 2, $D = 8$ and $M = 2$--4 were optimal, and there was a tendency to overfit with $D = 16$ or $M = 8$.
In contrast, \Dt{} in Experiment 1 improved as $M$ was increased (0.228 at $M = 8$).
However, the conclusions did not change under any of the settings.
We also tried a version in which the stimulus embeddings were restricted to be non-negative (softplus), but it lowered performance by 0.02--0.03 in both experiments and was therefore not adopted.
For Jester, we also evaluated $D = 16$, $M = 8$ and $D = 32$, and the gain was almost saturated at $D = 16$.

We also checked whether the results of the image version change with the choice of image features.
Instead of CLIP ViT-B/32, we used the self-supervised DINOv2 ViT-B/14\cite{oquab2024} and the larger CLIP ViT-L/14, and re-analysed the image versions of LAPIS, EVA, AADB and London with the same raters, folds, seeds, $D$, $M$ and principal component analysis procedure.
The qualitative conclusions did not change with any of the image features (Supplementary Note~6).

\subsection*{Conduct of the analyses and use of AI}
The implementation and execution of the analysis code were carried out on the author's server by an AI coding agent (Claude Code, Anthropic; model Claude Fable 5.1) following the research plan and execution specifications prepared by the author, and all judgments on the reports of each stage and all recording of decisions were made by the author.
The author verified all numbers with a script that regenerates the numbers of this paper from the result files.
The same AI was used as an aid in writing the manuscript, and the author checked and revised the entire text.

\subsection*{Statistical analyses and reproducibility}
All tests are two-sided.
For comparisons between models, we report the $P$ values of the $t$-test on the rater-level paired differences, with those of the Wilcoxon signed-rank test alongside in Supplementary Table~\ref{stab:stats}.
As the effect size we used Cohen's $d_z$ for paired differences (the mean of the differences divided by their standard deviation), and confidence intervals are 95\% intervals from the $t$ distribution.
The test statistics, effect sizes and confidence intervals of the main comparisons are collected in Supplementary Table~\ref{stab:stats}.
This study is a reanalysis of public data, and the hypotheses and analysis plan were not preregistered.
Sample sizes were determined by the public datasets, and no power analysis was performed in advance.
Small $P$ values are given in the form $10^{-n}$.
Sample sizes are stated with each analysis, and the exclusion criteria for raters are as stated in the dataset paragraphs above.
The software was Python 3.12 (numpy, scipy, scikit-learn, PyTorch, rsatoolbox 0.3.2, open\_clip, sentence-transformers, RDKit).
All random seeds were fixed.
Therefore, all numbers can be recomputed from the public data and the public code (the numbers based on XPASS-Vis require the password obtained from its authors).

\section*{Acknowledgements}
We thank Xu et al., DeBruine and Jones, Andrews et al., Hofmann et al., Liang et al., Maerten et al., Kong et al., Kang et al., Goldberg et al., and Keller and Vosshall for making their data publicly available, and Hayashi et al.\ for responding to our request for access to XPASS-Vis.

\section*{Funding}
H.H. discloses support for the research of this work from JSPS KAKENHI [grant numbers JP24K15047 and JP22H05163].

\section*{Data availability}
All data used in this study are publicly available.
The XPASS-Vis files, however, are password-protected, and their use requires a request to the original authors.
Xu et al.\ 2026: OSF \url{https://osf.io/nm7y8/}.
London Set: figshare \url{https://doi.org/10.6084/m9.figshare.5047666}.
AVFS: Zenodo \url{https://doi.org/10.5281/zenodo.7878655}; FFHQ: \url{https://github.com/NVlabs/ffhq-dataset}.
FaceSim3D: OSF \url{https://osf.io/4a2jf/}, Zenodo \url{https://doi.org/10.5281/zenodo.19236583}.
SCUT-FBP5500: \url{https://github.com/HCIILAB/SCUT-FBP5500-Database-Release}.
LAPIS: \url{https://huggingface.co/datasets/asmaerten/LAPIS}.
AADB: \url{https://github.com/aimerykong/deepImageAestheticsAnalysis}.
EVA: \url{https://github.com/kang-gnak/eva-dataset}.
Jester: \url{https://eigentaste.berkeley.edu/dataset/}.
Keller \& Vosshall 2016: Additional file 1 of the article.
XPASS-Vis: \url{https://github.com/Takato0884/proj-XPASS-Vis} (the link to the data is in the README of this repository; the zip files are password-protected, and the password is obtained by request to the original authors\cite{hayashi2026}).
Participant-level data and the perceivers' latent representations are not redistributed.
The one exception is the source data of Supplementary Fig.~\ref{sfig:effect}c (for each of the 2,501 raters of the London Set, the male-face preference index and its predicted value, with no identifiers), which are released as figure source data in the repository below.

\section*{Code availability}
The analysis code, run logs, aggregate-level results and the source data of the figures are available on GitHub (\url{https://github.com/labbip/face_id}) and archived on Zenodo (\url{https://doi.org/10.5281/zenodo.24302873}).
The code is released under the MIT licence, and the aggregate results and figures under CC BY 4.0.
For the analyses that use XPASS-Vis, only the code is released; their outputs and run logs are not included.
The results based on XPASS-Vis are reported only in the text and tables of this paper.

\section*{Competing interests}
The author declares no competing interests.

\section*{Author contributions}
H.H.: Conceptualization, Methodology, Software, Formal analysis, Investigation, Writing -- original draft, Writing -- review \& editing, Visualization, Funding acquisition.

\bibliography{refs}
\fi %

\ifsipart
\ifmainpart
\clearpage
\section*{Supplementary Information}
\else
\begin{center}
{\LARGE\bfseries Supplementary Information\par}
\vspace{3mm}
{\large The shape of individual differences in judgment: shared axes, personal weights, and when their spread crosses the consensus\par}
\vspace{3mm}
{\normalsize Hiroshi Higashi\par}
\end{center}
\vspace{5mm}
\fi
\renewcommand{\figurename}{Supplementary Fig.}
\renewcommand{\tablename}{Supplementary Table}
\setcounter{figure}{0}
\setcounter{table}{0}

\FloatBarrier
\subsection*{Supplementary Note 1: Details of the recovery in Xu et al.\ Experiment 2}
This Note takes the recovery in Experiment 2 reported in the main text and shows, in turn, that individualisation extends to unseen face pairs, that it holds with few trials, that the latent representation $\bm z$ is stable, and that $\bm z$ is linked to other people's judgments and to the participant's own self-description.

\textbf{Retest task and unseen face pairs.}
The retest task also includes pairs of faces, among the participant's 20 faces, that were never arranged together in the estimation trials.
In this retest task as well, $\rho$ rose from 0.329 for the population RDM to 0.367 for \Do{} (difference from \Dz{} (0.334): $+0.033$, $P < 10^{-40}$; Supplementary Table~\ref{stab:xu}).
In contrast, the shrinkage-blend baseline, which blends the participant's own estimation RDM with the population RDM, was the strongest for face pairs that the participant had actually arranged (0.351), but was no different from the population for pairs that did not appear in the estimation trials (0.329).
That is, unlike a method that uses the participant's own data as they are, individualisation by the model extends to unseen face pairs.

\textbf{Number of trials required.}
Next, we varied the number of trials used for estimation while keeping the held-out set fixed, and examined how many trials are needed for individualisation.
The held-out $\rho$ of \Do{} increased from 0.325 with 5 trials to 0.333 with 10 trials, 0.337 with 20 trials and 0.338 with 40 trials, and exceeded the population model \Dz{} even with the minimum of 5 trials (Supplementary Fig.~\ref{sfig:exp2}a).
The learning curve thus saturated quickly and had almost reached its ceiling at 20 trials.
In addition, when amortised inference was compared with direct optimisation of $\bm z$ by gradient descent on the same estimation data, the two were equivalent at 20 trials or more, and amortised inference was better with fewer trials (Methods).

\textbf{Relation to reliability.}
The individualisation gain was larger the higher the reliability of the participant's own judgments.
When participants were divided into quartiles of split-half reliability, \dOZ{} increased monotonically from $+0.020$ in the lowest quartile to $+0.053$ in the highest (Supplementary Fig.~\ref{sfig:exp2}b; all $P < 10^{-4}$).

\begin{figure}[htbp]
\centering
\begin{minipage}[t]{0.48\textwidth}
\panel{a}\\[1mm]
\includegraphics[width=\textwidth]{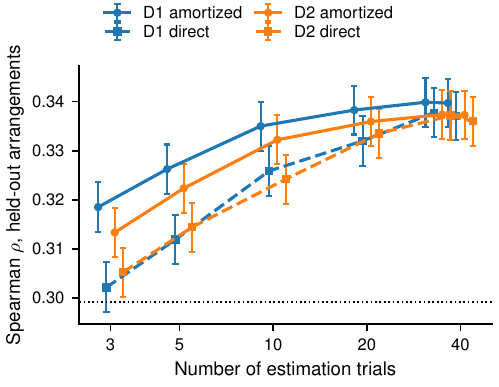}
\end{minipage}\hfill
\begin{minipage}[t]{0.48\textwidth}
\panel{b}\\[1mm]
\includegraphics[width=\textwidth]{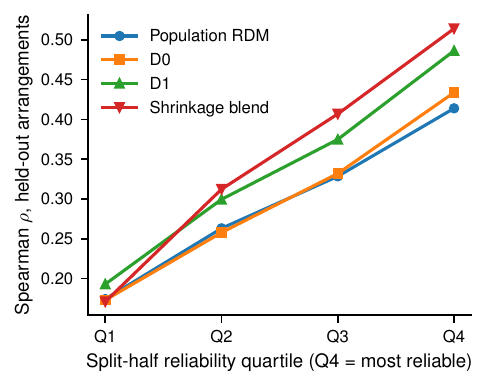}
\end{minipage}
\caption{\textbf{Effects of the number of trials and of judgment reliability in Xu et al.\ Experiment 2.}
\panel{a} Learning curve.
Held-out $\rho$ for \Do{} and \Dt{} (both $M = 4$) when the number of trials used for estimation is varied while the held-out set is kept fixed (mean $\pm$ s.e.m.).
Solid lines show amortised inference and dashed lines show direct inference.
The points at 34 on the horizontal axis use all estimation trials (mean 34 trials; the points at 40 use at most 40 trials).
The dotted line is the population model \Dz{}.
For visibility, the series at the same number of trials are slightly offset horizontally.
\panel{b} Held-out $\rho$ in each quartile when participants are divided into quartiles of split-half reliability (population RDM, \Dz{}, \Do{}, shrinkage blend).
}
\label{sfig:exp2}
\end{figure}

\textbf{Stability of the latent representation.}
The estimated $\bm z$ was stable within the same session.
The intraclass correlation\cite{shrout1979} between the $\bm z$ estimated separately from the estimation trials and from the held-out trials was 0.77 averaged over dimensions (0.67 for \Dt{} in Experiment 1).
The intraclass correlation with the $\bm z$ estimated from the retest task was lower, at 0.49.
Even so, when the held-out trials of the main task were predicted with the $\bm z$ estimated from the retest task (mostly unseen faces in Experiment 1), $\rho$ exceeded that of the population (Experiment 2: 0.316 versus 0.290; Experiment 1: 0.204 versus 0.193).
Therefore, the same person's latent representation is obtained even when it is estimated from a different set of faces.

\textbf{Latent distance and transfer between individuals.}
Distance in the space of $\bm z$ explained how well one person's $\bm z$ predicts another person's judgments (Supplementary Fig.~\ref{sfig:cross}).
To examine this, for all pairs of the 150 fold-out participants, we predicted the held-out trials of participant $k$ using the $\bm z$ of another participant $l$.
The $\rho$ of this prediction decreased monotonically as the distance between the two, $\|\bm z_k - \bm z_l\|$, increased (mixed-model slope $-0.078$/SD, Spearman $-0.32$).
On average, prediction with another person's $\bm z$ (0.259) was worse than the population (0.302), and prediction with the participant's own $\bm z$ (0.352) was the best.
That is, individualisation transferred only to others whose $\bm z$ was close, and for distant others it became negative transfer, worse than predicting with the population.

\begin{figure}[htbp]
\centering
\begin{minipage}[t]{0.48\textwidth}
\panel{a}\\[1mm]
\includegraphics[width=\textwidth]{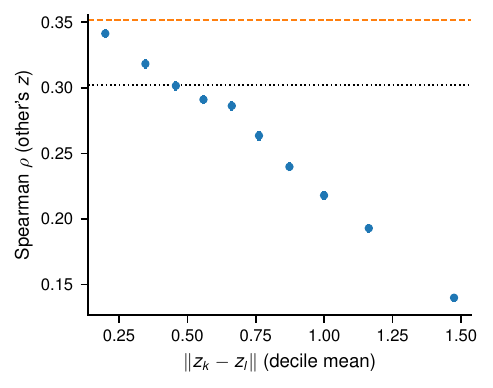}
\end{minipage}\hfill
\begin{minipage}[t]{0.48\textwidth}
\panel{b}\\[1mm]
\includegraphics[width=\textwidth]{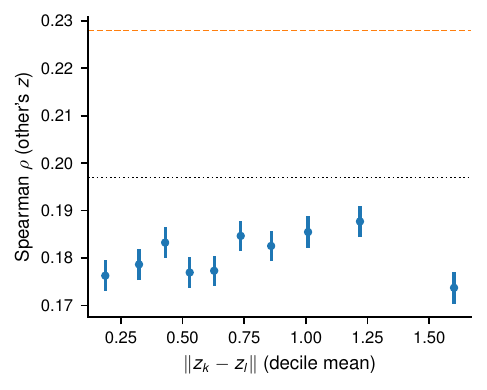}
\end{minipage}
\caption{\textbf{Latent distance explains transfer between individuals.}
For all pairs of the 150 fold-out participants of fold 0, the $\rho$ obtained when the held-out trials of participant $k$ are predicted with the $\bm z$ of another participant $l$ is shown for each decile of the distance between the two latent representations, $\|\bm z_k - \bm z_l\|$ (mean $\pm$ s.e.m.).
The larger the distance, the lower $\rho$.
The dotted line is the prediction by the population RDM, and the dashed line is the prediction by the participant's own $\bm z$.
\panel{a} Experiment 2, \Do{}.
\panel{b} Experiment 1, \Dt{}.
}
\label{sfig:cross}
\end{figure}

\textbf{Correspondence with verbal self-description.}
We further examined whether $\bm z$ is linked to the participant's own verbal account.
The free-text descriptions of strategy that participants wrote after the task were converted into sentence embeddings, and a regression from the sentence embeddings to $\bm z$ was trained on the participants of the training folds.
The arrangements of fold-out participants could then be predicted better than by the population from the free-text descriptions alone (Experiment 2: $+0.017$, s.e.m.\ 0.001, $P = 4 \times 10^{-35}$; Experiment 1: $+0.008$, $P = 6 \times 10^{-15}$).
Although the effect is small, this result shows that verbal self-description and arrangement behaviour are linked through the same latent representation.

\textbf{External validation of the face embedding.}
Finally, we checked whether the learned face embedding corresponds to known dimensions of face impressions.
The learned face embedding predicted the four factors of Lin et al.\cite{lin2021} (warmth, competence, femininity and youth), which were obtained from ratings of the same 100 faces on 100 trait words.
The accuracy of the prediction was 0.58--0.68 in $R^2$ under 10-fold cross-validation, and 0.96, 0.83, 0.78 and 0.68 in canonical correlation (Methods).

\begin{table}[htbp]
\centering
\caption{\textbf{Recovery of individual face-impression spaces in the two experiments of Xu et al.\ (10-fold cross-validation, fold-out participants, mean over seeds).}
Held-out $\rho$ is the mean over participants of Spearman $\rho$ for the participant's own held-out trials.
Retest $\rho$ is the same value for the retest task (the retest task consists mostly of unseen faces in Experiment 1, and of faces among the participant's 20 faces in Experiment 2).
The models used $M = 4$, and the latent representation was obtained by amortised inference.
The bottom three rows are differences between adjacent models, with the participant-level differences tested by a paired $t$-test.}
\label{stab:xu}
\footnotesize
\setlength{\tabcolsep}{2pt}
\begin{tabular}{p{3.9cm}cccc}
\toprule
 & \multicolumn{2}{c}{Experiment 1 (100 White faces, $n = 1{,}751$)} & \multicolumn{2}{c}{Experiment 2 (96 multiracial faces, $n = 1{,}682$)} \\
 & Held-out $\rho$ & Retest $\rho$ & Held-out $\rho$ & Retest $\rho$ \\
\midrule
Population RDM (training folds) & 0.190 & 0.217 & 0.295 & 0.329 \\
Shrinkage blend of own + population & 0.247 & 0.217 & 0.351 & 0.329 \\
\Dz{} shared only & 0.201 & 0.226 & 0.299 & 0.334 \\
\Do{} same-sign reweighting & 0.202 & 0.227 & \textbf{0.338} & \textbf{0.367} \\
\Dt{} linear transform & \textbf{0.220} & \textbf{0.235} & 0.341 & 0.369 \\
\Dh{} face-specific offsets & 0.221 & 0.231 & 0.337 & 0.366 \\
\midrule
\dOZ{} & $+0.001$ ($P = 0.59$) & $+0.001$ & $+0.039$ ($P < 10^{-50}$) & $+0.033$ ($P < 10^{-40}$) \\
\dTO{} & $+0.018$ ($P < 10^{-10}$) & $+0.008$ & $+0.002$ ($P = 0.31$) & $+0.002$ ($P = 0.07$) \\
\dHT{} & $+0.001$ ($P = 0.65$) & $-0.004$ & $-0.004$ ($P = 0.10$) & $-0.003$ ($P = 0.02$) \\
\bottomrule
\end{tabular}
\end{table}

\textbf{Comparison with baselines that do not use the model.}
The individual models outperformed the prediction from the population (Fig.~\ref{fig:xu}).
Figure~\ref{fig:xu} shows the held-out-trial $\rho$ of each of the methods described below.
The held-out-trial $\rho$ rose from 0.295 for the population RDM (the population distances computed from the participants in the training folds) to 0.338 for \Do{}.
The participant-level difference was $+0.043$ relative to the population RDM and $+0.039$ relative to the \Dz{} population model (0.299) (s.e.m.\ 0.003 and 0.002, $P < 10^{-39}$ in both a paired $t$-test and a Wilcoxon test).
Moreover, this accuracy of \Do{} exceeded the agreement obtained when the person's own estimation trials were split in half and compared with each other (split-half reliability, 0.309; ratio 1.09).
For comparison, we also examined a shrinkage-blend baseline that blends the person's own estimation RDM with the population RDM.
This baseline was the strongest on the very face pairs that had been arranged in the estimation trials (0.351; ``Shrinkage blend'' in Fig.~\ref{fig:xu}), but was no better than the population on pairs that did not appear in the estimation trials (0.329).
Figure~\ref{fig:xu} also shows the 5-dimensional MDS of the population (0.299), the prediction from the person's own estimation RDM alone (0.314), and the prediction obtained by reweighting the five population dimensions with the person's own data (0.327).
All of these fell below \Do{}.
On the other hand, advancing the model to \Dt{} (0.341) or \Dh{} (0.337) left the accuracy almost unchanged from \Do{}.
This point is examined in detail across datasets in the main text.

\FloatBarrier
\subsection*{Supplementary Note 2: Detailed results for each dataset (faces, artworks, photographs)}
This Note describes in detail, together with the learning curves (Supplementary Figs.~\ref{sfig:gen} and \ref{sfig:curves}), the results for each dataset that the main text gave only in outline.

\begin{table}[htbp]
\centering
\caption{\textbf{Summary across datasets: individualisation gain and the shape of individual differences.}
The values in each row were obtained for fold-out raters from held-out judgments of the rater's own stimuli (unseen photographs for EVA; mean over seeds).
``Gain'' is \dOZ{} and represents the improvement in individualisation from reweighting the shared dimensions.
``Shape'' is \dTO{} and represents the improvement from adding a linear transform that allows reversal.
``Population'' is the prediction by the population's mean judgment (the population RDM for Xu), and differs from the prediction of \Dz{}.
For this reason, the difference in the ``Population $\to$ \Do{}'' column does not equal the gain \dOZ{} (for example, \Dz{} for Xu Experiments 2 and 1 is 0.299 and 0.201).
``Loss under the \Do{} constraint'' is the measure from the sign check (Fig.~\ref{fig:sign}): the predictive power of a free linear readout minus that of a readout under the \Do{} constraint.
For both readouts, the weights were obtained from the rater's ratings of all training stimuli, and predictive power was evaluated by split-half cross-validation.
If this difference is positive, imposing the \Do{} constraint worsens prediction.
In parentheses is the proportion of raters whose cosine with the population weight is negative, that is, raters whose weights as a whole point in the direction opposite to the population.
Tied-pair accuracy is the proportion of stimulus pairs, not ordered by the population mean, for which the rater's own ordering was predicted correctly (population/\Do{}/\Dt{}; Supplementary Fig.~\ref{sfig:effect}b).
n.s.: $P > 0.05$.}
\label{tab:gen}
\footnotesize
\setlength{\tabcolsep}{3pt}
\begin{tabular}{p{2.6cm}p{1.45cm}p{2.3cm}p{1.9cm}p{1.8cm}p{1.9cm}p{1.9cm}}
\toprule
Data (measure, number used for estimation) & Raters & Population $\to$ \Do{} & Gain \dOZ{} & Shape \dTO{} & Loss under the \Do{} constraint & Tied-pair accuracy \\
\midrule
Xu Experiment 2 ($\rho$, all) & 1,682 & 0.295 $\to$ 0.338 & $+0.039$ & $+0.002$ n.s. & --- & --- \\
Xu Experiment 1 ($\rho$, all) & 1,751 & 0.190 $\to$ 0.202$^{a}$ & $+0.001$ n.s.$^{a}$ & $+0.018$ & --- & --- \\
London ($r$, 20 faces) & 2,501 & 0.590 $\to$ 0.622 & $+0.033$ & $+0.001$ n.s. & $-0.045$ (1.6\%) & 0.51/0.57/0.59 \\
AVFS novel (accuracy, 20) & 504 & 0.588 $\to$ 0.650 & $+0.061$ & $-0.009$ & --- & --- \\
AVFS main (accuracy, 20/all) & 1,283 & 0.603 $\to$ 0.622/0.630 & $+0.019$/$+0.028$ & $-0.002$/$+0.003$ & --- & --- \\
FaceSim3D 2D/3D (accuracy, 20) & 1,219/1,053 & 0.590 $\to$ 0.597/0.599 $\to$ 0.609 & $+0.007$/$+0.009$ & $-0.003$/$-0.005$ & --- & --- \\
SCUT ($r$, 500 faces) & 60 & 0.777 $\to$ 0.778 & $+0.001$ n.s. & $+0.006$ & $+0.002$ (0\%) & 0.54/0.55/0.58 \\
LAPIS artworks ($r$, 20/100) & 553/551 & 0.435 $\to$ 0.523/0.554 & $+0.092$/$+0.123$ & $+0.011$/$+0.021$ & $+0.046$ (9.6\%) & 0.51/0.62/0.63 \\
AADB photographs ($r$, 100) & 68$^{b}$ & 0.476 $\to$ 0.497 & $+0.029$ & $+0.003$ n.s. & --- & 0.50/0.66/0.66 \\
EVA photographs ($r$, 100, unseen) & 175 & 0.425 $\to$ 0.443 & $+0.017$ & $+0.020$ & $-0.006$ (0\%) & 0.50/0.56/0.59 \\
Jester jokes ($r$, 20/50) & 11,067 & 0.341 $\to$ 0.381/0.400 & $+0.039$/$+0.056$ & $+0.004$ & $-0.033$ (10\%, noise) & 0.50/0.54/0.54 \\
Keller olfaction ($r$, 100) & 55 & 0.513 $\to$ 0.512 & $-0.002$ n.s. & --- & --- & --- \\
\bottomrule
\end{tabular}
\par\smallskip
\parbox{\linewidth}{\footnotesize $^{a}$ In Xu Experiment 1, \Do{} was no different from the population and \Dt{} exceeded it.
For this reason, the values for \Dt{} are also given for this row: held-out $\rho$ 0.220, \Dt{} $-$ \Dz{} $= +0.019$.\\
$^{b}$ The AADB values are for the 55 of the 68 raters for whom 100 estimation ratings were available.}
\end{table}

\begin{table}[p]
\centering
\caption{\textbf{Test statistics for the model comparisons.}
For each row of Supplementary Table~\ref{tab:gen} and of Table~\ref{tab:xpass} of the main text, the statistics of the rater-level paired differences are shown.
$n$ is the number of raters, the difference is the mean over raters, CI is the 95\% confidence interval from the $t$ distribution, $t$ is the statistic of the paired $t$-test ($n - 1$ degrees of freedom), and $d_z$ is Cohen's effect size (the mean of the differences divided by their standard deviation).
$P_t$ and $P_W$ are the two-sided $P$ values of the paired $t$-test and the Wilcoxon signed-rank test.
The differences are means of unrounded values, so their last digit can differ from the difference of two rounded values (\dOZ{} in London is 0.0325, whereas the difference of the rounded $r$ values is 0.032).
The two tests disagreed at the 5\% level only for \dTO{} in London (a difference of $+0.001$).
\panel{a} \dOZ{} (gain from reweighting the shared dimensions).
\panel{b} \dTO{} (difference from adding a linear transform that allows reversal).
}
\label{stab:stats}
\fontsize{6}{7.5}\selectfont
\setlength{\tabcolsep}{4pt}
\begin{tabular}{>{\raggedright\arraybackslash}p{4.6cm}rllrrll}
\toprule
\panel{a} Data (measure, estimation size) & $n$ & Difference & 95\% CI & $t$ & $d_z$ & $P_t$ & $P_W$ \\
\midrule
Xu Experiment 2 ($\rho$, all) & 1,682 & $+0.039$ & [$+0.035$, $+0.044$] & $16.34$ & $0.40$ & $9 \times 10^{-56}$ & $10^{-52}$ \\
Xu Experiment 1 ($\rho$, all) & 1,751 & $+0.001$ & [$-0.003$, $+0.005$] & $0.53$ & $0.01$ & 0.59 & 0.58 \\
London ($r$, 20 faces) & 2,501 & $+0.033$ & [$+0.030$, $+0.035$] & $25.29$ & $0.51$ & $8 \times 10^{-126}$ & $2 \times 10^{-122}$ \\
AVFS novel (accuracy, 20) & 504 & $+0.061$ & [$+0.054$, $+0.068$] & $16.98$ & $0.76$ & $2 \times 10^{-51}$ & $4 \times 10^{-47}$ \\
AVFS main (accuracy, 20) & 1,283 & $+0.019$ & [$+0.017$, $+0.022$] & $14.90$ & $0.42$ & $2 \times 10^{-46}$ & $3 \times 10^{-38}$ \\
AVFS main (accuracy, all) & 1,283 & $+0.028$ & [$+0.025$, $+0.030$] & $22.42$ & $0.63$ & $3 \times 10^{-94}$ & $5 \times 10^{-91}$ \\
FaceSim3D 2D (accuracy, 20) & 1,219 & $+0.006$ & [$+0.004$, $+0.009$] & $6.19$ & $0.18$ & $8 \times 10^{-10}$ & $3 \times 10^{-10}$ \\
FaceSim3D 3D (accuracy, 20) & 1,053 & $+0.009$ & [$+0.007$, $+0.012$] & $8.35$ & $0.26$ & $2 \times 10^{-16}$ & $3 \times 10^{-15}$ \\
SCUT ($r$, 500 faces) & 60 & $+0.001$ & [$-0.001$, $+0.003$] & $1.12$ & $0.14$ & 0.27 & 0.80 \\
LAPIS artworks ($r$, 20) & 553 & $+0.092$ & [$+0.077$, $+0.107$] & $12.18$ & $0.52$ & $2 \times 10^{-30}$ & $2 \times 10^{-37}$ \\
LAPIS artworks ($r$, 100) & 551 & $+0.123$ & [$+0.108$, $+0.138$] & $16.05$ & $0.68$ & $8 \times 10^{-48}$ & $10^{-75}$ \\
AADB photographs ($r$, 100) & 55 & $+0.029$ & [$+0.017$, $+0.041$] & $5.01$ & $0.68$ & $6 \times 10^{-6}$ & $7 \times 10^{-6}$ \\
EVA photographs ($r$, 100, unseen) & 175 & $+0.017$ & [$+0.010$, $+0.024$] & $4.70$ & $0.36$ & $5 \times 10^{-6}$ & $2 \times 10^{-6}$ \\
Jester jokes ($r$, 20) & 10,849 & $+0.039$ & [$+0.037$, $+0.041$] & $37.77$ & $0.36$ & $2 \times 10^{-293}$ & $4 \times 10^{-275}$ \\
Jester jokes ($r$, 50) & 5,517 & $+0.056$ & [$+0.053$, $+0.059$] & $37.37$ & $0.50$ & $10^{-272}$ & $< 10^{-300}$ \\
Keller olfaction ($r$, 100) & 55 & $-0.002$ & [$-0.006$, $+0.003$] & $-0.76$ & $-0.10$ & 0.45 & 0.32 \\
\midrule
LAPIS thinned (129 raters, all ratings) & 227 & $+0.116$ & [$+0.093$, $+0.139$] & $10.08$ & $0.67$ & $6 \times 10^{-20}$ & $10^{-28}$ \\
LAPIS thinned (129 raters $\times$ 227 ratings) & 224 & $+0.093$ & [$+0.072$, $+0.113$] & $8.83$ & $0.59$ & $3 \times 10^{-16}$ & $3 \times 10^{-23}$ \\
XPASS-Vis art (own models) & 129 & $+0.028$ & [$+0.017$, $+0.039$] & $5.14$ & $0.45$ & $10^{-6}$ & $10^{-8}$ \\
Transfer: XPASS-Vis raters $\leftarrow$ LAPIS models & 129 & $+0.043$ & [$+0.027$, $+0.059$] & $5.25$ & $0.46$ & $6 \times 10^{-7}$ & $10^{-9}$ \\
Transfer: LAPIS raters $\leftarrow$ XPASS-Vis models (50) & 507 & $+0.085$ & [$+0.073$, $+0.097$] & $13.74$ & $0.61$ & $9 \times 10^{-37}$ & $3 \times 10^{-44}$ \\
XPASS-Vis fashion & 129 & $+0.002$ & [$-0.002$, $+0.006$] & $1.02$ & $0.09$ & 0.31 & 0.25 \\
XPASS-Vis landscapes & 129 & $+0.009$ & [$+0.004$, $+0.015$] & $3.43$ & $0.30$ & $8 \times 10^{-4}$ & $9 \times 10^{-4}$ \\
\bottomrule
\end{tabular}

\vspace{2mm}
\begin{tabular}{>{\raggedright\arraybackslash}p{4.6cm}rllrrll}
\toprule
\panel{b} Data (measure, estimation size) & $n$ & Difference & 95\% CI & $t$ & $d_z$ & $P_t$ & $P_W$ \\
\midrule
Xu Experiment 2 ($\rho$, all) & 1,682 & $+0.002$ & [$-0.002$, $+0.006$] & $1.01$ & $0.02$ & 0.31 & 0.64 \\
Xu Experiment 1 ($\rho$, all) & 1,751 & $+0.018$ & [$+0.014$, $+0.023$] & $8.76$ & $0.21$ & $4 \times 10^{-18}$ & $2 \times 10^{-16}$ \\
London ($r$, 20 faces) & 2,501 & $+0.001$ & [$0.000$, $+0.002$] & $1.54$ & $0.03$ & 0.12 & 0.002 \\
AVFS novel (accuracy, 20) & 504 & $-0.009$ & [$-0.011$, $-0.006$] & $-7.23$ & $-0.32$ & $2 \times 10^{-12}$ & $9 \times 10^{-14}$ \\
AVFS main (accuracy, 20) & 1,283 & $-0.002$ & [$-0.003$, $-0.001$] & $-2.90$ & $-0.08$ & 0.004 & $9 \times 10^{-4}$ \\
AVFS main (accuracy, all) & 1,283 & $+0.003$ & [$+0.002$, $+0.004$] & $5.49$ & $0.15$ & $5 \times 10^{-8}$ & $2 \times 10^{-10}$ \\
FaceSim3D 2D (accuracy, 20) & 1,219 & $-0.003$ & [$-0.005$, $-0.001$] & $-2.58$ & $-0.07$ & 0.010 & 0.012 \\
FaceSim3D 3D (accuracy, 20) & 1,053 & $-0.005$ & [$-0.007$, $-0.003$] & $-4.68$ & $-0.14$ & $3 \times 10^{-6}$ & $2 \times 10^{-6}$ \\
SCUT ($r$, 500 faces) & 60 & $+0.006$ & [$+0.003$, $+0.008$] & $4.32$ & $0.56$ & $6 \times 10^{-5}$ & $2 \times 10^{-7}$ \\
LAPIS artworks ($r$, 20) & 553 & $+0.011$ & [$+0.007$, $+0.015$] & $5.61$ & $0.24$ & $3 \times 10^{-8}$ & $3 \times 10^{-5}$ \\
LAPIS artworks ($r$, 100) & 551 & $+0.021$ & [$+0.018$, $+0.024$] & $12.16$ & $0.52$ & $3 \times 10^{-30}$ & $3 \times 10^{-40}$ \\
AADB photographs ($r$, 100) & 55 & $+0.003$ & [$-0.001$, $+0.007$] & $1.47$ & $0.20$ & 0.15 & 0.30 \\
EVA photographs ($r$, 100, unseen) & 175 & $+0.020$ & [$+0.013$, $+0.028$] & $5.67$ & $0.43$ & $6 \times 10^{-8}$ & $5 \times 10^{-8}$ \\
Jester jokes ($r$, 20) & 10,849 & $+0.004$ & [$+0.003$, $+0.004$] & $10.51$ & $0.10$ & $10^{-25}$ & $10^{-10}$ \\
Jester jokes ($r$, 50) & 5,517 & $+0.004$ & [$+0.003$, $+0.004$] & $8.28$ & $0.11$ & $10^{-16}$ & $6 \times 10^{-4}$ \\
\midrule
LAPIS thinned (129 raters, all ratings) & 227 & $+0.019$ & [$+0.013$, $+0.025$] & $6.27$ & $0.42$ & $2 \times 10^{-9}$ & $6 \times 10^{-10}$ \\
LAPIS thinned (129 raters $\times$ 227 ratings) & 224 & $+0.022$ & [$+0.015$, $+0.028$] & $6.25$ & $0.42$ & $2 \times 10^{-9}$ & $10^{-13}$ \\
XPASS-Vis art (own models) & 129 & $+0.007$ & [$+0.003$, $+0.011$] & $3.84$ & $0.34$ & $2 \times 10^{-4}$ & $3 \times 10^{-4}$ \\
Transfer: XPASS-Vis raters $\leftarrow$ LAPIS models & 129 & $+0.008$ & [$+0.003$, $+0.012$] & $3.53$ & $0.31$ & $6 \times 10^{-4}$ & $9 \times 10^{-4}$ \\
Transfer: LAPIS raters $\leftarrow$ XPASS-Vis models (50) & 507 & $+0.015$ & [$+0.011$, $+0.018$] & $7.41$ & $0.33$ & $5 \times 10^{-13}$ & $10^{-13}$ \\
XPASS-Vis fashion & 129 & $+0.010$ & [$+0.004$, $+0.015$] & $3.55$ & $0.31$ & $5 \times 10^{-4}$ & 0.005 \\
XPASS-Vis landscapes & 129 & $+0.011$ & [$+0.007$, $+0.015$] & $5.18$ & $0.46$ & $8 \times 10^{-7}$ & $6 \times 10^{-7}$ \\
\bottomrule
\end{tabular}
\end{table}

\begin{figure}[htbp]
\centering
\begin{minipage}[t]{0.48\textwidth}
\panel{a}\\[1mm]
\includegraphics[width=\textwidth]{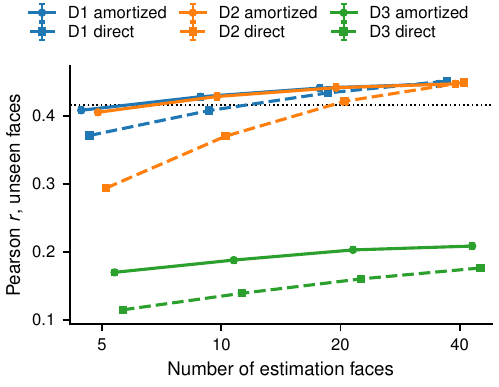}
\end{minipage}\hfill
\begin{minipage}[t]{0.48\textwidth}
\panel{b}\\[1mm]
\includegraphics[width=\textwidth]{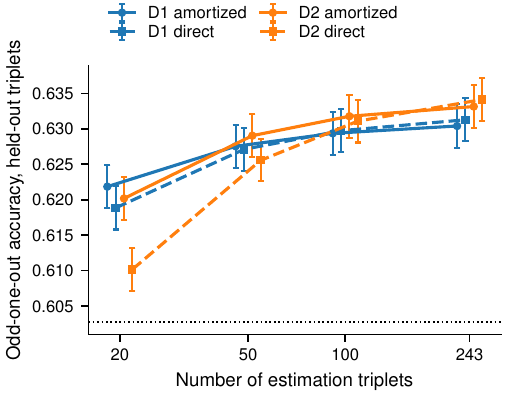}
\end{minipage}\\[3mm]
\begin{minipage}[t]{0.48\textwidth}
\panel{c}\\[1mm]
\includegraphics[width=\textwidth]{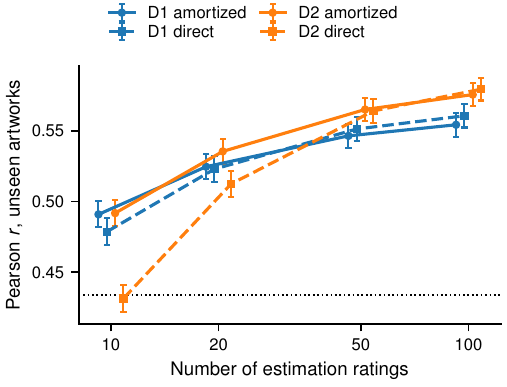}
\end{minipage}\hfill
\begin{minipage}[t]{0.48\textwidth}
\panel{d}\\[1mm]
\includegraphics[width=\textwidth]{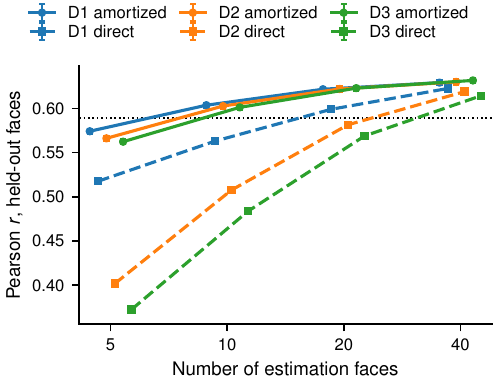}
\end{minipage}\\[3mm]
\begin{minipage}[t]{0.48\textwidth}
\panel{e}\\[1mm]
\includegraphics[width=\textwidth]{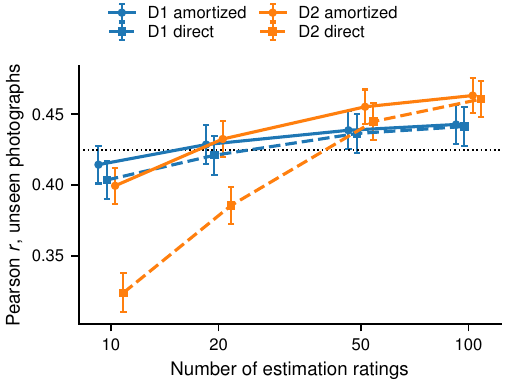}
\end{minipage}\hfill
\begin{minipage}[t]{0.48\textwidth}
\panel{f}\\[1mm]
\includegraphics[width=\textwidth]{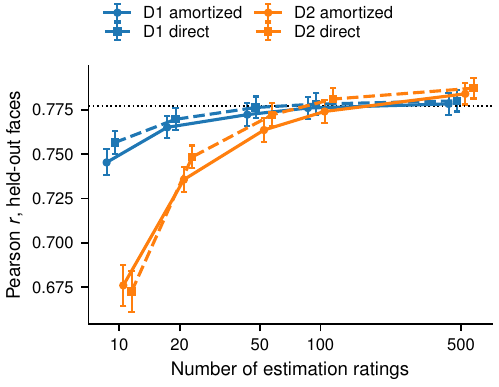}
\end{minipage}
\caption{\textbf{Learning curves for the main datasets (fold-out raters, mean $\pm$ s.e.m., mean over seeds).}
Each panel shows how predictive power changes as the number of judgments used for estimation increases.
For visibility, the series at the same number of judgments are slightly offset horizontally.
\panel{a} London Set, CLIP image-embedding version.
Pearson $r$ for faces unseen by both the model and the rater.
The dotted line is the population ridge regression from image features.
\panel{b} AVFS main set (4,921 faces, 1,283 raters).
Accuracy on the rater's own held-out triplets (the rightmost point uses all estimation triplets, mean 243).
\Dh{} (face-specific offsets for 4,921 faces) overfits and falls below \Dz{}.
\panel{c} LAPIS artworks (11,723 works, 553 raters).
Pearson $r$ for works unseen by both the model and the rater.
\panel{d} London Set, learned-embedding version.
Pearson $r$ for the remaining faces rated by the rater (training stimuli of the model).
The dotted line is the population mean of the training raters.
\panel{e} EVA photographs (4,070 photographs, 175 raters).
Pearson $r$ for photographs unseen by both the model and the rater.
\Dt{} exceeds \Do{} at 50 ratings or more.
\panel{f} SCUT-FBP5500 (5,500 faces, 60 raters).
Shared preferences dominate, and there is no individualisation gain.
\Dt{} exceeds \Do{} only at 500 faces, and the difference is 0.006.
}
\label{sfig:gen}
\end{figure}

\begin{figure}[htbp]
\centering
\begin{minipage}[t]{0.48\textwidth}
\panel{a}\\[1mm]
\includegraphics[width=\textwidth]{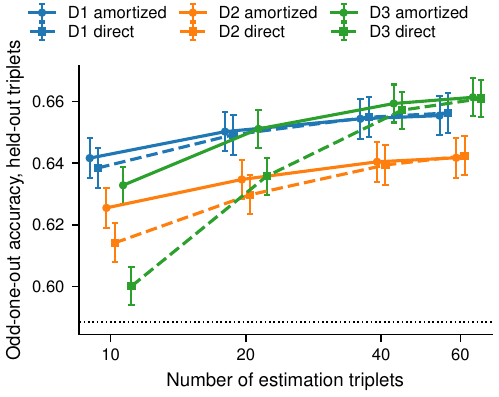}
\end{minipage}\hfill
\begin{minipage}[t]{0.48\textwidth}
\panel{b}\\[1mm]
\includegraphics[width=\textwidth]{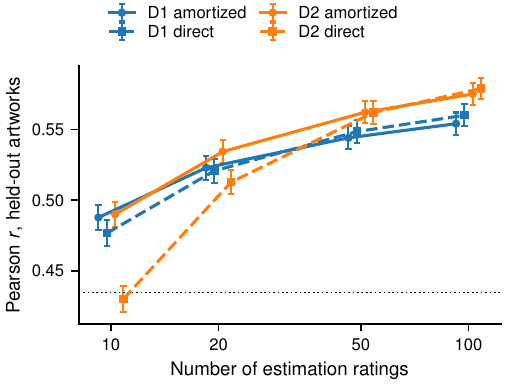}
\end{minipage}\\[3mm]
\begin{minipage}[t]{0.48\textwidth}
\panel{c}\\[1mm]
\includegraphics[width=\textwidth]{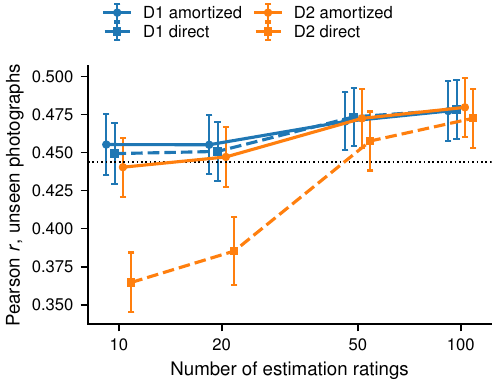}
\end{minipage}\hfill
\begin{minipage}[t]{0.48\textwidth}
\panel{d}\\[1mm]
\includegraphics[width=\textwidth]{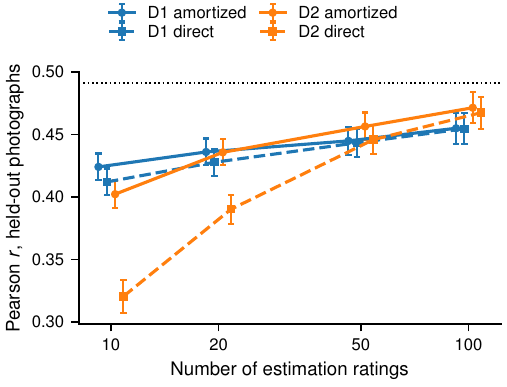}
\end{minipage}\\[3mm]
\begin{minipage}[t]{0.48\textwidth}
\panel{e}\\[1mm]
\includegraphics[width=\textwidth]{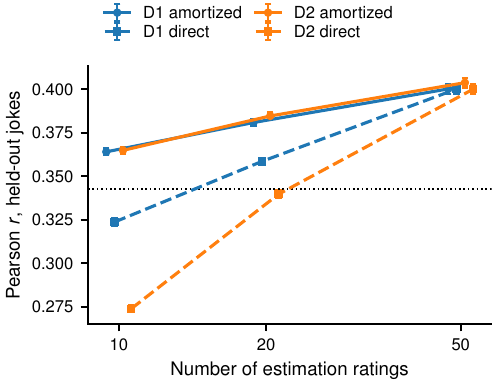}
\end{minipage}\hfill
\begin{minipage}[t]{0.48\textwidth}
\mbox{}
\end{minipage}
\caption{\textbf{Learning curves for the other datasets (fold-out raters, mean $\pm$ s.e.m., mean over seeds).}
Each panel shows how predictive power changes as the number of judgments used for estimation increases.
For visibility, the series at the same number of judgments are slightly offset horizontally.
The learning curve for the London Set (learned-embedding version) is shown in Supplementary Fig.~\ref{sfig:gen}d.
\panel{a} AVFS novel 56-face set (504 raters).
\panel{b} LAPIS (evaluated on the rater's remaining ratings of training works).
\panel{c} AADB (evaluated on photographs unseen by both the model and the rater; the dotted line is the image ridge regression).
\panel{d} EVA (evaluated on the rater's remaining ratings of training photographs; the dotted line is the per-image mean of 30 raters, which is stronger than the 8-dimensional image model).
\panel{e} Jester (150 jokes, 11,067 raters).
}
\label{sfig:curves}
\end{figure}

\textbf{Faces: gain when the task and the embedding are changed.}
For faces, an individualisation gain was obtained even when the task (ratings or triplets) and the way the embedding was constructed were changed.
In the learned-embedding version of the London Set, predictive power rose as the number of faces used for estimation increased, reaching 0.622 with 20 faces from 0.590 for the population mean (Supplementary Fig.~\ref{sfig:gen}d).
In contrast, the AVFS main set (4,921 faces, 1,283 raters) has extremely few judgments per person relative to the number of faces.
We therefore used CLIP features of the FFHQ images as the embedding instead of learning the embedding from the judgments.
The gain in this case was $+0.019$ with 20 triplets and $+0.028$ with all estimation judgments (mean 243) (Supplementary Fig.~\ref{sfig:gen}b).

\textbf{Transfer to unseen stimuli.}
The individualisation gain was also obtained for stimuli used for neither training nor estimation (Supplementary Fig.~\ref{sfig:gen}a; panel c for artworks and panel e for photographs).
For the London Set, the faces were divided into five folds, and ratings of faces used for neither training nor estimation were predicted.
The baseline is a ridge regression that predicts the population mean from image features.
Predictive power rose from 0.416 for this ridge regression to 0.441 when the $\bm z$ estimated from 20 faces was used ($+0.025$, $P < 10^{-9}$).
For AVFS, $\bm z$ was estimated from judgments on the main set, and triplets of the 56 novel faces, which never appear in training, were predicted.
In this case too, accuracy rose from 0.574 to 0.629 ($+0.055$, $P < 10^{-30}$).

Furthermore, the individual model predicted a person's responses better than the majority vote of other people.
AVFS contains 1,000 triplets shared by 24 people.
On these triplets, accuracy was 0.642 for the \Do{} model estimated from about 130 of the person's own judgments and 0.625 for the majority vote of the other approximately 23 people, so the \Do{} model was higher ($P = 4 \times 10^{-4}$; Supplementary Note~8).

\textbf{\Dt{} and \Dh{} for faces.}
In the face data, \Dt{} gave almost no improvement outside Xu Experiment 2 as well.
In the London Set, \dTO{} was only $+0.001$ with 20 faces ($P = 0.12$) and $+0.001$ with 40 faces ($P = 0.09$), and adding \Dh{} made no difference.
In AVFS, \Dt{}, which allows a general metric, was if anything worse than \Do{} (novel set $-0.009$, main set with 20 triplets $-0.002$).
In addition, \Dh{} overfitted on the main set and fell 0.05 below \Dz{} (Supplementary Fig.~\ref{sfig:gen}b).
In FaceSim3D too, \Dt{} was worse than \Do{} (see the next paragraph).

\Dh{}, which has stimulus-specific offsets, did not help in predicting unseen faces either.
In the image versions of the London Set and AVFS, \Dh{} was the best on the rater's training faces, but its prediction collapsed on unseen faces (0.203 on unseen faces of the London Set, against 0.441 for \Do{}).
That is, as in Xu Experiment 1, the face-specific component did not transfer to unseen faces.

\textbf{Face data dominated by consensus (FaceSim3D, SCUT-FBP5500).}
In the following two face datasets, raters' judgments were determined almost entirely by the population consensus, and the individualisation gain was small.
The first, FaceSim3D, consists of judgments of 100 faces photographed under controlled conditions (CFD) by participants of whom 98\% were White.
In these data, in which both the stimuli and the participants are homogeneous, the individualisation gain with 20 triplets was $+0.007$ (2D)/$+0.009$ (3D), which, although significant ($P < 10^{-9}$), was an order of magnitude smaller than in AVFS.
Moreover, \Dt{} was if anything worse than \Do{} ($-0.003$/$-0.005$).

In the second, SCUT-FBP5500, 60 Chinese raters rated all 5,500 faces.
In these data the predictive power of the population mean was high, at 0.777, and there was no gain even when the individual latent representation was estimated from 500 faces ($+0.001$, $P = 0.27$).
\Dt{} exceeded \Do{} only when estimation used 500 faces ($+0.006$, $P = 10^{-4}$), and there was no difference at 100 faces or fewer (Supplementary Fig.~\ref{sfig:gen}f).
We also examined transfer in which a model trained on Asian faces was evaluated on White faces, and transfer in the reverse direction.
However, because there is no gain in the first place even within faces of the same race, this transfer analysis gave no information about the shape of individual differences.
What SCUT-FBP5500 contributes to this study is rather the precision of the sign check (Fig.~\ref{fig:sign} in the main text).
This is because each rater's weights can be determined from as many as 4,400 face ratings, so the direction of the weights can be judged reliably.

\textbf{Difference between the two photograph datasets (EVA and AADB).}
As described in the main text, in EVA \Dt{} exceeded \Do{} (at 50 ratings or more; Supplementary Fig.~\ref{sfig:gen}e), whereas in AADB the advantage of \Dt{} was small and did not stably exceed 0.
The difference between these two datasets could not be explained by the difference in the number of raters.
This is because \dTO{} remained, at $+0.010$ (training photographs, $P = 0.006$)/$+0.017$ (unseen photographs, $P < 10^{-4}$), even when the EVA raters were reduced to 68 for each seed.
Besides the number of raters, AADB and EVA differ in the number of ratings per photograph (5 versus 30 or more), the rating scale (integers 0--5 versus 0--10) and the composition of the stimuli (miscellaneous photographs from Flickr versus photographs balanced across six AVA categories).
Which of these causes the difference cannot be determined from the present data.

\FloatBarrier
\subsection*{Supplementary Note 3: Jokes (Jester)}
Jester (11,067 raters, 150 jokes) is the dataset in which the predictive power of the population mean is the lowest of all the datasets (0.34) and shared preferences are the weakest.
We used these data to check whether \Dt{} gains an advantage when shared preferences are weak.
As a result, the advantage of \Dt{} was small in these data too.

The individualisation gain was intermediate in size between faces and artworks ($+0.039$ with 20 jokes, $+0.056$ with 50 jokes).
When the dimensionality of the model was increased, this gain grew to $+0.047$/$+0.070$ at $D = 16$, $M = 8$, but was almost unchanged at $+0.048$/$+0.073$ at $D = 32$, and thus saturated (Supplementary Fig.~\ref{sfig:curves}e).
In contrast, \dTO{}, although significant ($P < 10^{-16}$), was small, at $+0.004$--$0.007$.
Note that transfer to unseen jokes did not hold, but this is because the sentence embedding predicts the population mean of the jokes only weakly ($r = 0.21$).

From the sign check, we can say that a tail exists for jokes too, but that it could be identified only weakly in these data.
In Jester, for the 10,849 raters included in the sign check, the ratings available for estimating the weights were few, a median of 70 jokes per rater, and the estimate of each rater's weights was unstable (split-half reliability 0.26).
After correction for this noise, the mean reversed energy (the size of the component of the weights that points opposite to the population) did not exceed 0 ($-0.17$).
However, among the 5,517 raters for whom \dTO{} of the amortised models was available (those with 50 estimation ratings), in the top quartile of reversed energy (1,378 raters) the loss under the \Do{} constraint was $+0.065$ and \dTO{} was $+0.009$, and this \dTO{} was larger than in the other quartiles ($+0.001$--$0.002$).

Low-dimensional models ($D = 8$--32) do not fully capture the individual differences in jokes.
This is because the gain of model-free collaborative filtering reached $+0.06$ with 20 jokes, $+0.10$ with 50 jokes and $+0.13$ with 100 jokes, exceeding the gain of the model in this study.
In addition, when restricted to jokes on which raters' opinions are divided, the gain increased from $+0.056$ to $+0.077$.
The proportion of tied pairs, not ordered by the population mean, for which the rater's own ordering was predicted correctly was 0.54 (Supplementary Fig.~\ref{sfig:effect}).

\FloatBarrier
\subsection*{Supplementary Note 4: Olfaction (Keller \& Vosshall)}
The odour pleasantness ratings (55 raters $\times$ 480 molecules, 2 concentrations) are data in which no individualisation gain was obtained even though individual differences appear large.
The residual after removing the rater means and the molecule means accounted for 68\% of the variance.
However, the retest correlation of the residuals when the same molecule was presented repeatedly was only 0.31.

Indeed, the gain of the two-factor model (hybrid embedding based on RDKit descriptors) was at most $+0.003$ on the rater's training molecules (100 ratings).
This result was the same when the settings of the model were changed.
Even when the regularisation of the item embedding was corrected, the gain was $-0.002$ (training molecules)/$+0.006$ (unseen molecules, $P = 0.15$), no different from the population, and the result did not change when the two concentrations were treated as separate stimuli.
Furthermore, even with collaborative filtering that does not depend on the model of this study (ridge regression of the rater's residuals on the residual profiles of the other 49 raters\cite{koren2009}), the gain was only $+0.010$ with 200 ratings.
Transfer to unseen molecules was capped by the fact that the descriptors predict the population mean of unseen molecules only at $r = 0.27$.
From the above, we can say that in these data, with 55 raters and single ratings, low-dimensional individual differences that can be predicted from other people are either small or undetectable.

\FloatBarrier
\subsection*{Supplementary Note 5: Details of Xu et al.\ Experiment 1 and evaluation of statistical power with simulated participants}
Experiment 1 (1,751 participants), which used only White faces, is, together with the 500 faces of SCUT, an exception among the face data in which \Dt{}, which allows a linear transform, exceeded the same-sign reweighting model \Do{} (main text).
This Note first presents that result in detail and then uses simulated participants to examine whether the difference between Experiment 1 and Experiment 2 can be explained by a difference in statistical power (the ability of the model to find individual differences that really exist).

In Experiment 1, \Dt{} exceeded \Do{} by $+0.018$ on the participant's own held-out trials (trials not used for estimation) ($P < 10^{-10}$).
This advantage remained, although reduced to $+0.008$, for the unseen face pairs included in the retest task ($P < 10^{-5}$; Supplementary Table~\ref{stab:xu}, Supplementary Fig.~\ref{sfig:exp1}).
In contrast, \Dh{}, which allows a specific deviation for each face, was equivalent to \Dt{} on the held-out trials of Experiment 1, but instead became worse in transfer to unseen faces ($-0.020$ with direct optimisation at $M = 16$).
That is, the face-specific component did not transfer to faces that the person had not yet seen.

\begin{figure}[htbp]
\centering
\begin{minipage}[t]{\textwidth}
\panel{a}\\[1mm]
\includegraphics[width=\textwidth]{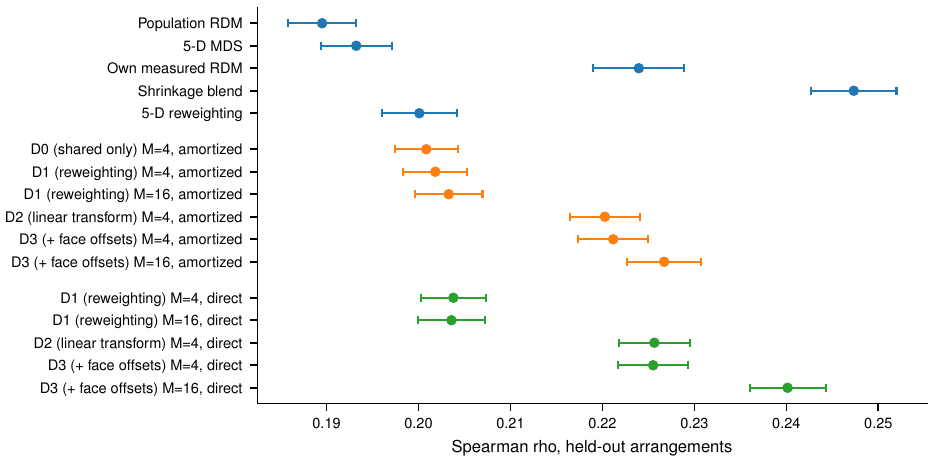}
\end{minipage}\\[3mm]
\begin{minipage}[t]{0.48\textwidth}
\panel{b}\\[1mm]
\includegraphics[width=\textwidth]{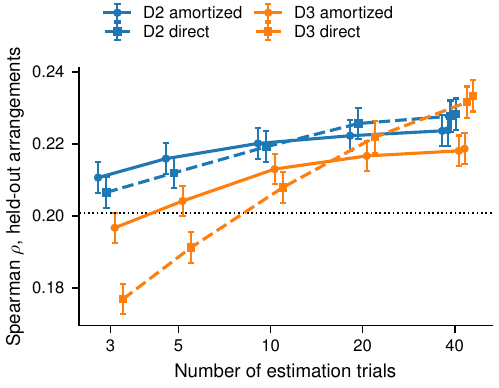}
\end{minipage}\hfill
\begin{minipage}[t]{0.48\textwidth}
\panel{c}\\[1mm]
\includegraphics[width=\textwidth]{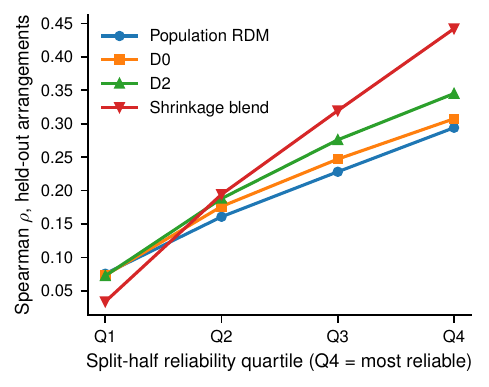}
\end{minipage}
\caption{\textbf{Xu et al.\ Experiment 1 (100 White faces, 1,751 participants).}
\panel{a} Accuracy with which each model predicted the held-out trials of fold-out participants (Spearman $\rho$, mean $\pm$ s.e.m.).
\panel{b} Learning curve when the number of trials used for estimation is varied (\Dt{}, $M = 4$, and \Dh{}, $M = 16$; mean $\pm$ s.e.m.).
Solid lines show amortised inference, dashed lines show direct inference, and the dotted line is \Dz{}.
The points at 42 on the horizontal axis use all estimation trials (mean 42 trials; the points at 40 use at most 40 trials).
For visibility, the series at the same number of trials are slightly offset horizontally.
\panel{c} Held-out $\rho$ in each quartile when participants are divided into quartiles of split-half reliability (population RDM, \Dz{}, \Dt{}, shrinkage blend).
}
\label{sfig:exp1}
\end{figure}

Next, we tested with simulated participants whether the difference between the results of Experiment 1 and Experiment 2 can be explained by a difference in statistical power (Methods).
The simulated participants were generated by imposing individual differences of known content on the same trial structure as the real data.
The imposed individual differences were of two types: \Do{}-type (dimension weights following a log-normal distribution) and \Dt{}-type (a continuous linear transform of rank 4).
To compare the two types fairly, the size of the individual differences was calibrated so that the correlation between the true individual RDM and the population RDM was the same for both types.

In simulated participants whose split-half reliability was matched to the value measured in Experiment 2 (0.32; 169 fold-out participants), \Dt{}-type individual differences could be detected.
On \Dt{}-type data, the \Dt{} model exceeded the population by $+0.013$ (95\% CI [0.010, 0.017], $t(168) = 6.83$, $d_z = 0.53$, $P = 10^{-10}$), whereas the \Do{} model exceeded it by only $+0.004$.
That is, if \Dt{}-type individual differences are present, they appear as a \dTO{} of $+0.009$ with amortised inference (95\% CI [0.006, 0.013], $t(168) = 5.19$, $d_z = 0.40$, $P = 6 \times 10^{-7}$) and of $+0.013$ with direct optimisation.
Conversely, on \Do{}-type data, \dTO{} with direct optimisation was $+0.001$ ($P = 0.79$), and no overfitting due to the extra degrees of freedom occurred.
With amortised inference, however, \dTO{} was positive on \Do{}-type data as well ($+0.028$).
This was because, for some seeds, the amortised inference of the \Do{} model stopped using $\bm z$.
The two types are therefore distinguished by \dTO{} with direct optimisation and by the gain of the \Do{} model, and not by the amortised \dTO{} alone.
In the real data of Experiment 2, the gain of \Do{} was large ($+0.039$) and \dTO{} was $+0.002$ (95\% CI [$-0.002$, 0.006]).
This combination does not fit the assumption that Experiment 2 contains continuous \Dt{}-type individual differences of a size comparable to the \Do{}-type ones, and is hard to explain by a lack of statistical power.
This control also has a limitation.
Even after the reliability was matched, the held-out $\rho$ of the population RDM was higher in the simulated data than in the real data (0.45 versus 0.30).
The simulated data are thus easier to predict than the real data, and the statistical power shown here is an estimate towards the upper end.
Note that the extent to which a case in which only some raters reverse discretely can be detected is examined with the sign check and the LAPIS-type control in the main text.

In contrast, in simulated participants whose split-half reliability was matched to the value measured in Experiment 1 (0.22; 176 fold-out participants), individual differences were difficult to recover.
This is because the \Do{} model did not exceed the population on either type of data ($+0.000$), the gain of the \Dt{} model was $+0.014$ or less, and even the oracle given the true $\bm z$ reached only $+0.05$--$0.06$.
Two things can be said about Experiment 1 from this result.
First, because the absence of a \Do{} gain in Experiment 1 lies within the limits of statistical power, we cannot conclude that reweighting-type individual differences are scarce in Experiment 1.
Second, this control cannot determine which type of individual difference gives rise to the \dTO{} $= +0.018$ (95\% CI [0.014, 0.023]) observed in Experiment 1.
On \Dt{}-type data, \dTO{} was $+0.008$ (95\% CI [0.006, 0.011]), smaller than the observed value, but on \Do{}-type data the amortised \dTO{} was also $+0.015$ (95\% CI [0.011, 0.018]), because the amortised inference of the \Do{} model did not use $\bm z$.
With direct optimisation, \dTO{} was $+0.012$ on \Dt{}-type data and $+0.003$ on \Do{}-type data ($P = 0.29$).
However, the content of the \Dt{}-type structure in Experiment 1 (a small mixing of dimensions in which the diagonal components dominate) cannot be identified from the present data, and its replication and interpretation remain for future work.

These conclusions did not change whichever setting of $D \in \{5, 8, 16\}$ and $M \in \{2, 4, 8\}$ was chosen (Methods).
We also confirmed with simulated participants that the encoder correctly recovers the weights used for generation (canonical correlation 0.89/0.77).

\FloatBarrier
\subsection*{Supplementary Note 6: Auxiliary analyses for the sign check}
This Note presents analyses that supplement the sign check in the main text (the analysis that examines, without relying on amortised inference, whether each rater's weights point in the direction opposite to the population), together with numbers not given in the main text.

\textbf{Absorption of reversals by \Do{} with a learned basis.}
A small \dTO{} alone does not show that there is no reversal (a weight whose sign is opposite to that of the population).
This is because \Do{} learns its basis, so it can duplicate an axis with the same content as two dimensions of opposite sign and thereby absorb discrete reversals of some raters into same-sign reweighting (Methods).
In LAPIS, this absorption actually occurred.
The embeddings of the \Do{} fold models for LAPIS had more capacity to read out the abstract direction in the sense opposite to the population (0.22) than in the same sense (0.09).
And for 16\% of the fold-out raters, the readout of the abstract direction by \Do{} was in fact opposite to the population (17\% for \Dt{}).
Therefore, \dTO{} $\approx 0$ does not mean that there is no reversal.
For this reason, the presence or absence of reversals was judged by the sign check, which does not rely on the comparison of amortised models.
In addition, because \Do{} has already absorbed part of the reversals, the size of \dTO{} should be regarded as a lower bound on the spread of the weights.
Note that the loadings of the first variation axis in LAPIS agreed with the correlations between each dimension and abstract preference at $r = -0.98$, which confirmed that this axis is the abstract--figurative direction.

\textbf{The small advantage of \Dt{} with 500 faces in SCUT-FBP5500.}
The small advantage of \Dt{} that appeared in SCUT when estimating from 500 faces (\dTO{} $+0.006$) was not due to reversals.
In SCUT, each rater's weights can be determined from 4,400 training faces, so the weight estimates were very stable (split-half reliability of the weights 0.99), and the reversed energy was 0.007.
What this advantage corresponded to was a continuous deviation of the weight direction from the population (the correlation between \dTO{} and the cosine with the population weight was $-0.67$).
Indeed, only 3\% of raters had a loss under the \Do{} constraint exceeding 0.02.

\textbf{Details of the simulated raters.}
We give the numbers for the two kinds of simulated raters constructed in the synthetic control of the main text.
One kind consisted of simulated raters whose weights spread continuously, and these qualitatively reproduced all the features seen in LAPIS (the sizes were comparable but not identical).
That is, \dOZ{} was $+0.130$ and \dTO{} was $+0.032$; \dTO{} increased monotonically across quartiles of reversed energy from $+0.013$ to $+0.047$; the loss under the \Do{} constraint was $+0.037$; and 5\% of raters had a negative cosine.
In addition, \dTO{} was $+0.030$ on unseen artworks as well, and 42.8\% of raters had a loss under the \Do{} constraint exceeding 0.02.
The corresponding values in LAPIS are \dOZ{} $+0.123$, \dTO{} $+0.021$ (from $+0.009$ to $+0.035$ across quartiles), a loss of $+0.046$, 9.6\% of raters with a negative cosine and 53\% with a loss exceeding 0.02.
The other kind consisted of simulated raters in which 10\% of the raters are discretely reflected (the signs of their weights are reversed).
In this case, \dOZ{} was $+0.058$, and \dTO{} was small at $+0.005$ and flat regardless of the quartile of reversed energy.
In contrast, 10.4\% of raters had a loss under the \Do{} constraint exceeding 0.02, which nearly matched the true value of 10\%, so the reflected group was correctly detected.
Comparing the reflected group with the remaining raters, there was almost no difference in \dTO{} ($-0.004$), whereas the loss under the \Do{} constraint was 0.25 larger in the reflected group.
In other words, what \dTO{} detects is a continuous spread; discrete reversals hardly appear in \dTO{} and instead appear in the loss under the \Do{} constraint of the sign check.

\textbf{Content of the tail in EVA and age.}
In EVA, individual weights were not concentrated on a single axis but varied along several axes.
The first variation axis accounted for only 40\% of the variance, and the effective number of axes was 3.9.
The $|\cos|$ between this axis and the population weight was 0.30, and the projection of the population weight onto this axis was 0.63 times the standard deviation of the raters' scores.
Consequently, even if one of these axes is reversed, the overall direction of the weights does not change.
Indeed, although 29\% of raters were on the opposite side from the population on the first variation axis, no rater had a negative cosine with the population weight.
Nevertheless, among the 144 raters for whom \dTO{} was available, in the top quartile of reversed energy (36 raters) the reversed energy reached 0.17 (0.13 after noise correction).

Next, we examined what the raters in this tail prefer.
The first variation axis corresponded to a preference for natural landscapes ($\rho = 0.49$), and the raters in the tail disliked the natural landscapes that the majority likes (residuals after removing rater means and image means, $-0.33$ versus $+0.07$, Mann--Whitney $P < 10^{-6}$).
In addition, they also disliked architecture and urban scenes ($-0.20$ versus $+0.01$, $P = 7 \times 10^{-4}$) and liked still lifes and close-ups ($+0.08$ versus $-0.10$, $P = 0.007$).
As for the relation with rater attributes, reversed energy was negatively correlated with year of birth (Spearman $-0.22$, $P = 0.003$), so that older raters were more likely to be in the tail.
In contrast, it was unrelated to experience with photography.

\textbf{How the figure of the first variation axis (Fig.~\ref{fig:sign}c,d) was made.}
The rater scores shown in the figure are the per-rater weights (ridge estimates that do not rely on amortised inference) projected onto the first eigenvector of the between-rater covariance of the weights.
These scores were computed in the basis of each \Dt{} fold model, scaled to SD units within each model, and then pooled over 20 models for LAPIS and 30 models for EVA.
The sign was chosen so that the projection of the population weight is positive.
The proportion of raters on the opposite side from the population is 27\% for LAPIS and 31\% for EVA in the pooled distributions.
The values in the main text (28\%, 29\%) differ slightly from these because the main text reports the median of the per-fold-model values.

\textbf{How much the loss under the \Do{} constraint and the cosine change with the choice of basis.}
The \Do{}-constrained readout is the population weight with each dimension multiplied by a non-negative scale, and it is defined on the standardised embedding basis of each \Dt{} fold model.
Because this constraint concerns the sign of each dimension, the value of the loss can change with the choice of basis even within the same eight-dimensional space.
We therefore recomputed the loss, keeping the same data and the same split halves, after replacing the basis with random orthogonal rotations (five of them) and with the principal component axes of the stimulus features.
The mean loss changed from $+0.046$ to $+0.055$--$0.061$ in LAPIS, from $-0.006$ to $-0.012$--$-0.008$ in EVA, and from $-0.045$ to $-0.041$--$-0.034$ in London, and in SCUT from $+0.002$ to at most $+0.003$.
The proportion of raters with a loss exceeding 0.02 changed from 53\% to 54--60\% in LAPIS, from 25\% to 12--21\% in EVA, from 16\% to 15--21\% in London, and from 3\% to 0--2\% in SCUT.
Thus the values move with the basis, and the proportion for EVA in particular moves considerably, but the ordering did not change: at or below 0 or near 0 for faces, EVA in between, and positive for LAPIS.

Furthermore, as a separate constraint that does not depend on the basis, we also computed the loss for a readout in which the whole population weight is multiplied by a single non-negative scalar.
This readout forbids any deviation from the direction of the population weight, so it is a different and stronger constraint than \Do{}.
The mean loss in this case (proportion exceeding 0.02) was $+0.143$ (82\%) for LAPIS, $+0.004$ (42\%) for EVA, $-0.034$ (23\%) for London, and $+0.014$ (13\%) for SCUT, and the ordering was the same.

In contrast, the cosine with the population weight did not change under orthogonal rotation of the basis (differences below $10^{-6}$).
When the way of standardising the dimensions was changed (re-standardising after rotation, or whitening), the proportion of raters with a negative cosine changed from 9.6\% to 7.1--10.3\% in LAPIS and from 1.6\% to 1.1--1.2\% in London, and remained 0\% in EVA and SCUT.

\textbf{Results with different image features.}
Because the image-based analyses rest on CLIP ViT-B/32 image features, we checked whether the results are specific to these features (Methods).
All numbers listed below are in the order CLIP ViT-B/32 (main analysis), DINOv2 ViT-B/14, CLIP ViT-L/14.

In LAPIS, the predictive performance of \Dz{} was 0.432, 0.411 and 0.436.
\dTO{} was $+0.011$, $+0.008$ and $+0.007$ with 20 ratings and $+0.021$, $+0.016$ and $+0.016$ with 100 ratings, so \Dt{} exceeded \Do{} with every feature set.
In the sign check, raters with a negative cosine were 9.6\%, 8.9\% and 9.6\%, the loss under the \Do{} constraint was $+0.046$, $+0.051$ and $+0.075$, and raters with a loss exceeding 0.02 were 53\%, 60\% and 73\%.
For the first variation axis, the proportion of variance was 64\%, 61\% and 63\%, raters on the opposite side from the population were 28\%, 30\% and 24\%, the bimodality coefficient was 0.49, 0.49 and 0.50, and the Spearman $\rho$ with abstract--figurative preference was $-0.93$, $-0.93$ and $-0.94$.
In other words, the conclusion of a continuous, unimodal spread along the abstract--figurative direction did not change, but the numbers describing the size of the tail moved with the features (raters on the opposite side 24--30\%, raters with a loss exceeding 0.02 53--73\%).
For the latter proportion, the value in the main analysis is the smallest of the three feature sets.

In EVA, the predictive performance of \Dz{} was 0.437, 0.319 and 0.440, so the DINOv2 features predicted the photograph ratings poorly.
\dTO{} with 100 ratings (the rater's remaining ratings of training photographs) was $+0.016$, $+0.011$ and $+0.013$, positive with every feature set.
In the sign check, raters with a negative cosine were 0\% with every feature set, the loss under the \Do{} constraint was $-0.006$, $-0.017$ and $-0.008$, and raters with a loss exceeding 0.02 were 25\%, 18\% and 23\%.
\dTO{} in the top quartile of reversed energy was $+0.043$, $+0.021$ and $+0.024$, and raters on the opposite side from the population on the first variation axis were 29\%, 24\% and 28\%.

In AADB, \dTO{} with 100 ratings was $+0.003$, $+0.004$ and $+0.005$ ($P = 0.15$, 0.05 and 0.009).
The relation of being smaller than in EVA was preserved, but because it became nominally significant with CLIP ViT-L/14, the advantage of \Dt{} in AADB is appropriately regarded not as ``absent'' but as ``small and not reliably above 0''.

In the image-based version of London as well, the conclusion for faces did not change.
\dTO{} with 40 faces was $-0.0004$, $-0.0003$ and $-0.0002$, the loss under the \Do{} constraint was $-0.045$, $-0.048$ and $-0.038$, and raters with a negative cosine were 1.6\%, 1.2\% and 1.3\%.

\textbf{Status of the proportion of raters with a negative cosine.}
The proportion of raters whose cosine with the population weight is negative was treated as a secondary measure.
This proportion was 0--2\% for faces, 0\% for EVA and 9.6\% for LAPIS.
It was treated as secondary because, as described in the paragraph on the tail in EVA, the overall direction of the weights does not change even if one axis is reversed, so this proportion can miss reversals.

\textbf{Fashion and landscapes in XPASS-Vis.}
For fashion and landscapes in XPASS-Vis, the eight-dimensional linear model of CLIP features (\Dz{}) did not reach the baseline given by the mean rating of each image (fashion 0.35 versus 0.43, landscapes 0.44 versus 0.48).
In these two domains \dOZ{} was also only $+0.002$ (n.s.) / $+0.009$.
However, because the stimulus representation itself is inadequate, we cannot distinguish whether the gain is small because individual differences are small or because of the limits of the stimulus representation.

\textbf{Sign-check results for each dataset (additions to the main text).}
We add the following numbers to the account of the sign check in the main text.
In the London Set (82 training faces), the reversed energy corrected for estimation noise did not exceed 0 ($-0.14$).
Imposing the \Do{} constraint improved predictive power by 0.045 because the constraint acted as regularisation.
The \dTO{} of the amortised models was also 0 in every quartile of reversed energy (green in Fig.~\ref{fig:sign}b).
In LAPIS, the \dTO{} of the amortised models correlated with the raters' reversed energy (Spearman $+0.19$, $P < 10^{-5}$) and, by quartile of reversed energy, increased monotonically as $+0.009$, $+0.017$, $+0.023$ and $+0.035$ (blue in Fig.~\ref{fig:sign}b).
The correspondence between the scores on the first variation axis and the abstract--figurative preference (the difference obtained by subtracting the ratings of figurative art from the ratings of abstract art) was Spearman $\rho = -0.96$ to $-0.88$ across all 20 fold models.
In EVA, the rater-by-rater distribution of the loss was spread to both sides around 0 (orange in Fig.~\ref{fig:sign}a).
In the quartiles other than the top quartile of reversed energy (36 raters), \dTO{} was $+0.002$--$0.011$.
For jokes (Jester), the tail could be identified only weakly, because the number of ratings per rater was small and the weight estimates were unstable.

\textbf{The same artworks rated by a different sample of raters (XPASS-Vis; additions to the main text).}
The numbers for the individualisation gain, the sign check, the subsampling and the transfer are given in the main text and Table~\ref{tab:xpass}.
Here we add details on the difference between the two rater samples and on the correlations across domains.
The two rater samples showed different properties on the same 2,345 artworks.
The XPASS-Vis raters varied less in their correlation with the population mean and were homogeneous with one another (standard deviation 0.15 versus 0.25 for LAPIS).
Raters who preferred abstract artworks were also few, at 7.0\% (17.0\% in LAPIS), and the within-rater retest correlation was also lower (0.59 versus 0.71).
The size of the tail observed in LAPIS (28\% on the opposite side, 9.6\% with a negative cosine, loss under the \Do{} constraint $+0.046$) is therefore not a constant intrinsic to the artworks.
It is the size of the spread observed in one particular rater sample, predominantly European and recruited by open call from the general public.

The main text shows that the indices of the tail were uncorrelated across domains.
The quantities that correlated across domains were the prediction accuracy of \Do{} ($\rho = 0.29$--$0.40$), the cosine with the population weight (0.19--0.29), the split-half reliability of the weights (0.19--0.28) and the retest correlation (0.26--0.29).
Note that the partial transfer across domains reported by Hayashi and colleagues\cite{hayashi2026} was replicated in this study as transfer of the between-rater similarity structure (Mantel $\rho = 0.31$--$0.46$) and of predictability, but the shape of individual differences did not transfer across domains.

\FloatBarrier
\subsection*{Supplementary Note 7: Details of the interpretation of the latent representation and the decoding of attributes}
We examined what the latent representation $\bm z$ represents, using the information that accompanies each dataset (Supplementary Fig.~\ref{sfig:effect}c, Methods).
In every dataset, $\bm z$ represented well what the person prefers but hardly represented attributes such as gender and age.

In the London Set, we first examined the largest axis of individual differences.
The first principal component of the rater residuals (ratings after removing rater means and face means) corresponded to the gender of the face at $r = -0.91$ (9.5\% of the residual variance).
That is, the largest axis of individual differences was whether a rater prefers male or female faces (Supplementary Fig.~\ref{sfig:supp}).
The $\bm z$ of \Do{} estimated from 40 faces predicted this index of male-face preference with a cross-validated $R^2$ of 0.71.
This value is the mean of the values evaluated within each of the 50 fold models with $D = 8$ and $M = 4$ (range 0.63--0.81).
In contrast, when raters were divided into four groups by gender and sexual orientation and classified from $\bm z$, the balanced accuracy was only 0.33 (chance level 0.25).
Classification is difficult because the two majority groups (women attracted to men and men attracted to women) have similar levels of male-face preference.
Thus, the self-reported attributes could not explain the main axis of individual differences.

\begin{figure}[htbp]
\centering
\begin{minipage}[t]{0.36\textwidth}
\panel{a}\\[1mm]
\includegraphics[width=\textwidth]{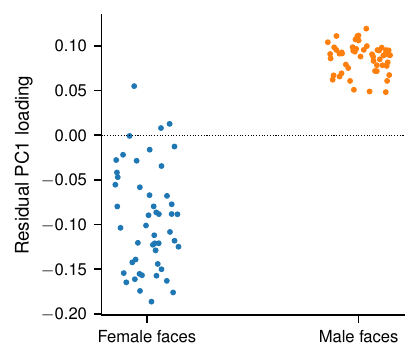}
\end{minipage}\hfill
\begin{minipage}[t]{0.60\textwidth}
\panel{b}\\[1mm]
\includegraphics[width=0.78\textwidth]{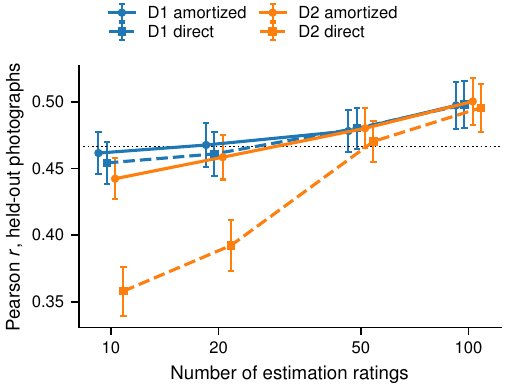}
\end{minipage}
\caption{\textbf{The largest axis of individual differences in the London Set, and the learning curve for AADB on the rater's remaining ratings of training photographs.}
\panel{a} Loadings of each face on the first principal component of the rater $\times$ face residual matrix (9.5\% of the residual variance), shown separately by the gender of the face ($|r| = 0.91$ with the gender of the face).
Because the loadings separate clearly by the gender of the face, this axis represents whether a rater prefers male or female faces.
\panel{b} Pearson $r$ for the rater's remaining ratings of training photographs in AADB (the dotted line is the photograph mean of the training raters; the series at the same number of judgments are slightly offset horizontally).}
\label{sfig:supp}
\end{figure}

The same tendency was seen in LAPIS.
The $\bm z$ estimated from 100 ratings predicted the preference for abstract artworks with an $R^2$ of 0.82 (\Do{}) / 0.85 (\Dt{}) and the preference for portraits with 0.78--0.81.
In contrast, $\bm z$ did not encode attributes or interests ($R^2$ of 0.02 for interest in art (the VAIAK scale) and 0 for age; classification of gender was 0.54 against a chance level of 0.5).

In Xu Experiment 2, the balanced accuracy of classifying the perceiver's race (White, Black, Asian) from $\bm z$ was 0.348, no different from the chance level (chance 0.335, permutation test $P = 0.27$).
Nor did we detect, in the decoded dissimilarities, an in-group effect that compresses or expands pairs of faces of the perceiver's own race ($P = 0.56$).
Gender (0.559, $P = 0.11$) and participant source (university students or online, 0.594, $P = 0.08$) were weakly encoded.
However, this encoding almost disappeared when split-half reliability was removed as a covariate, so it was mediated by differences in data quality.

In AVFS as well, the annotator's gender (0.513), ancestral region (0.410, chance 0.333) and age ($R^2 \approx 0$) could hardly be decoded from the $\bm z$ estimated from 60 triplets.

Across these four datasets, $\bm z$ captured what the person prefers and did not capture demographic attributes or self-reported interests.

\textbf{What the latent representation captures and what it does not.}
The latent representation is a small set of numbers, estimated from each person's judgments, that describes that person's tendencies in judgment.
Consistently across datasets, this latent representation captured what the person prefers (male or female faces, abstract or figurative art, portraits).
In contrast, it did not capture who the person is (race, gender, age, interest in art).
This result reinforces the report of Xu and colleagues from the side of the perceiver.
Xu and colleagues reported that the structure of stereotypes about the race of faces depends on the stimulus set\cite{xu2026}.
In the present study, the perceiver's race could not be decoded from the latent representation, and no in-group effect, in which pairs of own-race faces are treated specially, appeared.
This result does not, however, contradict previous studies that reported relationships between demographic attributes and face impressions\cite{xie2019,hehman2017}.
What our result means is that the variance explained by attributes is small and that the main axes of individual differences are orthogonal to attributes.
Indeed, the largest axis of individual differences in the London Set was a preference for male faces, but this preference was of similar magnitude in the two majority groups defined by rater gender and sexual orientation.
Research on individual differences should therefore probably take as its unit the latent representation estimated from judgments, rather than dividing people into groups by self-reported attributes.

\FloatBarrier
\subsection*{Supplementary Note 8: Comparison with the crowd consensus, and effect sizes on stimuli that the population cannot distinguish}
Because the gains in the main text are expressed as differences in correlation, it is hard to see what their size means.
In this Note we therefore present the gains from two other perspectives.
One is to compare the individual model with the crowd consensus, and the other is to measure the effect only on stimuli that cannot be distinguished by the population mean.

\textbf{Comparison with the crowd consensus.}
AVFS contains a subset in which 24 people judged 1,000 common triplets, and this allows the individual model to be compared directly with the crowd consensus.
Here the crowd consensus is the majority vote of the roughly 23 people other than the person in question (leave-one-out), and the rate at which it agreed with the person's own response was 0.625.
In contrast, the rate at which the \Do{} model estimated from judgments in the main set predicted the person's own responses on the common triplets was 0.642, exceeding the majority vote ($+0.017$, $P = 4 \times 10^{-4}$, $n = 331$).
Meanwhile, the population model \Dz{}, which contains no individual differences (0.609), fell short of the majority vote.
In other words, the latent representation estimated from about 130 of the person's own judgments predicted that person's face similarity judgments better than the crowd consensus did.

\textbf{Effect sizes on stimuli that the population cannot distinguish.}
A gain measured by correlation includes in its denominator the shared component that the population mean can explain, so the effect of individualisation looks small.
We therefore report two measures restricted to the part that the population mean cannot predict, alongside the values for all stimuli (Supplementary Fig.~\ref{sfig:effect}, Methods).

\begin{figure}[htbp]
\centering
\begin{minipage}[t]{\textwidth}
\panel{a}\\[1mm]
\includegraphics[width=\textwidth]{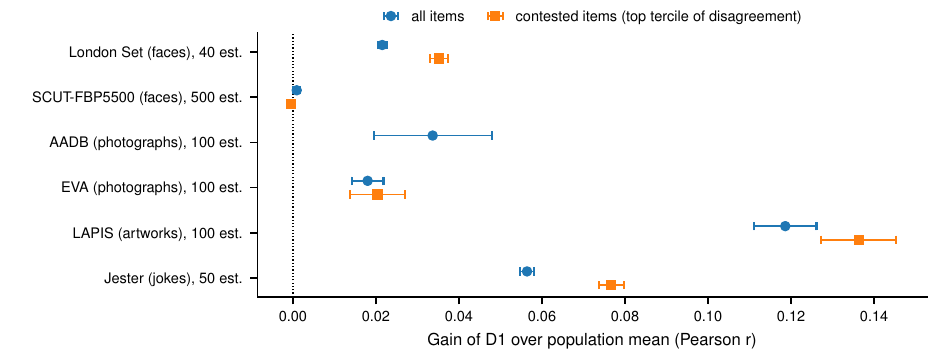}\\[2mm]
\panel{b}\\[1mm]
\includegraphics[width=\textwidth]{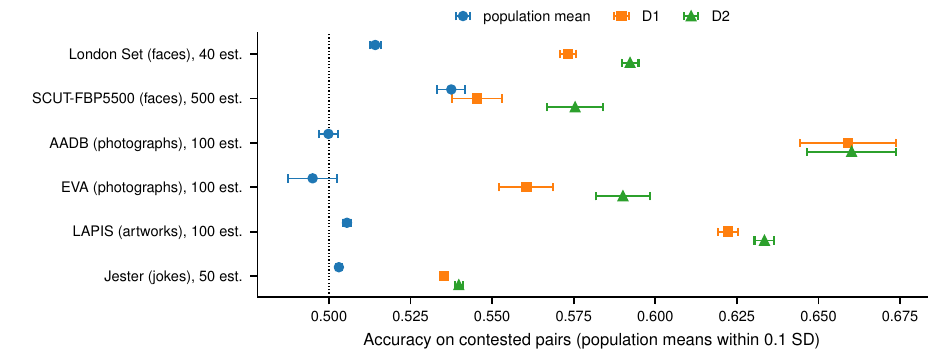}
\end{minipage}\\[3mm]
\begin{minipage}[t]{0.40\textwidth}
\panel{c}\\[1mm]
\includegraphics[width=\textwidth]{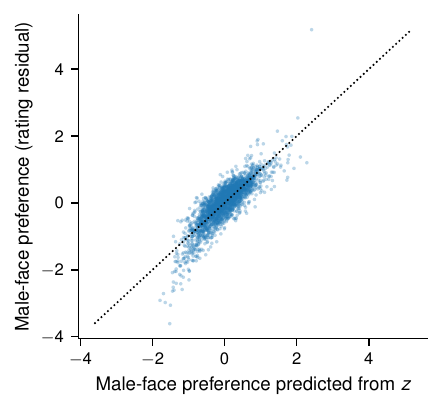}
\end{minipage}
\caption{\textbf{Effect sizes on stimuli that the population cannot distinguish, and what the latent representation encodes.}
\panel{a} How much the \Do{} model exceeds the population mean, shown as the per-rater gain (mean $\pm$ s.e.m., averaged over seeds).
Circles are the gain on all held-out stimuli, and squares are the gain restricted to contested stimuli (stimuli in the top third of between-rater variance among the training raters).
For estimation, the largest number of judgments in each dataset was used (London 40 faces, SCUT 500, AADB/EVA/LAPIS 100, Jester 50).
EVA shows the values for unseen photographs.
AADB was omitted because it has only 5 ratings per photograph, so contested stimuli cannot be defined.
\panel{b} Tied-pair accuracy, that is, the proportion of pairs for which the population mean, \Do{} and \Dt{} predicted the person's own ordering (mean $\pm$ s.e.m.).
The pairs considered are those pairs of the person's held-out stimuli for which the difference in population mean is less than 0.1 SD and the person's own ratings differ (at most 4,000 pairs per person, raters with at least 20 pairs; a tied prediction was scored as 0.5).
\panel{c} Prediction of male-face preference from the latent representation in the London Set.
The vertical axis is the rater's male-face preference index (the difference between the residual ratings of male and female faces).
The horizontal axis is the value of this index predicted by 10-fold ridge regression from the $\bm z$ of \Do{} estimated from 40 faces ($n = 2{,}501$; the dotted line is the identity line).
Prediction was performed separately for each fold model, and for the 10 fold models of seed 0, each rater is shown as one point from the model in which that rater is in the held-out fold.
}
\label{sfig:effect}
\end{figure}

The first measure is the gain restricted to stimuli that are contested among the training raters (stimuli in the top third of between-rater variance).
For this measure only, the gain is the difference of \Do{} from the population mean.
This gain was 1.1--1.6 times the value for all stimuli (London $+0.022 \to +0.035$, LAPIS $+0.119 \to +0.136$, Jester $+0.056 \to +0.077$, EVA unseen $+0.018 \to +0.020$; SCUT remained at 0).

The second measure is tied-pair accuracy.
This is the proportion of pairs for which the model predicted the person's own ordering, among those pairs of the person's held-out stimuli for which the difference in population mean is less than 0.1 SD (that is, the population model cannot order them) and the person's own ratings differ.
This proportion was 0.50--0.54 for the population mean, close to the chance level, whereas for \Do{} it was 0.57 for faces (London), 0.56 (EVA)--0.66 (AADB) for photographs, 0.62 for artworks (LAPIS) and 0.54 for jokes (Jester).
The advantage of \Dt{} for photographs and artworks also appeared in this measure (EVA 0.59, LAPIS 0.63, London 0.59, SCUT 0.58).
In other words, of two stimuli that the population cannot distinguish, the \Do{} model predicted the one the person prefers with a probability of 57\% for faces, 56--66\% for photographs and 62\% for artworks (59\%, 59--66\% and 63\% for \Dt{}).

\textbf{Size of the effects.}
The individualisation gain, that is, the amount by which the individual model exceeds the population model, is small in units of correlation (0.02--0.04 for faces).
However, when evaluated on scales with a clear meaning, this gain is not small.
The predictions of the individual model exceeded the rater's own split-half reliability (Xu et al.\ Experiment 2) and also exceeded the crowd consensus (AVFS).
In addition, even for pairs of stimuli that the population mean cannot rank, the model identified the person's own preference with a probability of 56--66\%.
These effect sizes are reported alongside the values for all stimuli, and the claims are not based only on favourable subsets.

\FloatBarrier
\subsection*{Supplementary Note 9: Relation to previous work (details)}
The question and method of this study lie at the intersection of four lines of research (Supplementary Table~\ref{stab:related}).
In this Note, for each line, we describe what previous work has done and what this study adds.

The first line is three-way multidimensional scaling.
The reweighting hypothesis of this study (\Do{}) corresponds to INDSCAL\cite{carroll1970}, and the linear-transform hypothesis (\Dt{}) to IDIOSCAL and three-mode factor analysis\cite{carroll1972,tucker1972}, and the relative merits of the two have been discussed since the 1970s\cite{carroll1974}.
However, these models took the complete dissimilarity matrix of each individual as input, were fitted to samples of a few dozen people, and were compared by goodness of fit.
They therefore could not be applied to thousands of people with sparse data per person, nor could a hypothesis be selected by how well it predicts individuals not used in training.
In contrast, this study rewrote both models as two-factor generative models consisting of stimulus embeddings and a latent representation of the individual.
We then made them applicable even from 20 judgments per person by amortised inference and compared the two hypotheses by cross-validation.
Furthermore, we investigated the cause of the advantage of \Dt{} by the per-rater sign check and the analysis of the axes of between-rater weight variation.
As a result, this advantage could be explained by ``individual weights spreading continuously along a shared direction, with the tail of the distribution crossing 0'', and the same shape was reproduced in a synthetic control whose structure is known.
In this way, this study gave the result of the model comparison a mechanistic explanation of why it arises.

The second line is research on psychological embeddings.
The attention weights of the generalised context model\cite{nosofsky1986} and PsiZ\cite{roads2019,roads2021} estimate INDSCAL-type weights, but the unit is not the individual but the group (for example, novices and experts).
SPoSE and VICE\cite{zheng2019,hebart2020,muttenthaler2022} aggregate the judgments of about 10,000 people into a single population-level embedding and leave individual differences as future work\cite{hebart2023}.
Closest to this study is the work of Andrews et al.\cite{andrews2023}.
For face similarity judgments, they learned embeddings conditioned on annotator ID and obtained per-individual dimension weights.
However, their method assumes that the annotators are known at training time, and it does not consider shapes of individual differences other than reweighting.
In addition, the weights obtained were used only to examine whether they differ between groups defined by annotator attributes such as gender and race.
In contrast, this study used the same AVFS data, estimated the latent representations of annotators who do not appear in training from a small number of judgments, and showed the following three points.
Namely, that the resulting prediction exceeds the majority vote of the other annotators, that allowing a general metric does not improve prediction, and that demographic attributes can hardly be decoded from the latent representation.

The third line is personalised aesthetic assessment and personalised facial attractiveness prediction in computer vision.
This field has competed on the accuracy of predicting a user's ratings from 10--100 rated images.
The methods used for this include learning an individual's deviation from a generic model\cite{ren2017,whitehill2008}, the use of personality traits and attributes\cite{li2020,yang2022}, meta-learning that treats each user as one task\cite{zhu2022,lin2023}, and task vectors or linear maps of the features of vision--language models\cite{yun2024,ryu2026,maerten2026}.
Research measuring the size of individual differences is also advancing.
For example, it has been reported that artworks are more subjective than photographs and that educational level and experience with photography and art are the main factors behind this difference\cite{chen2025}, and that individual preferences transfer partially between art, fashion and landscapes\cite{hayashi2026}.
This study shares data with these studies (LAPIS, AADB, EVA, SCUT-FBP5500, XPASS-Vis), but the question differs.
Whereas these studies ask about prediction accuracy or the size of individual differences, this study asks about the shape of individual differences.
To this end, this study made the model of the individual a four-dimensional latent representation rather than a fine-tuned network or a high-dimensional linear map, and compared nested hypotheses on the same data.
This allowed us to identify the reason why artworks are more ``subjective'' than photographs not as a difference in the size of the variance but as a difference in the shape of individual differences.
That shape is that individual weights spread along the abstract--figurative direction and that, in the tail of the distribution, the sign of the weight is opposite to that of the population.
In addition, applying the same model to faces, artworks, photographs, jokes and olfaction and comparing the shape of individual differences across domains is also something not found in existing studies that compete on accuracy within a single domain.
Note that collaborative filtering in recommender systems and meta-learning for new users\cite{koren2009,lee2019} rest on the same idea as the amortised inference of this study, in that they obtain a representation of the user from a small number of ratings.
This study used that idea to learn the stimulus axes simultaneously and to compare psychological hypotheses.

The fourth line is psychological research on individual differences in face impressions and aesthetic judgments.
Variance decomposition\cite{honekopp2006,hehman2017,xie2019,martinez2020}, twin studies\cite{germine2015,sutherland2020} and cross-domain comparisons of shared and personal taste\cite{vessel2010,vessel2018,leder2016} have quantified ``how much'' individual difference there is.
This study answered the next question, ``what shape does it take''.
And it showed that weak shared taste in itself does not determine the shape of individual differences (the counterexample being that the advantage of \Dt{} was small for jokes, where shared taste is weakest).
In addition, Xu et al.\cite{xu2026} explicitly stated the lack of ``complete individual-level data'' as a limitation of their own study.
In contrast, this study showed that an individual's face-impression space can be recovered from sparse arrangements even without complete data.
Moreover, the recovered space had an accuracy exceeding the reproducibility of the participant's own measurements.

\begin{table}[htbp]
\centering
\caption{\textbf{Comparison of this study with previous work that models individual differences.}
``Shape'' indicates the form in which individual differences are represented.
Weights: same-sign reweighting of shared dimensions; transform: rotation or linear transform of dimensions; free: fine-tuned network or high-dimensional linear map; none: population-level embedding only.
``New individuals'' indicates how the method is applied to individuals who do not appear in training.
``Hypothesis comparison'' indicates whether hypotheses of different shapes are compared by how well they predict unseen individuals on the same data.}
\label{stab:related}
\footnotesize
\setlength{\tabcolsep}{3pt}
\begin{tabular}{p{3.1cm}p{1.9cm}p{1.5cm}p{2.3cm}p{1.3cm}p{1.4cm}p{2.2cm}}
\toprule
Study & Domain & Shape & New individuals & Judgments per person & Hypothesis comparison & Unseen stimuli \\
\midrule
INDSCAL \cite{carroll1970} & Similarity & Weights & No (all individuals fitted jointly) & Complete matrix & --- & No \\
IDIOSCAL, three-mode factor analysis \cite{carroll1972,tucker1972} & Similarity & Transform & No & Complete matrix & Compared by fit \cite{carroll1974} & No \\
Generalised context model \cite{nosofsky1986} & Categorisation & Weights & Fitted per individual & Hundreds of trials & --- & No \\
PsiZ \cite{roads2019,roads2021} & Similarity (triplets) & Weights (per group) & Assigned to a group & Many, as a group & --- & No \\
SPoSE, VICE \cite{zheng2019,hebart2020,muttenthaler2022} & Similarity (triplets) & None & --- & Few (aggregated over the population) & --- & No \\
A View From Somewhere \cite{andrews2023} & Face similarity (triplets) & Weights & No (conditioned on annotator ID) & Median 260 & --- & Yes, via image features \\
Personalised facial attractiveness prediction \cite{whitehill2008,lebedeva2021,lin2023} & Facial attractiveness & Free & Fitted per individual / meta-learning & 15--100 images & --- & Yes, via image features \\
Personalised aesthetic assessment \cite{ren2017,li2020,zhu2022,yun2024,ryu2026} & Photographs, art & Free & Fine-tuning / linear map & 10--100 images & --- & Yes, via image features \\
Collaborative filtering, MeLU \cite{koren2009,lee2019} & Ratings in general & Free (latent factors) & Amortised / meta-learning & A few or more & --- & No (no item features) \\
Individuality transfer (EIDT) \cite{higashi2025} & Decision-making tasks & Transform (policy) & Amortised & Tens of trials & --- & --- \\
This study & Faces, art, photographs, jokes, olfaction (arrangement, triplets, ratings) & Weights / transform / stimulus-specific (nested) & Amortised (10--20 judgments) & From 20 & Compared by prediction of unseen individuals; sign check, variation axes, synthetic control & Yes, via image, text and molecular features \\
\bottomrule
\end{tabular}
\end{table}

\textbf{Background: the size of individual differences.}
The following is known about the size of individual differences.
For example, studies that decomposed the variance of face impressions into face, perceiver and their interaction found that, for some kinds of impression judged from faces (traits), such as how trustworthy or how dominant a face looks, the variance arising from perceivers rivals the variance arising from faces\cite{hehman2017,xie2019}.
In judgments of attractiveness, too, about half of the variance is attributable to personal taste\cite{honekopp2006}.
Twin studies further show that these individual differences in face preferences are shaped mainly by experience rather than by genes\cite{germine2015,sutherland2020}.
Similarly, for aesthetic judgments beyond faces, preferences for faces and landscapes are known to be highly shared across people, whereas preferences for artworks and architecture are shared to a lesser degree\cite{vessel2010,vessel2018}.
Studies that directly compared faces and artworks in the same participants also found that the size of individual differences differed.
Leder et al.\ had 95 women rate 100 abstract paintings and 100 male faces, and showed that the proportion of judgments attributable to personal taste was about 40\% for faces but reached about 75\% for abstract paintings\cite{leder2016}.
In data from about 1,500 twins, the proportion of rating variance arising from individuals was likewise 44\% for faces, compared with 69\% for abstract images (such as fractals; not artworks)\cite{bignardi2024}.
Within artworks, agreement between raters is lower for abstract than for figurative art\cite{schepman2015}.
All of these findings describe ``how much'' individual difference there is, as a proportion of variance.

\textbf{Relation to Xu et al.}
Xu et al.\cite{xu2026} used a multi-arrangement task without word labels and had more than 3,400 participants arrange faces.
They aggregated the participants' arrangements over the population to obtain a low-dimensional structure of face impressions, and showed that this structure (in particular, stereotypes concerning race) changes with the stimulus set.
The question of that paper was thus what structure the population-level face impression space has.
Because each participant arranged only 20 of the 100 faces, obtaining individual-level impression spaces is explicitly stated in that paper as an ideal for the future.
In contrast, we treat the same data at the level of individuals and ask about the shape of individual differences rather than the population space.

\textbf{Origin of the framework.}
The design of our model extends the individuality transfer framework that the author proposed for decision-making tasks\cite{higashi2025,higashi2026}.
The original framework infers a low-dimensional individual representation from behaviour and generates an individualised model with a hypernetwork\cite{ha2017}.
Here we extend this framework to a two-factor design in which the stimulus axis is learned at the same time.
An encoder that infers an individual's representation from a few observations follows the same idea as meta-learning for new users in recommender systems\cite{lee2019} and neural processes\cite{garnelo2018}, but here we use it to compare psychological hypotheses.

\FloatBarrier
\subsection*{Supplementary Note 10: Limitations (details)}
The main text lists four limitations.
This Note describes two limitations not covered in the main text and adds details to the limitations given there.

The first limitation not covered in the main text concerns effect sizes and the upper bound on prediction.
Because the effect sizes in units of correlation are small, we showed their meaning by comparing them with three benchmarks: the ceiling, the crowd and tied pairs.
For practical prediction, however, the image features (CLIP) set the upper bound on transfer to unseen stimuli (with population ridge regression: London 0.42, olfactory descriptors 0.27, joke sentence embeddings 0.21).
In addition, the main analyses used a single type of image feature, CLIP ViT-B/32.
When the image features were replaced by two other types, the qualitative conclusions held, but the numerical values describing the size of the tail shifted (in LAPIS, 24--30\% of raters on the opposite side from the population and 53--73\% of raters with a loss under the \Do{} constraint exceeding 0.02; Supplementary Note~6).
Moreover, the low-dimensional models ($D = 8$--32) do not capture all of the individual differences in jokes.
Indeed, collaborative filtering reached a larger gain than the model of this study ($+0.10$ with 50 jokes, against $+0.07$ for the model).

To the fourth limitation in the main text (the scope of the data), we add two points.
Olfaction is a dataset of 55 people and yields only a null result, in which no individualisation gain is seen.
SCUT-FBP5500 is a dataset from 60 Chinese raters, and its contribution is that the weights can be estimated densely, not an estimate of the gain.

To the third limitation in the main text (model comparison alone cannot determine whether reversals are present), we add the following.
Why \dTO{} $\approx 0$ does not show that there are no reversals is explained in Supplementary Note~6.
Conversely, \dTO{} $> 0$ gives a lower bound, indicating that there is at least that much structure that cannot be expressed by same-sign reweighting.
For this reason, we judged the presence of reversals not by comparing amortised models but by the proportion of negative cosines and the loss under the \Do{} constraint, and the results of these two indicators agreed.
Of these, the cosine does not change when the basis is rotated.
The loss under the \Do{} constraint is a per-dimension constraint, so its value shifts with the choice of basis, but the ordering of the datasets did not change whether the basis was randomly rotated or replaced by the principal component axes (Supplementary Note~6).
Revising the model so that it takes the variation axes as its basis remains for future work.

The second limitation not covered in the main text concerns the interpretation of the stimulus-specific offsets (\Dh{}).
In low-noise data, \Dh{} can show a gain even when there are no individual differences.
For this reason, making the positive claim that \Dh{} yields a gain requires a same-individual control.
However, the statement about \Dh{} in this study (that it does not transfer for faces) is not such a positive claim, and so does not depend on this issue.

\FloatBarrier
\subsection*{Supplementary Note 11: Established conclusions and exploratory observations}
The conclusions of this study can be divided into two kinds according to the nature of the evidence.
One kind comprises conclusions that replicated across multiple datasets and were confirmed with controls or a separate sample.
The other kind remains at the level of exploratory observation.
First, the confirmed conclusions are the following five.
(1) A person's judgment space can be recovered from 10--20 judgments, and this recovery transfers to unseen stimuli (four face datasets, artworks, two photograph datasets).
(2) For the face judgments analysed, same-sign reweighting of shared axes is sufficient.
The sign check, which does not rely on amortised inference, also detects no reversals (London Set and the dense SCUT-FBP5500).
(3) For the artworks of LAPIS, individual weights show a continuous, unimodal spread along the abstract--figurative direction, and in its tail the weights of some raters are opposite in sign to the population.
The advantage of the reversal-capable model corresponds to this shape (analysis of variation axes, diagnosis of the \Do{} basis, synthetic control on real image features).
(4) The size of this tail differs between rater samples even for the same artworks (LAPIS and XPASS-Vis).
(5) The latent representation encodes what a person prefers and does not encode demographic attributes (four datasets).

In contrast, the following points are exploratory observations and go no further than generating hypotheses.
First, the proportion of raters whose loss under the \Do{} constraint exceeds 0.02, and \dTO{}, increased in the order faces, photographs, artworks.
However, this ordering rests only on one artwork dataset (and one separate sample) and two photograph datasets (whose results diverged).
This ordering is therefore not established as a general property of the domains.
In addition, the following four points are each observations obtained post hoc from a single dataset.
These are the cause of the difference between rater samples (the origin of the homogeneity), the content of the raters in the tail of EVA (who dislike landscapes and architecture) and its relation to age, the small advantage of the linear transform in Xu et al.\ Experiment 1, and the weak tail for jokes.
Note that this study did not test how the shape of individual differences is formed through experience.

\textbf{The divergence between the two photograph datasets.}
Finally, we report as a fact, without an explanation, that the advantage of the reversal-capable model diverged between the two photograph datasets.
The difference in the number of raters is not the cause; the remaining candidates are the number of ratings per photograph, the rating scale and the composition of the stimuli.
In future work, attempting a replication with a photograph dataset accompanied by Big Five scores (PARA\cite{yang2022}) would allow this discrepancy to be resolved and the correspondence between the shape of individual differences and personality to be tested.

\textbf{Exceptions among the face datasets.}
There were two exceptions among the face datasets.
In Xu et al.\ Experiment 1, with White faces only, the model that allows mixing of dimensions was slightly better ($+0.018$).
However, what this advantage reflects cannot be identified from the present data, and its replication and interpretation remain for future work.
In SCUT, too, the same model was slightly better when 500 faces were used ($+0.006$), but there was no difference with 100 faces or fewer.
In contrast, face-specific offsets (\Dh{}), which allow person-specific deviations for each face, sometimes improved prediction for faces that the person had judged, but that improvement did not transfer to unseen faces.
Therefore, in none of the datasets did we obtain evidence for face-specific personal evaluations that do not lie in the shared space.

\ifmainpart\else
\FloatBarrier
\renewcommand{\refname}{Supplementary References}
\bibliography{refs}
\fi
\fi %

\end{document}